\documentclass[aps,prd,amsmath,amssymb,superscriptaddress,preprintnumbers,twocolumn,10pt,nofootinbib]{revtex4-1}
\usepackage{verbatim}
\usepackage[T1]{fontenc}
\usepackage[utf8]{inputenc}
\usepackage[american]{babel}
\usepackage{enumitem}
\usepackage{transparent}
\usepackage{siunitx}
\usepackage{graphicx}
\usepackage{dcolumn}
\usepackage{bm}
\usepackage{amssymb}
\usepackage{latexsym}
\usepackage{booktabs}
\usepackage{amsmath}
\usepackage{multirow}
\usepackage{url}
\usepackage{footnote}
\usepackage{float}
\usepackage{threeparttable}
\usepackage{xcolor}
\definecolor{navyblue}{rgb}{0.0, 0.0, 0.5}
\definecolor{royalblue}{rgb}{0.25, 0.41, 0.88}
\definecolor{cadmiumgreen}{rgb}{0.0, 0.42, 0.24}
\definecolor{blue-violet}{rgb}{0.54, 0.17, 0.89}
\definecolor{darkviolet}{rgb}{0.58, 0.0, 0.83}
\definecolor{orange(colorwheel)}{rgb}{1.0, 0.5, 0.0}

\usepackage[normalem]{ulem}
\usepackage{color}
\usepackage{array}
\usepackage{enumerate}
\usepackage{adjustbox}

\usepackage{makecell}
\usepackage{diagbox}
\usepackage{epstopdf}
\usepackage{epsfig}
\usepackage{longtable}
\usepackage{supertabular}
\usepackage{algorithm}
\usepackage{pifont}
\usepackage{algorithmic}
\usepackage{changepage}
\usepackage{setspace}

\usepackage{hyperref}
\hypersetup{
    colorlinks=true,
    linkcolor=royalblue,
    citecolor=magenta
}

\begin{document}

\title{Robust Preference for Dark Sector Interactions} %\TN{My advisor's suggestion.}

\author{Tian-Nuo Li}%\email{litiannuo@stumail.neu.edu.cn}
\affiliation{Liaoning Key Laboratory of Cosmology and Astrophysics, College of Sciences, Northeastern University, Shenyang 110819, China}

\author{William Giar\`e}%\email{giare@hawaii.edu}
\affiliation{Department of Physics and Astronomy, University of Hawai‘i, Honolulu, HI 96822, USA}

\author{Guo-Hong Du}
\affiliation{Liaoning Key Laboratory of Cosmology and Astrophysics, College of Sciences, Northeastern University, Shenyang 110819, China}

\author{Yun-He Li}
\affiliation{Liaoning Key Laboratory of Cosmology and Astrophysics, College of Sciences, Northeastern University, Shenyang 110819, China}

\author{Eleonora Di Valentino}%\email{e.divalentino@sheffield.ac.uk}
\affiliation{School of Mathematical and Physical Sciences, University of Sheffield, Hounsfield Road, Sheffield S3 7RH, United Kingdom}

\author{\\Jing-Fei Zhang}%\thanks{Corresponding author}\email{ jfzhang@mail.neu.edu.cn}
\affiliation{Liaoning Key Laboratory of Cosmology and Astrophysics, College of Sciences, Northeastern University, Shenyang 110819, China}

\author{Xin Zhang}\thanks{Corresponding author}\email{zhangxin@neu.edu.cn}
\affiliation{Liaoning Key Laboratory of Cosmology and Astrophysics, College of Sciences, Northeastern University, Shenyang 110819, China}
\affiliation{MOE Key Laboratory of Data Analytics and Optimization for Smart Industry, Northeastern University, Shenyang 110819, China}
\affiliation{National Frontiers Science Center for Industrial Intelligence and Systems Optimization, Northeastern University, Shenyang 110819, China}

\begin{abstract}
\noindent

Recent DESI baryon acoustic oscillation (BAO) data reveal deviations from the standard $\Lambda$CDM cosmology, commonly interpreted as evidence for dynamical dark energy (DE). We show that these deviations can instead be explained by interactions within the dark sector, without invoking a phantom-crossing DE equation of state. Using a unified computational framework, we analyze two representative interacting dark energy (IDE) models---coupled quintessence and a coupled fluid---and confront them with the latest cosmic microwave background (Planck, ACT, SPT), DESI BAO, and type Ia supernova data, including the DES-Dovekie recalibration. In both IDE realizations, we find a robust preference for non-vanishing interactions, with marginalized constraints deviating from the non-interacting $\Lambda$CDM limit at the $3$-$5\sigma$ level. This preference persists under the DES-Dovekie recalibration, which significantly weakens the evidence for dynamical DE. For the same number of free parameters, IDE models fit current low- and high-redshift data at least as well as, and in some cases better than, the CPL dynamical DE parametrization. Our results identify dark-sector interactions as a viable and physically motivated alternative to dynamical DE, and highlight the key role of future cosmological perturbation probes, such as weak lensing and galaxy clustering, in discriminating between these scenarios.

\end{abstract}

\maketitle

\textit{\textbf{Introduction.}---} 
Recent high-precision baryon acoustic oscillation (BAO) measurements released by the Dark Energy Spectroscopic Instrument (DESI) collaboration~\cite{DESI:2025zgx} appear to challenge the baseline $\Lambda$CDM model of cosmology. When combined with cosmic microwave background (CMB) observations and type Ia supernova (SN) distance measurements, DESI BAO data indicate potential deviations from the $\Lambda$CDM paradigm, often interpreted as a preference for dynamical dark energy (DE). Within phenomenological parametrizations of an evolving equation of state (EoS), such as the Chevallier-Polarski-Linder (CPL) model~\cite{Chevallier:2000qy,Linder:2002et}, this preference reaches a statistical significance in the range $2.8$--$3.8\sigma$~\cite{DESI:2025zgx,DESI:2025fii,DES:2025sig}.

These results have catalyzed significant research activity within the cosmology community, giving rise to an extensive body of work exploring and interpreting the apparent preference for dynamical DE~\cite{DESI:2025fii,DESI:2024mwx,DESI:2025wyn,DESI:2025zgx,Cortes:2024lgw,Shlivko:2024llw,Luongo:2024fww,Gialamas:2024lyw,Wang:2024dka,Ye:2024ywg,Tada:2024znt,DESI:2024kob,Bhattacharya:2024hep,Ramadan:2024kmn,Malekjani:2024bgi,Giare:2024gpk,Reboucas:2024smm,Park:2024pew,Li:2024qus,Jiang:2024xnu,Yang:2024kdo,Li:2025vuh,Wolf:2025jed,Shajib:2025tpd,Giare:2025pzu,Chaussidon:2025npr,Pang:2025lvh,RoyChoudhury:2025dhe,Paliathanasis:2025cuc,Scherer:2025esj,Giare:2024oil,Liu:2025mub,Teixeira:2025czm,Specogna:2025guo,Cheng:2025lod,Herold:2025hkb,Ozulker:2025ehg,Ormondroyd:2025iaf,Silva:2025twg,Ishak:2025cay,Fazzari:2025lzd,Zhang:2025lam,Cheng:2025hug,Cai:2025mas,Song:2025bio,Li:2025ops,Lee:2025pzo,Wu:2025vfs,Santos:2025wiv,Khoury:2025txd,Li:2025ula,Kessler:2025kju,Smith:2025icl}, now widely regarded as the consensus explanation of the DESI findings. From a physical perspective, however, it is important to stress that late-time cosmological observations such as BAO and SN do not measure the DE EoS directly. Rather, they constrain the expansion history of the Universe and the distance-redshift relation through observables inferred from either the apparent angular size of a standard ruler or the flux of a source with known intrinsic brightness. The theoretical predictions for these observables depend crucially on the energy content of the Universe, which from recombination to the present epoch is dominated by dark matter (DM) and DE. Consequently, late-time cosmological observations are primarily sensitive to the redshift evolution of the dark-sector energy densities, rather than to the DE EoS \emph{per se}. Deviations from $\Lambda$CDM may therefore arise from physical mechanisms that modify this evolution, without necessarily requiring an intrinsic dynamics of DE alone. Interpreting such deviations in terms of an evolving DE component, instead, implicitly relies on specific assumptions about the dark sector, most notably that DM is non-interacting and that the energy-momentum tensors of DM and DE are separately conserved.

While these assumptions are commonly adopted in standard analyses and are well-founded in theories such as minimally coupled scalar fields, they do not represent the only possibility for dark-sector dynamics. As a well-motivated generalization of the standard non-interacting framework, a wide class of theoretically motivated scenarios also accommodates interactions within the dark sector. Such interactions can arise, for instance, in extended scalar-field frameworks \cite{Amendola:1999er,Wang:2016lxa}, conformal or disformal couplings \cite{Bettoni:2013diz,Zumalacarregui:2013pma}, or effective field theory descriptions of more fundamental high-energy theories \cite{Gubitosi:2012hu,Frusciante:2019xia}.

Allowing for an exchange of energy and momentum between DM and DE leads to a broad class of models commonly referred to as interacting dark energy (IDE); see Ref.~\cite{Wang:2024vmw} for a review. These scenarios have attracted increasing attention in recent years~\cite{Zhang:2007uh,Salvatelli:2013wra,Li:2015vla,Murgia:2016ccp,Kumar:2017dnp,DiValentino:2017iww,DiValentino:2020vnx,Gao:2021xnk,Pan:2023mie,Forconi:2023hsj,Pourtsidou:2016ico,Nunes:2021zzi,Wang:2021kxc,Lucca:2020zjb,Zhai:2023yny,Becker:2020hzj,Hoerning:2023hks,Giare:2024ytc,Escamilla:2023shf,DiValentino:2019ffd,Li:2024qso,Halder:2024uao,Castello:2023zjr,Yao:2023jau,Li:2023gtu,Mishra:2023ueo,Nunes:2016dlj,Silva:2025hxw,vanderWesthuizen:2025rip,Zhang:2025dwu,Wang:2025znm,Li:2025muv,Li:2025owk,Lyu:2025nsd,Yang:2025uyv,Pan:2025qwy}, motivated both by their potential to address theoretical issues such as the coincidence problem~\cite{Chimento:2003iea,Zhang:2005rg,Zhang:2005rj,Dutta:2017kch} and by their ability to alleviate cosmological tensions~\cite{DiValentino:2017iww,Yang:2018euj,Guo:2018ans,Feng:2019jqa,DiValentino:2019ffd,Li:2020gtk,Gao:2021xnk,Gao:2022ahg,DiValentino:2021izs,Lucca:2021dxo,Giare:2024smz}. From a phenomenological perspective, IDE models generically modify both the background expansion history and the growth of cosmological perturbations. At the background level, the net effect of a dark-sector interaction can mimic that of a dynamical DE component, leading to observational degeneracies between IDE models and time-dependent effective EoS parametrizations. Crucially, however, these scenarios remain physically distinct, as they can predict different signatures in the evolution of cosmological perturbations.

In this work, employing a unified computational approach, we investigate two representative realizations of IDE. The first is the coupled quintessence (CQ) scenario first introduced in Ref.~\cite{Amendola:1999er}, in which DE is described by a scalar field interacting with DM through a Lagrangian-defined coupling. Such couplings are not introduced arbitrarily, but can naturally arise in a broad class of scalar-tensor theories and can be equivalently reformulated, via a Weyl rescaling, in terms of modified gravity models such as $f(R)$ theories~\cite{Wetterich:1994bg,Pettorino:2008ez,Wetterich:2014bma}. The second is a phenomenological coupled fluid (CF) framework, in which the interaction is specified at the level of the energy-momentum transfer between the dark sector components. Using a broad set of cosmological observations, we show that both IDE models fit current data at least as well as, and in some cases better than, dynamical DE models, while involving the same number of free parameters. We also find a preference for a non-vanishing interaction in the dark sector, with a statistical significance comparable to, or larger than, that reported for dynamical DE. Importantly, this preference is not washed out by recent SN recalibrations that reduce the apparent evidence for dynamical DE. Overall, our results support an alternative physical interpretation of late-time cosmological observations, in which deviations from $\Lambda$CDM arise from interactions within the dark sector rather than from an intrinsic dynamics of DE. We stress that the CQ scenario does \textit{not} require a phantom crossing DE EoS.

\textit{\textbf{Interacting dark energy models.}---} 
\label{sec2}
%TN:NEW
In the framework of IDE cosmology, we assume a direct, non-gravitational coupling between DE and CDM. Consequently, the covariant conservation equations are modified as
\begin{equation}
  \label{eq:energyexchange} \nabla_\nu T^\nu_{\hphantom{j}\mu,\rm de} = -\nabla_\nu T^\nu_{\hphantom{j}\mu,\rm c}  = Q_\mu,
\end{equation} 
where $T^{\nu}_{\hphantom{j}\mu}$ denotes the energy-momentum tensor and $Q_{\mu}$ corresponds to the energy-momentum transfer vector. A given $Q_{\mu}$ uniquely specifies the energy transfer rate $Q$, its perturbation $\delta Q$, and the momentum transfer rate $f$. 

To systematically analyze these distinct interaction scenarios, we adopt a unified computational framework that standardizes the treatment of perturbations. Instead of deriving the evolution equations individually for each model, we parameterize the perturbations of the energy and momentum transfer rates as linear combinations of the density perturbation $\delta_I$ and the velocity perturbation $\theta_I$ (where $I = \mathrm{de}, \mathrm{c}$) of the dark sectors, given by~\cite{Li:2023fdk}
\begin{align}
  \delta Q & = C_1 \delta_{\mathrm{de}} + C_2 \delta_{\mathrm{c}} + C_3 \theta_{\mathrm{de}}, \label{eq:unified_dQ} \\
  f_k      & = -Q \theta + D_1 \theta_{\mathrm{de}} + D_2 \theta_{\mathrm{c}}. \label{eq:unified_fk}
\end{align}
Here, the coefficients $C_1$, $C_2$, $C_3$, $D_1$, and $D_2$ serve as time-dependent mapping functions. In this work, we study two typical IDE models, namely the CQ model and the CF model. 

In the CQ scenario, DE is described by a scalar field $\phi$, and the interaction is introduced at the Lagrangian level~\cite{Amendola:1999er}. The background energy transfer rate is derived as
\begin{equation}
Q = \beta \rho_{\mathrm{c}} \frac{\sqrt{\kappa}\phi'}{a},
\end{equation}
where $\kappa = 8\pi G$ and $\beta$ is the coupling constant. For the quintessence potential, we adopt the inverse power-law form,
\begin{equation}
U(\phi) = U_0 (\sqrt{\kappa}\phi)^{-\alpha}.
\end{equation}
Thus, this model introduces two additional free parameters, $\alpha$ and $\beta$, compared to the $\Lambda$CDM model. Here, $\alpha$ is the exponent of the inverse power-law potential, and we impose a prior condition of $\alpha > 0$ to ensure a physically viable quintessence behavior. For the CQ model, the parameterized functions in Eqs.~(\ref{eq:unified_dQ}) and (\ref{eq:unified_fk}) are given by $C_1 = \frac{Q}{1+w}$, $C_2 = D_1 = Q$, $C_3 = \frac{a^2 U_\phi}{-\phi'} Q$, and $D_2 = 0$.

In the CF scenario, we treat DE as a fluid with a constant EoS parameter $w$. Lacking a fundamental Lagrangian, the interaction term is constructed phenomenologically, typically by assuming the energy transfer rate $Q$ to be proportional to either $\rho_{\mathrm{de}}$ or $\rho_{\mathrm{c}}$~\cite{Amendola:1999qq,Billyard:2000bh}. We consider a form of $Q$ that has been extensively studied and is favored by observational data~\cite{Li:2024qso,Li:2025muv}, given by
\begin{equation}
Q = \beta H_0 \rho_{\mathrm{de}}.
\end{equation}
The model is characterized by two extra free parameters, $w$ and $\beta$, compared to the $\Lambda$CDM model. For the CF model, the parameterized functions in Eqs.~(\ref{eq:unified_dQ}) and (\ref{eq:unified_fk}) are realized by $C_1 = D_2 = Q$ and $C_2 = C_3 = D_1 = 0$. We employ the extended parameterized post-Friedmann framework to reliably calculate cosmological perturbations across the entire parameter space of IDE models~\cite{Li:2014eha}. Additional details on the theoretical models are provided in the \textit{Supplementary Material}.

\begin{table}[htbp]
\centering
\caption{Fitting results (at the $1\sigma$ confidence level) in the CQ and CF models from the CMB, DESI, and SN data.}
\label{tab1}
\setlength{\tabcolsep}{1.2mm}
\renewcommand{\arraystretch}{1.3}
% \resizebox{\linewidth}{!}{
% \scriptsize
\footnotesize
% \small
\begin{tabular}{lc c c }
\hline 
\hline
Model/Dataset & $\beta$ & $\alpha$ (CQ) or $w$ (CF)  \\
\hline
$\bm{\textbf{CQ}}$ &  &  &  \\
\texttt{CMB+DESI} & $0.0507^{+0.0091}_{-0.0075}$ & $< 0.39$  \\
\texttt{CMB+DESI+PantheonPlus} & $0.0517^{+0.0092}_{-0.0078}$ & $0.41\pm 0.19$  \\
\texttt{CMB+DESI+DESY5} & $0.0561^{+0.0094}_{-0.0068}$ & $0.62\pm 0.17$  \\
\texttt{CMB+DESI+DES-Dovekie} & $0.0527\pm 0.0087$ & $0.46\pm 0.18$  \\
\hline
$\bm{\textbf{CF}}$ &  &  &  \\
\texttt{CMB+DESI} & $-2.37^{+0.82}_{-0.93}$ & $-1.45^{+0.13}_{-0.15}$ \\
\texttt{CMB+DESI+PantheonPlus} & $-2.25^{+0.73}_{-0.63}$ & $-1.46^{+0.16}_{-0.13}$ \\
\texttt{CMB+DESI+DESY5} & $-3.08\pm 0.70$  & $-1.60\pm 0.15$ \\
\texttt{CMB+DESI+DES-Dovekie} & $-2.59^{+0.76}_{-0.62}$  & $-1.52^{+0.16}_{-0.13}$ \\

\hline
\hline
\end{tabular}
\end{table}

\textit{\textbf{Results and discussion.}---} 
We constrain two IDE scenarios (CQ and CF) and compare them with dynamical DE (CPL) and the baseline $\Lambda$CDM model using Bayesian MCMC inference. Theoretical predictions for background and perturbation evolution are computed using the \texttt{IDECAMB} package~\cite{Li:2023fdk}, a modified version of \texttt{CAMB}~\cite{Lewis:1999bs} tailored for IDE scenarios. The parameter inference is performed with \texttt{Cobaya}~\cite{Torrado:2020dgo}. Our analysis combines CMB temperature and polarization measurements from Planck 2018~\cite{Efstathiou:2019mdh,Rosenberg:2022sdy}, ACT DR6~\cite{ACT:2025fju}, and SPT-3G~\cite{SPT-3G:2025bzu}, together with the joint CMB lensing spectrum~\cite{ACT:2025qjh,Carron:2022eyg,ACT:2023dou,ACT:2023kun}. We further include BAO data from DESI DR2~\cite{DESI:2025zgx} and SN measurements from PantheonPlus~\cite{Brout:2022vxf}, DESY5~\cite{DES:2024jxu}, and DES-Dovekie~\cite{DES:2025sig}. For additional details on datasets and analysis methodology, we refer to the \textit{Supplementary Material}.

\begin{figure}[t]
\begin{minipage}{0.38\textwidth}
\centering
\includegraphics[width=\linewidth]{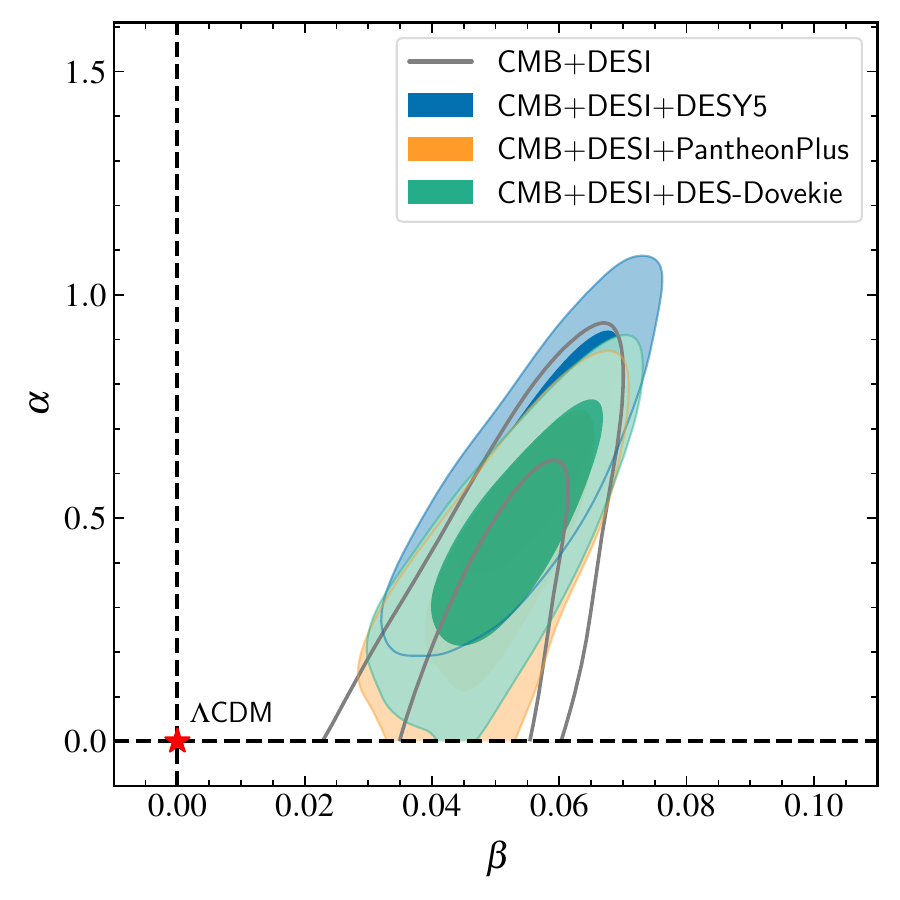}
\end{minipage}
\begin{minipage}{0.39\textwidth}
\centering
\includegraphics[width=\linewidth]{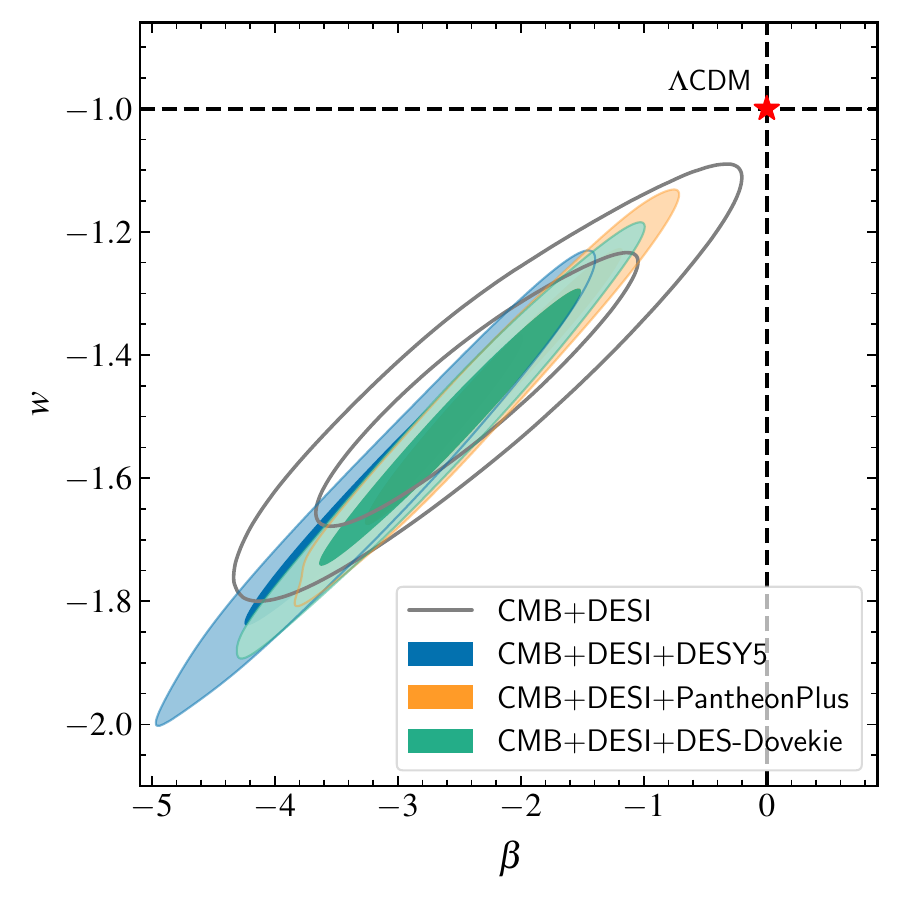}
\end{minipage}
\centering \caption{\label{fig:CQ_and_CF:params} Two-dimensional marginalized contours ($1\sigma$ and $2\sigma$ confidence levels) from DESI, CMB, and SN data. \textbf{\textit{Upper panel:}} Constraints in the $\beta$--$\alpha$ plane within the CQ model. 
\textbf{\textit{Lower panel:}} Constraints in the $\beta$--$w$ plane within the CF model.}
\end{figure}

The marginalized constraints on the coupling parameter $\beta$ and on the additional model parameters (the potential slope $\alpha$ for CQ and the DE EoS $w$ for CF) are summarized in Table~\ref{tab1}, while their two-dimensional posterior distributions are shown in Fig.~\ref{fig:CQ_and_CF:params}. Using CMB and DESI BAO data, both alone and in combination with different SN compilations, we find a consistent preference for a non-vanishing interaction in both IDE realizations. For all considered datasets, the inferred parameters deviate from the values corresponding to the non-interacting $\Lambda$CDM limit (i.e., $\beta = 0$, $\alpha = 0$, $w = -1$) at the $\sim 3-5\sigma$ level, indicating a statistically significant preference for interactions within the dark sector. The robustness of this preference from the data side is tested in the \textit{Supplementary Material}.

\begin{figure}[htpb]
\begin{minipage}{0.4\textwidth}
\centering
\includegraphics[width=\linewidth]{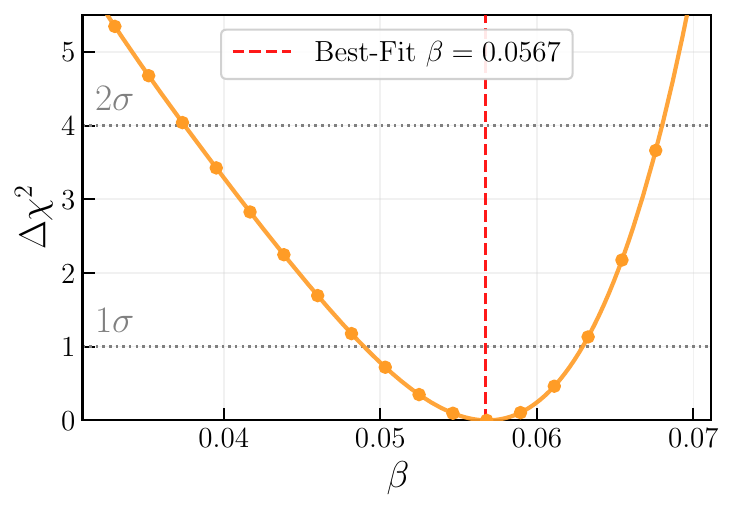}
\end{minipage}
\begin{minipage}{0.4\textwidth}
\centering
\includegraphics[width=\linewidth]{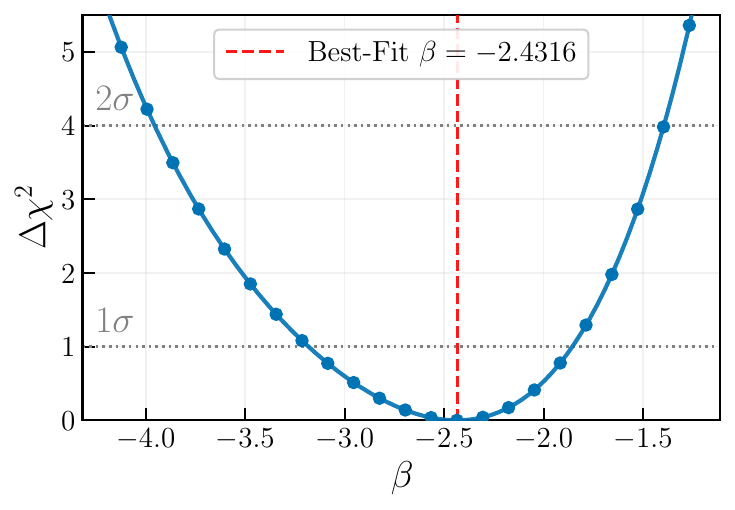}
\end{minipage}
\centering \caption{\label{fig:CQ_and_CF:profile} Using the CMB+DESI+DES-Dovekie data, the profile likelihood for the parameter $\beta$ is presented for the CQ (upper panel) and CF (lower panel) models.}
\end{figure}

To assess the robustness of this preference and to disentangle genuine likelihood-driven effects from possible prior-volume contributions, we compute the profile likelihood for the coupling parameter $\beta$. The results are shown in Fig.~\ref{fig:CQ_and_CF:profile} for the CQ and CF models, using the CMB+DESI+DES-Dovekie dataset. In both cases, the profile likelihood exhibits a well-defined minimum at a non-zero value of $\beta$, significantly displaced from the non-interacting limit. This behavior is fully consistent with the corresponding marginalized posterior of the coupling parameter, and the non-interacting scenario is disfavored at a significance comparable to that inferred from the marginalized posteriors. The associated $\Delta\chi^2$ profiles are approximately parabolic around the minimum, indicating that the preference for interactions is stable and well constrained. 

While a robust preference for interactions emerges in both IDE realizations, their physical origin and phenomenology are markedly different. In the CQ model, DE is described by a scalar field interacting with DM at the Lagrangian level. We find a statistically significant preference for a positive coupling, $\beta > 0$, exceeding the $5\sigma$ level in the marginalized constraints (see Table~\ref{tab1}), corresponding to a net transfer of energy-momentum from DM to the scalar-field component of DE. The constraints on the potential parameter $\alpha$ favor positive values, consistent with an inverse power-law potential that decreases monotonically with the field amplitude. This leads to a sufficiently shallow potential and supports a slowly evolving quintessence field. As a result of the interaction, the DM density dilutes faster than in $\Lambda$CDM, while the DE density is sustained by the continuous energy transfer and remains enhanced at intermediate redshifts, $0.5 \lesssim z \lesssim 2$. The resulting background evolution departs from $\Lambda$CDM in this redshift range and partially mimics the phenomenology of dynamical DE models such as CPL, which display an effective phantom-like behavior at earlier times followed by a present-day quintessence-like phase. Importantly, this similarity is purely effective: despite the modified background evolution, the EoS of the scalar field in the CQ model remains quintessence-like at all redshifts. A detailed illustration of the background evolution is provided in the \textit{Supplementary Material}.

The situation is qualitatively different in the CF scenario. In this case, we find a strong preference for a negative coupling, $\beta < 0$, and a phantom DE EoS, $w < -1$. This combination implies a net transfer of energy from DE to DM. As a result, DE is suppressed at intermediate and high redshifts, while DM is continuously fed by the interaction and dilutes more slowly than in $\Lambda$CDM. This leads to a progressive late-time enhancement of the matter abundance, with present-day values as large as $\Omega_{\mathrm{m}} \sim 0.6$, significantly higher than those inferred in the CQ, $\Lambda$CDM, and CPL scenarios. At the background level, the enhanced DM abundance is compensated by the strongly phantom DE EoS, allowing the model to reproduce the observed distance measurements despite its extreme late-time properties. At the perturbation level, this regime is associated with a strongly suppressed growth of structure, yielding very low values of $\sigma_8$ and $S_8$ compared to those typically obtained in the other models. While this indicates that the CF realization lies in an extreme corner of the IDE parameter space, the corresponding best-fit solutions still provide fits to the CMB, BAO, and SN data considered in this work that are better than those of $\Lambda$CDM and CPL. A detailed discussion of the background and perturbation evolution, together with a more quantitative assessment of the implications for structure formation, is provided in the \textit{Supplementary Material}.

\begin{table}[!htbp]
\centering
\caption{Summary of goodness-of-fit and model comparison metrics for the CPL, CQ, and CF models relative to $\Lambda$CDM. For each data combination we report the $\Delta\chi^2_{\rm MAP}$ contributions from CMB, BAO, and SN, the Total $\Delta\chi^2_{\rm MAP}$, and the $\Delta{\rm DIC}$. Negative values indicate an improvement with respect to $\Lambda$CDM.}
\label{tab:delta_chi2_dic}
\setlength{\tabcolsep}{0.6mm}
\renewcommand{\arraystretch}{1.1}
\small
\begin{tabular}{l|cccc|c}
\hline\hline
 &  \multicolumn{4}{c|}{$\Delta\chi^2_{\rm MAP}$} &  \\
Model/Data & CMB & BAO & SN & Total & $\Delta{\rm DIC}$\\
\hline
\multicolumn{1}{l}{\textbf{CPL}}\\
\texttt{CMB+DESI} & $-7.37$ & $-5.11$ & -- & $-12.48$ & $-11.39$\\
\texttt{CMB+DESI+PantheonPlus} & $-3.83$ & $-4.19$ & $-1.86$ & $-9.88$ & $-8.59$\\
\texttt{CMB+DESI+DESY5} & $-3.69$ & $-6.58$ & $-10.47$ & $-20.74$ & $-18.54$\\
\texttt{CMB+DESI+DES-Dovekie} & $-5.81$ & $-2.91$ & $-5.41$ & $-14.13$ & $-12.22$\\
\hline
\multicolumn{1}{l}{\textbf{CQ}}\\
\texttt{CMB+DESI} & $-5.09$ & $-2.05$ & -- & $-7.14$ & $-6.52$\\
\texttt{CMB+DESI+PantheonPlus} & $-1.93$ & $-4.65$ & $-2.53$ & $-9.11$ & $-7.01$\\
\texttt{CMB+DESI+DESY5} & $-1.29$ & $-5.88$ & $-7.56$ & $-14.73$ & $-12.41$\\
\texttt{CMB+DESI+DES-Dovekie} & $-5.26$ & $-4.08$ & $-4.32$ & $-13.66$ & $-11.31$\\
\hline
\multicolumn{1}{l}{\textbf{CF}}\\
\texttt{CMB+DESI} & $-8.79$ & $-4.12$ & -- & $-12.91$ & $-8.08$ \\
\texttt{CMB+DESI+PantheonPlus} & $-6.35 $ & $-4.41$ & $-2.32$ & $-13.08$ & $-10.17$ \\
\texttt{CMB+DESI+DESY5} & $-5.64$ & $-6.96$ & $-10.36$ & $-22.96$ & $-18.63$ \\
\texttt{CMB+DESI+DES-Dovekie} & $-6.82$ & $-3.59$ & $-5.23$ & $-15.64$ & $-12.89$\\ 
\hline\hline
\end{tabular}
\end{table}

To quantitatively compare the goodness of fit across models, we compute the $\Delta\chi^2_{\rm MAP}$ relative to $\Lambda$CDM and the Deviance Information Criterion (DIC) for a complementary model assessment. The results summarized in Table~\ref{tab:delta_chi2_dic} show that the CF model provides the largest improvement over $\Lambda$CDM, yielding the most favourable values of both $\Delta\chi^2_{\rm MAP}$ and $\Delta{\rm DIC}$ across almost all data combinations. The only exception is the CMB+DESI case, for which the $\Delta{\rm DIC}$ mildly favours CPL, while the $\Delta\chi^2_{\rm MAP}$ still prefers CF. The CPL parametrization consistently ranks as the second-best model, while the CQ scenario typically follows as the third. Importantly, however, despite ranking third, CQ still exhibits a clear and statistically significant improvement over $\Lambda$CDM and remains highly competitive with both CF and CPL. This is particularly noteworthy given that CQ is defined at the level of a well-posed Lagrangian theory, rather than as a purely phenomenological parametrization.

\begin{figure}[htbp!]
\includegraphics[width=\linewidth]{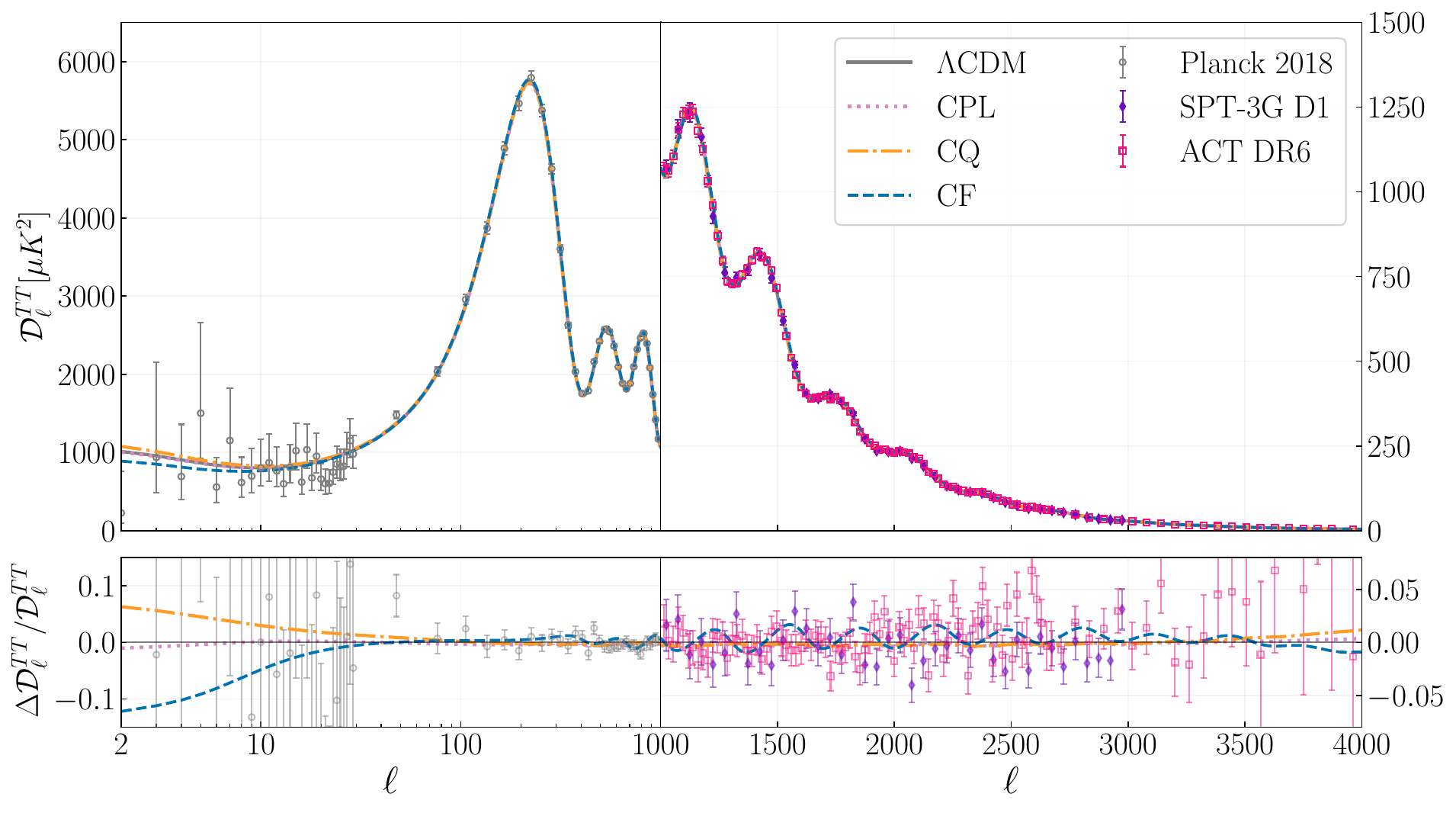}
\centering
\caption{\label{fig:CMB_fit} Comparison of the temperature power spectrum for the $\Lambda$CDM, CPL, CQ, and CF models, using the best-fit values inferred from the CMB+DESI+DES-Dovekie analysis, over the Planck, ACT, and SPT temperature power spectrum data points. We also present the fractional difference between the $\Lambda$CDM and CPL, CQ, or CF models.}
\end{figure}

To better understand the origin of these improvements, it is instructive to examine how the different models fit the underlying observables. In Fig.~\ref{fig:CMB_fit}, we show the CMB temperature angular power spectrum for the best-fit $\Lambda$CDM, CPL, CQ, and CF models, compared against Planck 2018 and ACT DR6 and SPT-3G D1 measurements. The lower panels display the relative residuals, $\Delta D_\ell^{TT}/D_\ell^{TT}$, with respect to $\Lambda$CDM. All models provide an excellent fit to the CMB temperature power spectrum across the full multipole range probed by current data. At large angular scales ($\ell \lesssim 30$), mild differences are visible, especially for the interacting scenarios, reflecting the impact of modified late-time dynamics on the Integrated Sachs-Wolfe effect, though these remain well within the cosmic-variance-dominated uncertainties. At intermediate and small angular scales ($\ell \gtrsim 1000$), all models closely follow the ACT and SPT measurements. In the CF case, the residuals display mildly enhanced oscillatory features in the range $1000 \lesssim \ell \lesssim 2000$ with respect to $\Lambda$CDM and CQ. The amplitude of these oscillations remains at the $\sim 2\%$ level and can be understood in terms of the late-time clustering properties of the CF scenario. The combination of a large present-day matter density and a suppressed growth of structure (low $\sigma_8$ and $S_8$) leads to reduced CMB lensing smoothing, mildly enhancing the contrast of the acoustic features at intermediate multipoles. At higher multipoles ($\ell \gtrsim 2000$), the oscillatory behavior is progressively damped, as photon diffusion suppresses the primary anisotropies and reduces the sensitivity of the temperature spectrum to late-time effects. As a result, differences in the late-time growth of structure and lensing efficiency among the models become negligible, and all scenarios converge to a very similar behavior. Consistent with this picture, both IDE models (CQ and CF), as well as dynamical DE, yield a modest but systematic improvement in the CMB fit with respect to $\Lambda$CDM, with $\Delta\chi^2_{\rm CMB}$ ranging approximately between $-4$ and $-7$, depending on the model and data combination; see also Table~\ref{tab:delta_chi2_dic}. Similarly, the IDE models provide an improved fit to the BAO and SN data relative to $\Lambda$CDM, with $\Delta\chi^2$ values ranging approximately between $-3$ and $-11$. A detailed presentation of the fits to these geometric observables, including the specific contributions to the improvement at different redshifts and the corresponding residual analysis, is provided in the \textit{Supplementary Material}.

Taken together, these results indicate that deviations from $\Lambda$CDM recently reported in the literature can naturally be interpreted in terms of interactions within the dark sector, realized through different interacting phenomenologies such as CQ and CF, without necessarily invoking an \textit{intrinsic} dynamics of DE.

\textit{\textbf{Conclusion.}---}
In this work, we have investigated IDE scenarios as an alternative physical interpretation of the growing evidence for deviations from the standard $\Lambda$CDM cosmology. Focusing on two distinct interacting phenomenologies, CQ and CF, we have found that in both cases the marginalized constraints and the profile likelihood analysis of the interaction parameter $\beta$ show a strong preference for a non-vanishing coupling, reaching the $5\sigma$ level. Importantly, this preference remains stable under recent SN recalibrations that have been argued to reduce the evidence for dynamical DE within the CPL framework.

We have demonstrated that IDE models provide a statistically competitive fit to current cosmological data, matching or even outperforming phenomenological dynamical DE parametrizations with the same number of free parameters. Given the significant degeneracy observed between IDE and dynamical DE models in fitting background and CMB observables, further characterization of their distinct signatures, particularly in the dynamics of cosmological perturbations probed by CMB and LSS surveys, is essential for differentiating between these two scenarios.

\textit{\textbf{Acknowledgments.}---}
We thank Yichao Li and Yi-Min Zhang for their helpful discussions. This work was supported by the National Natural Science Foundation of China (Grants Nos. 12533001, 12575049, and 12473001), the National SKA Program of China (Grants Nos. 2022SKA0110200 and 2022SKA0110203), the China Manned Space Program (Grant No. CMS-CSST-2025-A02), and the National 111 Project (Grant No. B16009). WG acknowledges support from the Lancaster-Sheffield Consortium for Fundamental Physics through the Science and Technology Facilities Council (STFC) grant ST/X000621/1 during the early stages of this project, and from the National Aeronautics and Space Administration (NASA) under Grant No. 80NSSC24K0898 at later stages. EDV is supported by a Royal Society Dorothy Hodgkin Research Fellowship. This article is based upon work from the COST Action CA21136 - ``Addressing observational tensions in cosmology with systematics and fundamental physics (CosmoVerse)'', supported by COST - ``European Cooperation in Science and Technology''.

\bibliography{main}

\begin{thebibliography}{167}%
\makeatletter
\providecommand \@ifxundefined [1]{%
 \@ifx{#1\undefined}
}%
\providecommand \@ifnum [1]{%
 \ifnum #1\expandafter \@firstoftwo
 \else \expandafter \@secondoftwo
 \fi
}%
\providecommand \@ifx [1]{%
 \ifx #1\expandafter \@firstoftwo
 \else \expandafter \@secondoftwo
 \fi
}%
\providecommand \natexlab [1]{#1}%
\providecommand \enquote  [1]{``#1''}%
\providecommand \bibnamefont  [1]{#1}%
\providecommand \bibfnamefont [1]{#1}%
\providecommand \citenamefont [1]{#1}%
\providecommand \href@noop [0]{\@secondoftwo}%
\providecommand \href [0]{\begingroup \@sanitize@url \@href}%
\providecommand \@href[1]{\@@startlink{#1}\@@href}%
\providecommand \@@href[1]{\endgroup#1\@@endlink}%
\providecommand \@sanitize@url [0]{\catcode `\\12\catcode `\$12\catcode `\&12\catcode `\#12\catcode `\^12\catcode `\_12\catcode `\%12\relax}%
\providecommand \@@startlink[1]{}%
\providecommand \@@endlink[0]{}%
\providecommand \url  [0]{\begingroup\@sanitize@url \@url }%
\providecommand \@url [1]{\endgroup\@href {#1}{\urlprefix }}%
\providecommand \urlprefix  [0]{URL }%
\providecommand \Eprint [0]{\href }%
\providecommand \doibase [0]{http://dx.doi.org/}%
\providecommand \selectlanguage [0]{\@gobble}%
\providecommand \bibinfo  [0]{\@secondoftwo}%
\providecommand \bibfield  [0]{\@secondoftwo}%
\providecommand \translation [1]{[#1]}%
\providecommand \BibitemOpen [0]{}%
\providecommand \bibitemStop [0]{}%
\providecommand \bibitemNoStop [0]{.\EOS\space}%
\providecommand \EOS [0]{\spacefactor3000\relax}%
\providecommand \BibitemShut  [1]{\csname bibitem#1\endcsname}%
\let\auto@bib@innerbib\@empty
%</preamble>
\bibitem [{\citenamefont {Abdul~Karim}\ \emph {et~al.}(2025)\citenamefont {Abdul~Karim} \emph {et~al.}}]{DESI:2025zgx}%
  \BibitemOpen
  \bibfield  {author} {\bibinfo {author} {\bibfnamefont {M.}~\bibnamefont {Abdul~Karim}} \emph {et~al.} (\bibinfo {collaboration} {DESI}),\ }\href {\doibase 10.1103/tr6y-kpc6} {\bibfield  {journal} {\bibinfo  {journal} {Phys. Rev. D}\ }\textbf {\bibinfo {volume} {112}},\ \bibinfo {pages} {083515} (\bibinfo {year} {2025})},\ \Eprint {http://arxiv.org/abs/2503.14738} {arXiv:2503.14738 [astro-ph.CO]} \BibitemShut {NoStop}%
\bibitem [{\citenamefont {Chevallier}\ and\ \citenamefont {Polarski}(2001)}]{Chevallier:2000qy}%
  \BibitemOpen
  \bibfield  {author} {\bibinfo {author} {\bibfnamefont {M.}~\bibnamefont {Chevallier}}\ and\ \bibinfo {author} {\bibfnamefont {D.}~\bibnamefont {Polarski}},\ }\href {\doibase 10.1142/S0218271801000822} {\bibfield  {journal} {\bibinfo  {journal} {Int. J. Mod. Phys. D}\ }\textbf {\bibinfo {volume} {10}},\ \bibinfo {pages} {213} (\bibinfo {year} {2001})},\ \Eprint {http://arxiv.org/abs/gr-qc/0009008} {arXiv:gr-qc/0009008} \BibitemShut {NoStop}%
\bibitem [{\citenamefont {Linder}(2003)}]{Linder:2002et}%
  \BibitemOpen
  \bibfield  {author} {\bibinfo {author} {\bibfnamefont {E.~V.}\ \bibnamefont {Linder}},\ }\href {\doibase 10.1103/PhysRevLett.90.091301} {\bibfield  {journal} {\bibinfo  {journal} {Phys. Rev. Lett.}\ }\textbf {\bibinfo {volume} {90}},\ \bibinfo {pages} {091301} (\bibinfo {year} {2003})},\ \Eprint {http://arxiv.org/abs/astro-ph/0208512} {arXiv:astro-ph/0208512} \BibitemShut {NoStop}%
\bibitem [{\citenamefont {Lodha}\ \emph {et~al.}(2025{\natexlab{a}})\citenamefont {Lodha} \emph {et~al.}}]{DESI:2025fii}%
  \BibitemOpen
  \bibfield  {author} {\bibinfo {author} {\bibfnamefont {K.}~\bibnamefont {Lodha}} \emph {et~al.} (\bibinfo {collaboration} {DESI}),\ }\href {\doibase 10.1103/w4c6-1r5j} {\bibfield  {journal} {\bibinfo  {journal} {Phys. Rev. D}\ }\textbf {\bibinfo {volume} {112}},\ \bibinfo {pages} {083511} (\bibinfo {year} {2025}{\natexlab{a}})},\ \Eprint {http://arxiv.org/abs/2503.14743} {arXiv:2503.14743 [astro-ph.CO]} \BibitemShut {NoStop}%
\bibitem [{\citenamefont {Popovic}\ \emph {et~al.}(2026)\citenamefont {Popovic} \emph {et~al.}}]{DES:2025sig}%
  \BibitemOpen
  \bibfield  {author} {\bibinfo {author} {\bibfnamefont {B.}~\bibnamefont {Popovic}} \emph {et~al.} (\bibinfo {collaboration} {DES}),\ }\href {\doibase 10.1093/mnras/stag632} {\bibfield  {journal} {\bibinfo  {journal} {Mon. Not. Roy. Astron. Soc.}\ }\textbf {\bibinfo {volume} {548}},\ \bibinfo {pages} {stag632} (\bibinfo {year} {2026})},\ \Eprint {http://arxiv.org/abs/2511.07517} {arXiv:2511.07517 [astro-ph.CO]} \BibitemShut {NoStop}%
\bibitem [{\citenamefont {Adame}\ \emph {et~al.}(2025)\citenamefont {Adame} \emph {et~al.}}]{DESI:2024mwx}%
  \BibitemOpen
  \bibfield  {author} {\bibinfo {author} {\bibfnamefont {A.~G.}\ \bibnamefont {Adame}} \emph {et~al.} (\bibinfo {collaboration} {DESI}),\ }\href {\doibase 10.1088/1475-7516/2025/02/021} {\bibfield  {journal} {\bibinfo  {journal} {JCAP}\ }\textbf {\bibinfo {volume} {02}},\ \bibinfo {pages} {021} (\bibinfo {year} {2025})},\ \Eprint {http://arxiv.org/abs/2404.03002} {arXiv:2404.03002 [astro-ph.CO]} \BibitemShut {NoStop}%
\bibitem [{\citenamefont {Gu}\ \emph {et~al.}(2025)\citenamefont {Gu} \emph {et~al.}}]{DESI:2025wyn}%
  \BibitemOpen
  \bibfield  {author} {\bibinfo {author} {\bibfnamefont {G.}~\bibnamefont {Gu}} \emph {et~al.} (\bibinfo {collaboration} {DESI}),\ }\href {\doibase 10.1038/s41550-025-02669-6} {\bibfield  {journal} {\bibinfo  {journal} {Nature Astron.}\ }\textbf {\bibinfo {volume} {9}},\ \bibinfo {pages} {1879} (\bibinfo {year} {2025})},\ \Eprint {http://arxiv.org/abs/2504.06118} {arXiv:2504.06118 [astro-ph.CO]} \BibitemShut {NoStop}%
\bibitem [{\citenamefont {Cort{\^e}s}\ and\ \citenamefont {Liddle}(2024)}]{Cortes:2024lgw}%
  \BibitemOpen
  \bibfield  {author} {\bibinfo {author} {\bibfnamefont {M.}~\bibnamefont {Cort{\^e}s}}\ and\ \bibinfo {author} {\bibfnamefont {A.~R.}\ \bibnamefont {Liddle}},\ }\href {\doibase 10.1088/1475-7516/2024/12/007} {\bibfield  {journal} {\bibinfo  {journal} {JCAP}\ }\textbf {\bibinfo {volume} {12}},\ \bibinfo {pages} {007} (\bibinfo {year} {2024})},\ \Eprint {http://arxiv.org/abs/2404.08056} {arXiv:2404.08056 [astro-ph.CO]} \BibitemShut {NoStop}%
\bibitem [{\citenamefont {Shlivko}\ and\ \citenamefont {Steinhardt}(2024)}]{Shlivko:2024llw}%
  \BibitemOpen
  \bibfield  {author} {\bibinfo {author} {\bibfnamefont {D.}~\bibnamefont {Shlivko}}\ and\ \bibinfo {author} {\bibfnamefont {P.~J.}\ \bibnamefont {Steinhardt}},\ }\href {\doibase 10.1016/j.physletb.2024.138826} {\bibfield  {journal} {\bibinfo  {journal} {Phys. Lett. B}\ }\textbf {\bibinfo {volume} {855}},\ \bibinfo {pages} {138826} (\bibinfo {year} {2024})},\ \Eprint {http://arxiv.org/abs/2405.03933} {arXiv:2405.03933 [astro-ph.CO]} \BibitemShut {NoStop}%
\bibitem [{\citenamefont {Luongo}\ and\ \citenamefont {Muccino}(2024)}]{Luongo:2024fww}%
  \BibitemOpen
  \bibfield  {author} {\bibinfo {author} {\bibfnamefont {O.}~\bibnamefont {Luongo}}\ and\ \bibinfo {author} {\bibfnamefont {M.}~\bibnamefont {Muccino}},\ }\href {\doibase 10.1051/0004-6361/202450512} {\bibfield  {journal} {\bibinfo  {journal} {Astron. Astrophys.}\ }\textbf {\bibinfo {volume} {690}},\ \bibinfo {pages} {A40} (\bibinfo {year} {2024})},\ \Eprint {http://arxiv.org/abs/2404.07070} {arXiv:2404.07070 [astro-ph.CO]} \BibitemShut {NoStop}%
\bibitem [{\citenamefont {Gialamas}\ \emph {et~al.}(2025)\citenamefont {Gialamas}, \citenamefont {H{\"u}tsi}, \citenamefont {Kannike}, \citenamefont {Racioppi}, \citenamefont {Raidal}, \citenamefont {Vasar},\ and\ \citenamefont {Veerm{\"a}e}}]{Gialamas:2024lyw}%
  \BibitemOpen
  \bibfield  {author} {\bibinfo {author} {\bibfnamefont {I.~D.}\ \bibnamefont {Gialamas}}, \bibinfo {author} {\bibfnamefont {G.}~\bibnamefont {H{\"u}tsi}}, \bibinfo {author} {\bibfnamefont {K.}~\bibnamefont {Kannike}}, \bibinfo {author} {\bibfnamefont {A.}~\bibnamefont {Racioppi}}, \bibinfo {author} {\bibfnamefont {M.}~\bibnamefont {Raidal}}, \bibinfo {author} {\bibfnamefont {M.}~\bibnamefont {Vasar}}, \ and\ \bibinfo {author} {\bibfnamefont {H.}~\bibnamefont {Veerm{\"a}e}},\ }\href {\doibase 10.1103/PhysRevD.111.043540} {\bibfield  {journal} {\bibinfo  {journal} {Phys. Rev. D}\ }\textbf {\bibinfo {volume} {111}},\ \bibinfo {pages} {043540} (\bibinfo {year} {2025})},\ \Eprint {http://arxiv.org/abs/2406.07533} {arXiv:2406.07533 [astro-ph.CO]} \BibitemShut {NoStop}%
\bibitem [{\citenamefont {Wang}\ and\ \citenamefont {Piao}(2026)}]{Wang:2024dka}%
  \BibitemOpen
  \bibfield  {author} {\bibinfo {author} {\bibfnamefont {H.}~\bibnamefont {Wang}}\ and\ \bibinfo {author} {\bibfnamefont {Y.-S.}\ \bibnamefont {Piao}},\ }\href {\doibase 10.1016/j.physletb.2026.140180} {\bibfield  {journal} {\bibinfo  {journal} {Phys. Lett. B}\ }\textbf {\bibinfo {volume} {873}},\ \bibinfo {pages} {140180} (\bibinfo {year} {2026})},\ \Eprint {http://arxiv.org/abs/2404.18579} {arXiv:2404.18579 [astro-ph.CO]} \BibitemShut {NoStop}%
\bibitem [{\citenamefont {Ye}\ \emph {et~al.}(2025)\citenamefont {Ye}, \citenamefont {Martinelli}, \citenamefont {Hu},\ and\ \citenamefont {Silvestri}}]{Ye:2024ywg}%
  \BibitemOpen
  \bibfield  {author} {\bibinfo {author} {\bibfnamefont {G.}~\bibnamefont {Ye}}, \bibinfo {author} {\bibfnamefont {M.}~\bibnamefont {Martinelli}}, \bibinfo {author} {\bibfnamefont {B.}~\bibnamefont {Hu}}, \ and\ \bibinfo {author} {\bibfnamefont {A.}~\bibnamefont {Silvestri}},\ }\href {\doibase 10.1103/PhysRevLett.134.181002} {\bibfield  {journal} {\bibinfo  {journal} {Phys. Rev. Lett.}\ }\textbf {\bibinfo {volume} {134}},\ \bibinfo {pages} {181002} (\bibinfo {year} {2025})},\ \Eprint {http://arxiv.org/abs/2407.15832} {arXiv:2407.15832 [astro-ph.CO]} \BibitemShut {NoStop}%
\bibitem [{\citenamefont {Tada}\ and\ \citenamefont {Terada}(2024)}]{Tada:2024znt}%
  \BibitemOpen
  \bibfield  {author} {\bibinfo {author} {\bibfnamefont {Y.}~\bibnamefont {Tada}}\ and\ \bibinfo {author} {\bibfnamefont {T.}~\bibnamefont {Terada}},\ }\href {\doibase 10.1103/PhysRevD.109.L121305} {\bibfield  {journal} {\bibinfo  {journal} {Phys. Rev. D}\ }\textbf {\bibinfo {volume} {109}},\ \bibinfo {pages} {L121305} (\bibinfo {year} {2024})},\ \Eprint {http://arxiv.org/abs/2404.05722} {arXiv:2404.05722 [astro-ph.CO]} \BibitemShut {NoStop}%
\bibitem [{\citenamefont {Lodha}\ \emph {et~al.}(2025{\natexlab{b}})\citenamefont {Lodha} \emph {et~al.}}]{DESI:2024kob}%
  \BibitemOpen
  \bibfield  {author} {\bibinfo {author} {\bibfnamefont {K.}~\bibnamefont {Lodha}} \emph {et~al.} (\bibinfo {collaboration} {DESI}),\ }\href {\doibase 10.1103/PhysRevD.111.023532} {\bibfield  {journal} {\bibinfo  {journal} {Phys. Rev. D}\ }\textbf {\bibinfo {volume} {111}},\ \bibinfo {pages} {023532} (\bibinfo {year} {2025}{\natexlab{b}})},\ \Eprint {http://arxiv.org/abs/2405.13588} {arXiv:2405.13588 [astro-ph.CO]} \BibitemShut {NoStop}%
\bibitem [{\citenamefont {Bhattacharya}\ \emph {et~al.}(2024)\citenamefont {Bhattacharya}, \citenamefont {Borghetto}, \citenamefont {Malhotra}, \citenamefont {Parameswaran}, \citenamefont {Tasinato},\ and\ \citenamefont {Zavala}}]{Bhattacharya:2024hep}%
  \BibitemOpen
  \bibfield  {author} {\bibinfo {author} {\bibfnamefont {S.}~\bibnamefont {Bhattacharya}}, \bibinfo {author} {\bibfnamefont {G.}~\bibnamefont {Borghetto}}, \bibinfo {author} {\bibfnamefont {A.}~\bibnamefont {Malhotra}}, \bibinfo {author} {\bibfnamefont {S.}~\bibnamefont {Parameswaran}}, \bibinfo {author} {\bibfnamefont {G.}~\bibnamefont {Tasinato}}, \ and\ \bibinfo {author} {\bibfnamefont {I.}~\bibnamefont {Zavala}},\ }\href {\doibase 10.1088/1475-7516/2024/09/073} {\bibfield  {journal} {\bibinfo  {journal} {JCAP}\ }\textbf {\bibinfo {volume} {09}},\ \bibinfo {pages} {073} (\bibinfo {year} {2024})},\ \Eprint {http://arxiv.org/abs/2405.17396} {arXiv:2405.17396 [astro-ph.CO]} \BibitemShut {NoStop}%
\bibitem [{\citenamefont {Ramadan}\ \emph {et~al.}(2024)\citenamefont {Ramadan}, \citenamefont {Sakstein},\ and\ \citenamefont {Rubin}}]{Ramadan:2024kmn}%
  \BibitemOpen
  \bibfield  {author} {\bibinfo {author} {\bibfnamefont {O.~F.}\ \bibnamefont {Ramadan}}, \bibinfo {author} {\bibfnamefont {J.}~\bibnamefont {Sakstein}}, \ and\ \bibinfo {author} {\bibfnamefont {D.}~\bibnamefont {Rubin}},\ }\href {\doibase 10.1103/PhysRevD.110.L041303} {\bibfield  {journal} {\bibinfo  {journal} {Phys. Rev. D}\ }\textbf {\bibinfo {volume} {110}},\ \bibinfo {pages} {L041303} (\bibinfo {year} {2024})},\ \Eprint {http://arxiv.org/abs/2405.18747} {arXiv:2405.18747 [astro-ph.CO]} \BibitemShut {NoStop}%
\bibitem [{\citenamefont {Malekjani}\ \emph {et~al.}(2025)\citenamefont {Malekjani}, \citenamefont {Davari},\ and\ \citenamefont {Pourojaghi}}]{Malekjani:2024bgi}%
  \BibitemOpen
  \bibfield  {author} {\bibinfo {author} {\bibfnamefont {M.}~\bibnamefont {Malekjani}}, \bibinfo {author} {\bibfnamefont {Z.}~\bibnamefont {Davari}}, \ and\ \bibinfo {author} {\bibfnamefont {S.}~\bibnamefont {Pourojaghi}} (\bibinfo {collaboration} {DESI}),\ }\href {\doibase 10.1103/PhysRevD.111.083547} {\bibfield  {journal} {\bibinfo  {journal} {Phys. Rev. D}\ }\textbf {\bibinfo {volume} {111}},\ \bibinfo {pages} {083547} (\bibinfo {year} {2025})},\ \Eprint {http://arxiv.org/abs/2407.09767} {arXiv:2407.09767 [astro-ph.CO]} \BibitemShut {NoStop}%
\bibitem [{\citenamefont {Giar{\`e}}\ \emph {et~al.}(2024{\natexlab{a}})\citenamefont {Giar{\`e}}, \citenamefont {Najafi}, \citenamefont {Pan}, \citenamefont {Di~Valentino},\ and\ \citenamefont {Firouzjaee}}]{Giare:2024gpk}%
  \BibitemOpen
  \bibfield  {author} {\bibinfo {author} {\bibfnamefont {W.}~\bibnamefont {Giar{\`e}}}, \bibinfo {author} {\bibfnamefont {M.}~\bibnamefont {Najafi}}, \bibinfo {author} {\bibfnamefont {S.}~\bibnamefont {Pan}}, \bibinfo {author} {\bibfnamefont {E.}~\bibnamefont {Di~Valentino}}, \ and\ \bibinfo {author} {\bibfnamefont {J.~T.}\ \bibnamefont {Firouzjaee}},\ }\href {\doibase 10.1088/1475-7516/2024/10/035} {\bibfield  {journal} {\bibinfo  {journal} {JCAP}\ }\textbf {\bibinfo {volume} {10}},\ \bibinfo {pages} {035} (\bibinfo {year} {2024}{\natexlab{a}})},\ \Eprint {http://arxiv.org/abs/2407.16689} {arXiv:2407.16689 [astro-ph.CO]} \BibitemShut {NoStop}%
\bibitem [{\citenamefont {Rebou{\c{c}}as}\ \emph {et~al.}(2025)\citenamefont {Rebou{\c{c}}as}, \citenamefont {de~Souza}, \citenamefont {Zhong}, \citenamefont {Miranda},\ and\ \citenamefont {Rosenfeld}}]{Reboucas:2024smm}%
  \BibitemOpen
  \bibfield  {author} {\bibinfo {author} {\bibfnamefont {J.}~\bibnamefont {Rebou{\c{c}}as}}, \bibinfo {author} {\bibfnamefont {D.~H.~F.}\ \bibnamefont {de~Souza}}, \bibinfo {author} {\bibfnamefont {K.}~\bibnamefont {Zhong}}, \bibinfo {author} {\bibfnamefont {V.}~\bibnamefont {Miranda}}, \ and\ \bibinfo {author} {\bibfnamefont {R.}~\bibnamefont {Rosenfeld}},\ }\href {\doibase 10.1088/1475-7516/2025/02/024} {\bibfield  {journal} {\bibinfo  {journal} {JCAP}\ }\textbf {\bibinfo {volume} {02}},\ \bibinfo {pages} {024} (\bibinfo {year} {2025})},\ \Eprint {http://arxiv.org/abs/2408.14628} {arXiv:2408.14628 [astro-ph.CO]} \BibitemShut {NoStop}%
\bibitem [{\citenamefont {Park}\ \emph {et~al.}(2025)\citenamefont {Park}, \citenamefont {de~Cruz~P{\'e}rez},\ and\ \citenamefont {Ratra}}]{Park:2024pew}%
  \BibitemOpen
  \bibfield  {author} {\bibinfo {author} {\bibfnamefont {C.-G.}\ \bibnamefont {Park}}, \bibinfo {author} {\bibfnamefont {J.}~\bibnamefont {de~Cruz~P{\'e}rez}}, \ and\ \bibinfo {author} {\bibfnamefont {B.}~\bibnamefont {Ratra}},\ }\href {\doibase 10.1142/S0218271825500580} {\bibfield  {journal} {\bibinfo  {journal} {Int. J. Mod. Phys. D}\ }\textbf {\bibinfo {volume} {34}},\ \bibinfo {pages} {2550058} (\bibinfo {year} {2025})},\ \Eprint {http://arxiv.org/abs/2410.13627} {arXiv:2410.13627 [astro-ph.CO]} \BibitemShut {NoStop}%
\bibitem [{\citenamefont {Li}\ \emph {et~al.}(2025{\natexlab{a}})\citenamefont {Li}, \citenamefont {Li}, \citenamefont {Du}, \citenamefont {Wu}, \citenamefont {Feng}, \citenamefont {Zhang},\ and\ \citenamefont {Zhang}}]{Li:2024qus}%
  \BibitemOpen
  \bibfield  {author} {\bibinfo {author} {\bibfnamefont {T.-N.}\ \bibnamefont {Li}}, \bibinfo {author} {\bibfnamefont {Y.-H.}\ \bibnamefont {Li}}, \bibinfo {author} {\bibfnamefont {G.-H.}\ \bibnamefont {Du}}, \bibinfo {author} {\bibfnamefont {P.-J.}\ \bibnamefont {Wu}}, \bibinfo {author} {\bibfnamefont {L.}~\bibnamefont {Feng}}, \bibinfo {author} {\bibfnamefont {J.-F.}\ \bibnamefont {Zhang}}, \ and\ \bibinfo {author} {\bibfnamefont {X.}~\bibnamefont {Zhang}},\ }\href {\doibase 10.1140/epjc/s10052-025-14279-7} {\bibfield  {journal} {\bibinfo  {journal} {Eur. Phys. J. C}\ }\textbf {\bibinfo {volume} {85}},\ \bibinfo {pages} {608} (\bibinfo {year} {2025}{\natexlab{a}})},\ \Eprint {http://arxiv.org/abs/2411.08639} {arXiv:2411.08639 [astro-ph.CO]} \BibitemShut {NoStop}%
\bibitem [{\citenamefont {Jiang}\ \emph {et~al.}(2024)\citenamefont {Jiang}, \citenamefont {Pedrotti}, \citenamefont {da~Costa},\ and\ \citenamefont {Vagnozzi}}]{Jiang:2024xnu}%
  \BibitemOpen
  \bibfield  {author} {\bibinfo {author} {\bibfnamefont {J.-Q.}\ \bibnamefont {Jiang}}, \bibinfo {author} {\bibfnamefont {D.}~\bibnamefont {Pedrotti}}, \bibinfo {author} {\bibfnamefont {S.~S.}\ \bibnamefont {da~Costa}}, \ and\ \bibinfo {author} {\bibfnamefont {S.}~\bibnamefont {Vagnozzi}},\ }\href {\doibase 10.1103/PhysRevD.110.123519} {\bibfield  {journal} {\bibinfo  {journal} {Phys. Rev. D}\ }\textbf {\bibinfo {volume} {110}},\ \bibinfo {pages} {123519} (\bibinfo {year} {2024})},\ \Eprint {http://arxiv.org/abs/2408.02365} {arXiv:2408.02365 [astro-ph.CO]} \BibitemShut {NoStop}%
\bibitem [{\citenamefont {Yang}\ \emph {et~al.}(2024)\citenamefont {Yang}, \citenamefont {Ren}, \citenamefont {Wang}, \citenamefont {Lu}, \citenamefont {Zhang}, \citenamefont {Cai},\ and\ \citenamefont {Saridakis}}]{Yang:2024kdo}%
  \BibitemOpen
  \bibfield  {author} {\bibinfo {author} {\bibfnamefont {Y.}~\bibnamefont {Yang}}, \bibinfo {author} {\bibfnamefont {X.}~\bibnamefont {Ren}}, \bibinfo {author} {\bibfnamefont {Q.}~\bibnamefont {Wang}}, \bibinfo {author} {\bibfnamefont {Z.}~\bibnamefont {Lu}}, \bibinfo {author} {\bibfnamefont {D.}~\bibnamefont {Zhang}}, \bibinfo {author} {\bibfnamefont {Y.-F.}\ \bibnamefont {Cai}}, \ and\ \bibinfo {author} {\bibfnamefont {E.~N.}\ \bibnamefont {Saridakis}},\ }\href {\doibase 10.1016/j.scib.2024.07.029} {\bibfield  {journal} {\bibinfo  {journal} {Sci. Bull.}\ }\textbf {\bibinfo {volume} {69}},\ \bibinfo {pages} {2698} (\bibinfo {year} {2024})},\ \Eprint {http://arxiv.org/abs/2404.19437} {arXiv:2404.19437 [astro-ph.CO]} \BibitemShut {NoStop}%
\bibitem [{\citenamefont {Li}\ \emph {et~al.}(2026{\natexlab{a}})\citenamefont {Li}, \citenamefont {Du}, \citenamefont {Zhou}, \citenamefont {Li}, \citenamefont {Zhang},\ and\ \citenamefont {Zhang}}]{Li:2025vuh}%
  \BibitemOpen
  \bibfield  {author} {\bibinfo {author} {\bibfnamefont {T.-N.}\ \bibnamefont {Li}}, \bibinfo {author} {\bibfnamefont {G.-H.}\ \bibnamefont {Du}}, \bibinfo {author} {\bibfnamefont {S.-H.}\ \bibnamefont {Zhou}}, \bibinfo {author} {\bibfnamefont {Y.-H.}\ \bibnamefont {Li}}, \bibinfo {author} {\bibfnamefont {J.-F.}\ \bibnamefont {Zhang}}, \ and\ \bibinfo {author} {\bibfnamefont {X.}~\bibnamefont {Zhang}},\ }\href {\doibase 10.1016/j.dark.2026.102254} {\bibfield  {journal} {\bibinfo  {journal} {Phys. Dark Univ.}\ }\textbf {\bibinfo {volume} {52}},\ \bibinfo {pages} {102254} (\bibinfo {year} {2026}{\natexlab{a}})},\ \Eprint {http://arxiv.org/abs/2511.22512} {arXiv:2511.22512 [astro-ph.CO]} \BibitemShut {NoStop}%
\bibitem [{\citenamefont {Wolf}\ \emph {et~al.}(2025)\citenamefont {Wolf}, \citenamefont {Garc{\'\i}a-Garc{\'\i}a}, \citenamefont {Anton},\ and\ \citenamefont {Ferreira}}]{Wolf:2025jed}%
  \BibitemOpen
  \bibfield  {author} {\bibinfo {author} {\bibfnamefont {W.~J.}\ \bibnamefont {Wolf}}, \bibinfo {author} {\bibfnamefont {C.}~\bibnamefont {Garc{\'\i}a-Garc{\'\i}a}}, \bibinfo {author} {\bibfnamefont {T.}~\bibnamefont {Anton}}, \ and\ \bibinfo {author} {\bibfnamefont {P.~G.}\ \bibnamefont {Ferreira}},\ }\href {\doibase 10.1103/jysf-k72m} {\bibfield  {journal} {\bibinfo  {journal} {Phys. Rev. Lett.}\ }\textbf {\bibinfo {volume} {135}},\ \bibinfo {pages} {081001} (\bibinfo {year} {2025})},\ \Eprint {http://arxiv.org/abs/2504.07679} {arXiv:2504.07679 [astro-ph.CO]} \BibitemShut {NoStop}%
\bibitem [{\citenamefont {Shajib}\ and\ \citenamefont {Frieman}(2025)}]{Shajib:2025tpd}%
  \BibitemOpen
  \bibfield  {author} {\bibinfo {author} {\bibfnamefont {A.~J.}\ \bibnamefont {Shajib}}\ and\ \bibinfo {author} {\bibfnamefont {J.~A.}\ \bibnamefont {Frieman}},\ }\href {\doibase 10.1103/kjpb-r698} {\bibfield  {journal} {\bibinfo  {journal} {Phys. Rev. D}\ }\textbf {\bibinfo {volume} {112}},\ \bibinfo {pages} {063508} (\bibinfo {year} {2025})},\ \Eprint {http://arxiv.org/abs/2502.06929} {arXiv:2502.06929 [astro-ph.CO]} \BibitemShut {NoStop}%
\bibitem [{\citenamefont {Giar{\`e}}\ \emph {et~al.}(2025{\natexlab{a}})\citenamefont {Giar{\`e}}, \citenamefont {Mahassen}, \citenamefont {Di~Valentino},\ and\ \citenamefont {Pan}}]{Giare:2025pzu}%
  \BibitemOpen
  \bibfield  {author} {\bibinfo {author} {\bibfnamefont {W.}~\bibnamefont {Giar{\`e}}}, \bibinfo {author} {\bibfnamefont {T.}~\bibnamefont {Mahassen}}, \bibinfo {author} {\bibfnamefont {E.}~\bibnamefont {Di~Valentino}}, \ and\ \bibinfo {author} {\bibfnamefont {S.}~\bibnamefont {Pan}},\ }\href {\doibase 10.1016/j.dark.2025.101906} {\bibfield  {journal} {\bibinfo  {journal} {Phys. Dark Univ.}\ }\textbf {\bibinfo {volume} {48}},\ \bibinfo {pages} {101906} (\bibinfo {year} {2025}{\natexlab{a}})},\ \Eprint {http://arxiv.org/abs/2502.10264} {arXiv:2502.10264 [astro-ph.CO]} \BibitemShut {NoStop}%
\bibitem [{\citenamefont {Chaussidon}\ \emph {et~al.}(2025)\citenamefont {Chaussidon} \emph {et~al.}}]{Chaussidon:2025npr}%
  \BibitemOpen
  \bibfield  {author} {\bibinfo {author} {\bibfnamefont {E.}~\bibnamefont {Chaussidon}} \emph {et~al.},\ }\href {\doibase 10.1103/xtql-wh3h} {\bibfield  {journal} {\bibinfo  {journal} {Phys. Rev. D}\ }\textbf {\bibinfo {volume} {112}},\ \bibinfo {pages} {063548} (\bibinfo {year} {2025})},\ \Eprint {http://arxiv.org/abs/2503.24343} {arXiv:2503.24343 [astro-ph.CO]} \BibitemShut {NoStop}%
\bibitem [{\citenamefont {Pang}\ \emph {et~al.}(2025)\citenamefont {Pang}, \citenamefont {Zhang},\ and\ \citenamefont {Huang}}]{Pang:2025lvh}%
  \BibitemOpen
  \bibfield  {author} {\bibinfo {author} {\bibfnamefont {Y.-H.}\ \bibnamefont {Pang}}, \bibinfo {author} {\bibfnamefont {X.}~\bibnamefont {Zhang}}, \ and\ \bibinfo {author} {\bibfnamefont {Q.-G.}\ \bibnamefont {Huang}},\ }\href {\doibase 10.1007/s11433-025-2713-8} {\bibfield  {journal} {\bibinfo  {journal} {Sci. China Phys. Mech. Astron.}\ }\textbf {\bibinfo {volume} {68}},\ \bibinfo {pages} {280410} (\bibinfo {year} {2025})},\ \Eprint {http://arxiv.org/abs/2503.21600} {arXiv:2503.21600 [astro-ph.CO]} \BibitemShut {NoStop}%
\bibitem [{\citenamefont {Roy~Choudhury}(2025)}]{RoyChoudhury:2025dhe}%
  \BibitemOpen
  \bibfield  {author} {\bibinfo {author} {\bibfnamefont {S.}~\bibnamefont {Roy~Choudhury}},\ }\href {\doibase 10.3847/2041-8213/ade1cc} {\bibfield  {journal} {\bibinfo  {journal} {Astrophys. J. Lett.}\ }\textbf {\bibinfo {volume} {986}},\ \bibinfo {pages} {L31} (\bibinfo {year} {2025})},\ \Eprint {http://arxiv.org/abs/2504.15340} {arXiv:2504.15340 [astro-ph.CO]} \BibitemShut {NoStop}%
\bibitem [{\citenamefont {Paliathanasis}(2025)}]{Paliathanasis:2025cuc}%
  \BibitemOpen
  \bibfield  {author} {\bibinfo {author} {\bibfnamefont {A.}~\bibnamefont {Paliathanasis}},\ }\href {\doibase 10.1016/j.dark.2025.101956} {\bibfield  {journal} {\bibinfo  {journal} {Phys. Dark Univ.}\ }\textbf {\bibinfo {volume} {48}},\ \bibinfo {pages} {101956} (\bibinfo {year} {2025})},\ \Eprint {http://arxiv.org/abs/2502.16221} {arXiv:2502.16221 [astro-ph.CO]} \BibitemShut {NoStop}%
\bibitem [{\citenamefont {Scherer}\ \emph {et~al.}(2025)\citenamefont {Scherer}, \citenamefont {Sabogal}, \citenamefont {Nunes},\ and\ \citenamefont {De~Felice}}]{Scherer:2025esj}%
  \BibitemOpen
  \bibfield  {author} {\bibinfo {author} {\bibfnamefont {M.}~\bibnamefont {Scherer}}, \bibinfo {author} {\bibfnamefont {M.~A.}\ \bibnamefont {Sabogal}}, \bibinfo {author} {\bibfnamefont {R.~C.}\ \bibnamefont {Nunes}}, \ and\ \bibinfo {author} {\bibfnamefont {A.}~\bibnamefont {De~Felice}},\ }\href {\doibase 10.1103/n86r-sjgm} {\bibfield  {journal} {\bibinfo  {journal} {Phys. Rev. D}\ }\textbf {\bibinfo {volume} {112}},\ \bibinfo {pages} {043513} (\bibinfo {year} {2025})},\ \Eprint {http://arxiv.org/abs/2504.20664} {arXiv:2504.20664 [astro-ph.CO]} \BibitemShut {NoStop}%
\bibitem [{\citenamefont {Giar{\`e}}(2025)}]{Giare:2024oil}%
  \BibitemOpen
  \bibfield  {author} {\bibinfo {author} {\bibfnamefont {W.}~\bibnamefont {Giar{\`e}}},\ }\href {\doibase 10.1103/ss37-cxhn} {\bibfield  {journal} {\bibinfo  {journal} {Phys. Rev. D}\ }\textbf {\bibinfo {volume} {112}},\ \bibinfo {pages} {023508} (\bibinfo {year} {2025})},\ \Eprint {http://arxiv.org/abs/2409.17074} {arXiv:2409.17074 [astro-ph.CO]} \BibitemShut {NoStop}%
\bibitem [{\citenamefont {Liu}\ \emph {et~al.}(2025)\citenamefont {Liu}, \citenamefont {Li},\ and\ \citenamefont {Wang}}]{Liu:2025mub}%
  \BibitemOpen
  \bibfield  {author} {\bibinfo {author} {\bibfnamefont {T.}~\bibnamefont {Liu}}, \bibinfo {author} {\bibfnamefont {X.}~\bibnamefont {Li}}, \ and\ \bibinfo {author} {\bibfnamefont {J.}~\bibnamefont {Wang}},\ }\href {\doibase 10.3847/1538-4357/aded8b} {\bibfield  {journal} {\bibinfo  {journal} {Astrophys. J.}\ }\textbf {\bibinfo {volume} {988}},\ \bibinfo {pages} {243} (\bibinfo {year} {2025})},\ \Eprint {http://arxiv.org/abs/2504.21373} {arXiv:2504.21373 [astro-ph.CO]} \BibitemShut {NoStop}%
\bibitem [{\citenamefont {Teixeira}\ \emph {et~al.}(2025)\citenamefont {Teixeira}, \citenamefont {Giar{\`e}}, \citenamefont {Hogg}, \citenamefont {Montandon}, \citenamefont {Poudou},\ and\ \citenamefont {Poulin}}]{Teixeira:2025czm}%
  \BibitemOpen
  \bibfield  {author} {\bibinfo {author} {\bibfnamefont {E.~M.}\ \bibnamefont {Teixeira}}, \bibinfo {author} {\bibfnamefont {W.}~\bibnamefont {Giar{\`e}}}, \bibinfo {author} {\bibfnamefont {N.~B.}\ \bibnamefont {Hogg}}, \bibinfo {author} {\bibfnamefont {T.}~\bibnamefont {Montandon}}, \bibinfo {author} {\bibfnamefont {A.}~\bibnamefont {Poudou}}, \ and\ \bibinfo {author} {\bibfnamefont {V.}~\bibnamefont {Poulin}},\ }\href {\doibase 10.1103/zzmp-rxrh} {\bibfield  {journal} {\bibinfo  {journal} {Phys. Rev. D}\ }\textbf {\bibinfo {volume} {112}},\ \bibinfo {pages} {023515} (\bibinfo {year} {2025})},\ \Eprint {http://arxiv.org/abs/2504.10464} {arXiv:2504.10464 [astro-ph.CO]} \BibitemShut {NoStop}%
\bibitem [{\citenamefont {Specogna}\ \emph {et~al.}(2026)\citenamefont {Specogna}, \citenamefont {Adil}, \citenamefont {Ozulker}, \citenamefont {Di~Valentino}, \citenamefont {Nunes}, \citenamefont {Akarsu},\ and\ \citenamefont {Sen}}]{Specogna:2025guo}%
  \BibitemOpen
  \bibfield  {author} {\bibinfo {author} {\bibfnamefont {E.}~\bibnamefont {Specogna}}, \bibinfo {author} {\bibfnamefont {S.~A.}\ \bibnamefont {Adil}}, \bibinfo {author} {\bibfnamefont {E.}~\bibnamefont {Ozulker}}, \bibinfo {author} {\bibfnamefont {E.}~\bibnamefont {Di~Valentino}}, \bibinfo {author} {\bibfnamefont {R.~C.}\ \bibnamefont {Nunes}}, \bibinfo {author} {\bibfnamefont {O.}~\bibnamefont {Akarsu}}, \ and\ \bibinfo {author} {\bibfnamefont {A.~A.}\ \bibnamefont {Sen}},\ }\href {\doibase 10.1103/b7ht-lx26} {\bibfield  {journal} {\bibinfo  {journal} {Phys. Rev. D}\ }\textbf {\bibinfo {volume} {113}},\ \bibinfo {pages} {103549} (\bibinfo {year} {2026})},\ \Eprint {http://arxiv.org/abs/2504.17859} {arXiv:2504.17859 [gr-qc]} \BibitemShut {NoStop}%
\bibitem [{\citenamefont {Cheng}\ \emph {et~al.}(2025)\citenamefont {Cheng}, \citenamefont {Di~Valentino}, \citenamefont {Escamilla}, \citenamefont {Sen},\ and\ \citenamefont {Visinelli}}]{Cheng:2025lod}%
  \BibitemOpen
  \bibfield  {author} {\bibinfo {author} {\bibfnamefont {H.}~\bibnamefont {Cheng}}, \bibinfo {author} {\bibfnamefont {E.}~\bibnamefont {Di~Valentino}}, \bibinfo {author} {\bibfnamefont {L.~A.}\ \bibnamefont {Escamilla}}, \bibinfo {author} {\bibfnamefont {A.~A.}\ \bibnamefont {Sen}}, \ and\ \bibinfo {author} {\bibfnamefont {L.}~\bibnamefont {Visinelli}},\ }\href {\doibase 10.1088/1475-7516/2025/09/031} {\bibfield  {journal} {\bibinfo  {journal} {JCAP}\ }\textbf {\bibinfo {volume} {09}},\ \bibinfo {pages} {031} (\bibinfo {year} {2025})},\ \Eprint {http://arxiv.org/abs/2505.02932} {arXiv:2505.02932 [astro-ph.CO]} \BibitemShut {NoStop}%
\bibitem [{\citenamefont {Herold}\ and\ \citenamefont {Karwal}(2025)}]{Herold:2025hkb}%
  \BibitemOpen
  \bibfield  {author} {\bibinfo {author} {\bibfnamefont {L.}~\bibnamefont {Herold}}\ and\ \bibinfo {author} {\bibfnamefont {T.}~\bibnamefont {Karwal}},\ }\href@noop {} {\  (\bibinfo {year} {2025})},\ \Eprint {http://arxiv.org/abs/2506.12004} {arXiv:2506.12004 [astro-ph.CO]} \BibitemShut {NoStop}%
\bibitem [{\citenamefont {{\"O}z{\"u}lker}\ \emph {et~al.}(2025)\citenamefont {{\"O}z{\"u}lker}, \citenamefont {Di~Valentino},\ and\ \citenamefont {Giar{\`e}}}]{Ozulker:2025ehg}%
  \BibitemOpen
  \bibfield  {author} {\bibinfo {author} {\bibfnamefont {E.}~\bibnamefont {{\"O}z{\"u}lker}}, \bibinfo {author} {\bibfnamefont {E.}~\bibnamefont {Di~Valentino}}, \ and\ \bibinfo {author} {\bibfnamefont {W.}~\bibnamefont {Giar{\`e}}},\ }\href@noop {} {\  (\bibinfo {year} {2025})},\ \Eprint {http://arxiv.org/abs/2506.19053} {arXiv:2506.19053 [astro-ph.CO]} \BibitemShut {NoStop}%
\bibitem [{\citenamefont {Ormondroyd}\ \emph {et~al.}(2025)\citenamefont {Ormondroyd}, \citenamefont {Handley}, \citenamefont {Hobson},\ and\ \citenamefont {Lasenby}}]{Ormondroyd:2025iaf}%
  \BibitemOpen
  \bibfield  {author} {\bibinfo {author} {\bibfnamefont {A.~N.}\ \bibnamefont {Ormondroyd}}, \bibinfo {author} {\bibfnamefont {W.~J.}\ \bibnamefont {Handley}}, \bibinfo {author} {\bibfnamefont {M.~P.}\ \bibnamefont {Hobson}}, \ and\ \bibinfo {author} {\bibfnamefont {A.~N.}\ \bibnamefont {Lasenby}},\ }\href@noop {} {\  (\bibinfo {year} {2025})},\ \Eprint {http://arxiv.org/abs/2503.17342} {arXiv:2503.17342 [astro-ph.CO]} \BibitemShut {NoStop}%
\bibitem [{\citenamefont {Silva}\ and\ \citenamefont {Nunes}(2025)}]{Silva:2025twg}%
  \BibitemOpen
  \bibfield  {author} {\bibinfo {author} {\bibfnamefont {E.}~\bibnamefont {Silva}}\ and\ \bibinfo {author} {\bibfnamefont {R.~C.}\ \bibnamefont {Nunes}},\ }\href {\doibase 10.1088/1475-7516/2025/11/078} {\bibfield  {journal} {\bibinfo  {journal} {JCAP}\ }\textbf {\bibinfo {volume} {11}},\ \bibinfo {pages} {078} (\bibinfo {year} {2025})},\ \Eprint {http://arxiv.org/abs/2507.13989} {arXiv:2507.13989 [astro-ph.CO]} \BibitemShut {NoStop}%
\bibitem [{\citenamefont {Ishak}\ and\ \citenamefont {Medina-Varela}(2025)}]{Ishak:2025cay}%
  \BibitemOpen
  \bibfield  {author} {\bibinfo {author} {\bibfnamefont {M.}~\bibnamefont {Ishak}}\ and\ \bibinfo {author} {\bibfnamefont {L.}~\bibnamefont {Medina-Varela}},\ }\href@noop {} {\  (\bibinfo {year} {2025})},\ \Eprint {http://arxiv.org/abs/2507.22856} {arXiv:2507.22856 [astro-ph.CO]} \BibitemShut {NoStop}%
\bibitem [{\citenamefont {Fazzari}\ \emph {et~al.}(2026)\citenamefont {Fazzari}, \citenamefont {Giar{\`e}},\ and\ \citenamefont {Di~Valentino}}]{Fazzari:2025lzd}%
  \BibitemOpen
  \bibfield  {author} {\bibinfo {author} {\bibfnamefont {E.}~\bibnamefont {Fazzari}}, \bibinfo {author} {\bibfnamefont {W.}~\bibnamefont {Giar{\`e}}}, \ and\ \bibinfo {author} {\bibfnamefont {E.}~\bibnamefont {Di~Valentino}},\ }\href {\doibase 10.3847/2041-8213/ae2917} {\bibfield  {journal} {\bibinfo  {journal} {Astrophys. J. Lett.}\ }\textbf {\bibinfo {volume} {996}},\ \bibinfo {pages} {L5} (\bibinfo {year} {2026})},\ \Eprint {http://arxiv.org/abs/2509.16196} {arXiv:2509.16196 [astro-ph.CO]} \BibitemShut {NoStop}%
\bibitem [{\citenamefont {Zhang}\ \emph {et~al.}(2026)\citenamefont {Zhang}, \citenamefont {Xu},\ and\ \citenamefont {Chen}}]{Zhang:2025lam}%
  \BibitemOpen
  \bibfield  {author} {\bibinfo {author} {\bibfnamefont {Z.}~\bibnamefont {Zhang}}, \bibinfo {author} {\bibfnamefont {T.}~\bibnamefont {Xu}}, \ and\ \bibinfo {author} {\bibfnamefont {Y.}~\bibnamefont {Chen}},\ }\href {\doibase 10.3847/1538-4357/ae4738} {\bibfield  {journal} {\bibinfo  {journal} {Astrophys. J.}\ }\textbf {\bibinfo {volume} {999}},\ \bibinfo {pages} {248} (\bibinfo {year} {2026})},\ \Eprint {http://arxiv.org/abs/2512.07281} {arXiv:2512.07281 [astro-ph.CO]} \BibitemShut {NoStop}%
\bibitem [{\citenamefont {Cheng}\ \emph {et~al.}(2026)\citenamefont {Cheng}, \citenamefont {Di~Valentino},\ and\ \citenamefont {Visinelli}}]{Cheng:2025hug}%
  \BibitemOpen
  \bibfield  {author} {\bibinfo {author} {\bibfnamefont {H.}~\bibnamefont {Cheng}}, \bibinfo {author} {\bibfnamefont {E.}~\bibnamefont {Di~Valentino}}, \ and\ \bibinfo {author} {\bibfnamefont {L.}~\bibnamefont {Visinelli}},\ }\href {\doibase 10.1016/j.jheap.2026.100610} {\bibfield  {journal} {\bibinfo  {journal} {JHEAp}\ }\textbf {\bibinfo {volume} {53}},\ \bibinfo {pages} {100610} (\bibinfo {year} {2026})},\ \Eprint {http://arxiv.org/abs/2505.22066} {arXiv:2505.22066 [astro-ph.CO]} \BibitemShut {NoStop}%
\bibitem [{\citenamefont {Cai}\ \emph {et~al.}(2025)\citenamefont {Cai}, \citenamefont {Ren}, \citenamefont {Qiu}, \citenamefont {Li},\ and\ \citenamefont {Zhang}}]{Cai:2025mas}%
  \BibitemOpen
  \bibfield  {author} {\bibinfo {author} {\bibfnamefont {Y.}~\bibnamefont {Cai}}, \bibinfo {author} {\bibfnamefont {X.}~\bibnamefont {Ren}}, \bibinfo {author} {\bibfnamefont {T.}~\bibnamefont {Qiu}}, \bibinfo {author} {\bibfnamefont {M.}~\bibnamefont {Li}}, \ and\ \bibinfo {author} {\bibfnamefont {X.}~\bibnamefont {Zhang}},\ }\href {\doibase 10.1093/nsr/nwag115} {\  (\bibinfo {year} {2025}),\ 10.1093/nsr/nwag115},\ \Eprint {http://arxiv.org/abs/2505.24732} {arXiv:2505.24732 [astro-ph.CO]} \BibitemShut {NoStop}%
\bibitem [{\citenamefont {Song}\ \emph {et~al.}(2026)\citenamefont {Song}, \citenamefont {Du}, \citenamefont {Li}, \citenamefont {Wang}, \citenamefont {Qi}, \citenamefont {Zhang},\ and\ \citenamefont {Zhang}}]{Song:2025bio}%
  \BibitemOpen
  \bibfield  {author} {\bibinfo {author} {\bibfnamefont {J.-Y.}\ \bibnamefont {Song}}, \bibinfo {author} {\bibfnamefont {G.-H.}\ \bibnamefont {Du}}, \bibinfo {author} {\bibfnamefont {T.-N.}\ \bibnamefont {Li}}, \bibinfo {author} {\bibfnamefont {L.-F.}\ \bibnamefont {Wang}}, \bibinfo {author} {\bibfnamefont {J.-Z.}\ \bibnamefont {Qi}}, \bibinfo {author} {\bibfnamefont {J.-F.}\ \bibnamefont {Zhang}}, \ and\ \bibinfo {author} {\bibfnamefont {X.}~\bibnamefont {Zhang}},\ }\href {\doibase 10.1007/s11433-025-2888-0} {\bibfield  {journal} {\bibinfo  {journal} {Sci. China Phys. Mech. Astron.}\ }\textbf {\bibinfo {volume} {69}},\ \bibinfo {pages} {240413} (\bibinfo {year} {2026})},\ \Eprint {http://arxiv.org/abs/2511.12017} {arXiv:2511.12017 [astro-ph.CO]} \BibitemShut {NoStop}%
\bibitem [{\citenamefont {Li}\ and\ \citenamefont {Wang}(2025)}]{Li:2025ops}%
  \BibitemOpen
  \bibfield  {author} {\bibinfo {author} {\bibfnamefont {J.-X.}\ \bibnamefont {Li}}\ and\ \bibinfo {author} {\bibfnamefont {S.}~\bibnamefont {Wang}},\ }\href {\doibase 10.1140/epjc/s10052-025-15065-1} {\bibfield  {journal} {\bibinfo  {journal} {Eur. Phys. J. C}\ }\textbf {\bibinfo {volume} {85}},\ \bibinfo {pages} {1308} (\bibinfo {year} {2025})},\ \Eprint {http://arxiv.org/abs/2506.22953} {arXiv:2506.22953 [astro-ph.CO]} \BibitemShut {NoStop}%
\bibitem [{\citenamefont {Lee}\ \emph {et~al.}(2026)\citenamefont {Lee}, \citenamefont {Yang}, \citenamefont {Di~Valentino}, \citenamefont {Pan},\ and\ \citenamefont {van~de Bruck}}]{Lee:2025pzo}%
  \BibitemOpen
  \bibfield  {author} {\bibinfo {author} {\bibfnamefont {D.~H.}\ \bibnamefont {Lee}}, \bibinfo {author} {\bibfnamefont {W.}~\bibnamefont {Yang}}, \bibinfo {author} {\bibfnamefont {E.}~\bibnamefont {Di~Valentino}}, \bibinfo {author} {\bibfnamefont {S.}~\bibnamefont {Pan}}, \ and\ \bibinfo {author} {\bibfnamefont {C.}~\bibnamefont {van~de Bruck}},\ }\href {\doibase 10.1103/z7y2-yvhg} {\bibfield  {journal} {\bibinfo  {journal} {Phys. Rev. D}\ }\textbf {\bibinfo {volume} {113}},\ \bibinfo {pages} {063554} (\bibinfo {year} {2026})},\ \Eprint {http://arxiv.org/abs/2507.11432} {arXiv:2507.11432 [astro-ph.CO]} \BibitemShut {NoStop}%
\bibitem [{\citenamefont {Wu}\ \emph {et~al.}(2026)\citenamefont {Wu}, \citenamefont {Li}, \citenamefont {Du},\ and\ \citenamefont {Zhang}}]{Wu:2025vfs}%
  \BibitemOpen
  \bibfield  {author} {\bibinfo {author} {\bibfnamefont {P.-J.}\ \bibnamefont {Wu}}, \bibinfo {author} {\bibfnamefont {T.-N.}\ \bibnamefont {Li}}, \bibinfo {author} {\bibfnamefont {G.-H.}\ \bibnamefont {Du}}, \ and\ \bibinfo {author} {\bibfnamefont {X.}~\bibnamefont {Zhang}},\ }\href {\doibase 10.1088/1674-1137/ae3be2} {\bibfield  {journal} {\bibinfo  {journal} {Chin. Phys. C}\ }\textbf {\bibinfo {volume} {50}},\ \bibinfo {pages} {045105} (\bibinfo {year} {2026})},\ \Eprint {http://arxiv.org/abs/2509.02945} {arXiv:2509.02945 [astro-ph.CO]} \BibitemShut {NoStop}%
\bibitem [{\citenamefont {Santos}\ \emph {et~al.}(2025)\citenamefont {Santos}, \citenamefont {Morais}, \citenamefont {Pan}, \citenamefont {Yang},\ and\ \citenamefont {Di~Valentino}}]{Santos:2025wiv}%
  \BibitemOpen
  \bibfield  {author} {\bibinfo {author} {\bibfnamefont {F.~B. M.~d.}\ \bibnamefont {Santos}}, \bibinfo {author} {\bibfnamefont {J.}~\bibnamefont {Morais}}, \bibinfo {author} {\bibfnamefont {S.}~\bibnamefont {Pan}}, \bibinfo {author} {\bibfnamefont {W.}~\bibnamefont {Yang}}, \ and\ \bibinfo {author} {\bibfnamefont {E.}~\bibnamefont {Di~Valentino}},\ }\href@noop {} {\  (\bibinfo {year} {2025})},\ \Eprint {http://arxiv.org/abs/2504.04646} {arXiv:2504.04646 [astro-ph.CO]} \BibitemShut {NoStop}%
\bibitem [{\citenamefont {Khoury}\ \emph {et~al.}(2025)\citenamefont {Khoury}, \citenamefont {Lin},\ and\ \citenamefont {Trodden}}]{Khoury:2025txd}%
  \BibitemOpen
  \bibfield  {author} {\bibinfo {author} {\bibfnamefont {J.}~\bibnamefont {Khoury}}, \bibinfo {author} {\bibfnamefont {M.-X.}\ \bibnamefont {Lin}}, \ and\ \bibinfo {author} {\bibfnamefont {M.}~\bibnamefont {Trodden}},\ }\href {\doibase 10.1103/w4qb-plk8} {\bibfield  {journal} {\bibinfo  {journal} {Phys. Rev. Lett.}\ }\textbf {\bibinfo {volume} {135}},\ \bibinfo {pages} {181001} (\bibinfo {year} {2025})},\ \Eprint {http://arxiv.org/abs/2503.16415} {arXiv:2503.16415 [astro-ph.CO]} \BibitemShut {NoStop}%
\bibitem [{\citenamefont {Li}\ and\ \citenamefont {Zhang}(2025)}]{Li:2025ula}%
  \BibitemOpen
  \bibfield  {author} {\bibinfo {author} {\bibfnamefont {Y.-H.}\ \bibnamefont {Li}}\ and\ \bibinfo {author} {\bibfnamefont {X.}~\bibnamefont {Zhang}},\ }\href {\doibase 10.1088/1475-7516/2025/12/018} {\bibfield  {journal} {\bibinfo  {journal} {JCAP}\ }\textbf {\bibinfo {volume} {12}},\ \bibinfo {pages} {018} (\bibinfo {year} {2025})},\ \Eprint {http://arxiv.org/abs/2506.18477} {arXiv:2506.18477 [astro-ph.CO]} \BibitemShut {NoStop}%
\bibitem [{\citenamefont {Kessler}\ \emph {et~al.}(2025)\citenamefont {Kessler}, \citenamefont {Escamilla}, \citenamefont {Pan},\ and\ \citenamefont {Di~Valentino}}]{Kessler:2025kju}%
  \BibitemOpen
  \bibfield  {author} {\bibinfo {author} {\bibfnamefont {D.~A.}\ \bibnamefont {Kessler}}, \bibinfo {author} {\bibfnamefont {L.~A.}\ \bibnamefont {Escamilla}}, \bibinfo {author} {\bibfnamefont {S.}~\bibnamefont {Pan}}, \ and\ \bibinfo {author} {\bibfnamefont {E.}~\bibnamefont {Di~Valentino}},\ }\href@noop {} {\  (\bibinfo {year} {2025})},\ \Eprint {http://arxiv.org/abs/2504.00776} {arXiv:2504.00776 [astro-ph.CO]} \BibitemShut {NoStop}%
\bibitem [{\citenamefont {Smith}\ \emph {et~al.}(2025)\citenamefont {Smith}, \citenamefont {{\"O}z{\"u}lker}, \citenamefont {Di~Valentino},\ and\ \citenamefont {van~de Bruck}}]{Smith:2025icl}%
  \BibitemOpen
  \bibfield  {author} {\bibinfo {author} {\bibfnamefont {A.}~\bibnamefont {Smith}}, \bibinfo {author} {\bibfnamefont {E.}~\bibnamefont {{\"O}z{\"u}lker}}, \bibinfo {author} {\bibfnamefont {E.}~\bibnamefont {Di~Valentino}}, \ and\ \bibinfo {author} {\bibfnamefont {C.}~\bibnamefont {van~de Bruck}},\ }\href@noop {} {\  (\bibinfo {year} {2025})},\ \Eprint {http://arxiv.org/abs/2510.21931} {arXiv:2510.21931 [astro-ph.CO]} \BibitemShut {NoStop}%
\bibitem [{\citenamefont {Amendola}(2000)}]{Amendola:1999er}%
  \BibitemOpen
  \bibfield  {author} {\bibinfo {author} {\bibfnamefont {L.}~\bibnamefont {Amendola}},\ }\href {\doibase 10.1103/PhysRevD.62.043511} {\bibfield  {journal} {\bibinfo  {journal} {Phys. Rev. D}\ }\textbf {\bibinfo {volume} {62}},\ \bibinfo {pages} {043511} (\bibinfo {year} {2000})},\ \Eprint {http://arxiv.org/abs/astro-ph/9908023} {arXiv:astro-ph/9908023} \BibitemShut {NoStop}%
\bibitem [{\citenamefont {Wang}\ \emph {et~al.}(2016)\citenamefont {Wang}, \citenamefont {Abdalla}, \citenamefont {Atrio-Barandela},\ and\ \citenamefont {Pavon}}]{Wang:2016lxa}%
  \BibitemOpen
  \bibfield  {author} {\bibinfo {author} {\bibfnamefont {B.}~\bibnamefont {Wang}}, \bibinfo {author} {\bibfnamefont {E.}~\bibnamefont {Abdalla}}, \bibinfo {author} {\bibfnamefont {F.}~\bibnamefont {Atrio-Barandela}}, \ and\ \bibinfo {author} {\bibfnamefont {D.}~\bibnamefont {Pavon}},\ }\href {\doibase 10.1088/0034-4885/79/9/096901} {\bibfield  {journal} {\bibinfo  {journal} {Rept. Prog. Phys.}\ }\textbf {\bibinfo {volume} {79}},\ \bibinfo {pages} {096901} (\bibinfo {year} {2016})},\ \Eprint {http://arxiv.org/abs/1603.08299} {arXiv:1603.08299 [astro-ph.CO]} \BibitemShut {NoStop}%
\bibitem [{\citenamefont {Bettoni}\ and\ \citenamefont {Liberati}(2013)}]{Bettoni:2013diz}%
  \BibitemOpen
  \bibfield  {author} {\bibinfo {author} {\bibfnamefont {D.}~\bibnamefont {Bettoni}}\ and\ \bibinfo {author} {\bibfnamefont {S.}~\bibnamefont {Liberati}},\ }\href {\doibase 10.1103/PhysRevD.88.084020} {\bibfield  {journal} {\bibinfo  {journal} {Phys. Rev. D}\ }\textbf {\bibinfo {volume} {88}},\ \bibinfo {pages} {084020} (\bibinfo {year} {2013})},\ \Eprint {http://arxiv.org/abs/1306.6724} {arXiv:1306.6724 [gr-qc]} \BibitemShut {NoStop}%
\bibitem [{\citenamefont {Zumalac{\'a}rregui}\ and\ \citenamefont {Garc{\'\i}a-Bellido}(2014)}]{Zumalacarregui:2013pma}%
  \BibitemOpen
  \bibfield  {author} {\bibinfo {author} {\bibfnamefont {M.}~\bibnamefont {Zumalac{\'a}rregui}}\ and\ \bibinfo {author} {\bibfnamefont {J.}~\bibnamefont {Garc{\'\i}a-Bellido}},\ }\href {\doibase 10.1103/PhysRevD.89.064046} {\bibfield  {journal} {\bibinfo  {journal} {Phys. Rev. D}\ }\textbf {\bibinfo {volume} {89}},\ \bibinfo {pages} {064046} (\bibinfo {year} {2014})},\ \Eprint {http://arxiv.org/abs/1308.4685} {arXiv:1308.4685 [gr-qc]} \BibitemShut {NoStop}%
\bibitem [{\citenamefont {Gubitosi}\ \emph {et~al.}(2013)\citenamefont {Gubitosi}, \citenamefont {Piazza},\ and\ \citenamefont {Vernizzi}}]{Gubitosi:2012hu}%
  \BibitemOpen
  \bibfield  {author} {\bibinfo {author} {\bibfnamefont {G.}~\bibnamefont {Gubitosi}}, \bibinfo {author} {\bibfnamefont {F.}~\bibnamefont {Piazza}}, \ and\ \bibinfo {author} {\bibfnamefont {F.}~\bibnamefont {Vernizzi}},\ }\href {\doibase 10.1088/1475-7516/2013/02/032} {\bibfield  {journal} {\bibinfo  {journal} {JCAP}\ }\textbf {\bibinfo {volume} {02}},\ \bibinfo {pages} {032} (\bibinfo {year} {2013})},\ \Eprint {http://arxiv.org/abs/1210.0201} {arXiv:1210.0201 [hep-th]} \BibitemShut {NoStop}%
\bibitem [{\citenamefont {Frusciante}\ and\ \citenamefont {Perenon}(2020)}]{Frusciante:2019xia}%
  \BibitemOpen
  \bibfield  {author} {\bibinfo {author} {\bibfnamefont {N.}~\bibnamefont {Frusciante}}\ and\ \bibinfo {author} {\bibfnamefont {L.}~\bibnamefont {Perenon}},\ }\href {\doibase 10.1016/j.physrep.2020.02.004} {\bibfield  {journal} {\bibinfo  {journal} {Phys. Rept.}\ }\textbf {\bibinfo {volume} {857}},\ \bibinfo {pages} {1} (\bibinfo {year} {2020})},\ \Eprint {http://arxiv.org/abs/1907.03150} {arXiv:1907.03150 [astro-ph.CO]} \BibitemShut {NoStop}%
\bibitem [{\citenamefont {Wang}\ \emph {et~al.}(2024)\citenamefont {Wang}, \citenamefont {Abdalla}, \citenamefont {Atrio-Barandela},\ and\ \citenamefont {Pav{\'o}n}}]{Wang:2024vmw}%
  \BibitemOpen
  \bibfield  {author} {\bibinfo {author} {\bibfnamefont {B.}~\bibnamefont {Wang}}, \bibinfo {author} {\bibfnamefont {E.}~\bibnamefont {Abdalla}}, \bibinfo {author} {\bibfnamefont {F.}~\bibnamefont {Atrio-Barandela}}, \ and\ \bibinfo {author} {\bibfnamefont {D.}~\bibnamefont {Pav{\'o}n}},\ }\href {\doibase 10.1088/1361-6633/ad2527} {\bibfield  {journal} {\bibinfo  {journal} {Rept. Prog. Phys.}\ }\textbf {\bibinfo {volume} {87}},\ \bibinfo {pages} {036901} (\bibinfo {year} {2024})},\ \Eprint {http://arxiv.org/abs/2402.00819} {arXiv:2402.00819 [astro-ph.CO]} \BibitemShut {NoStop}%
\bibitem [{\citenamefont {Zhang}\ \emph {et~al.}(2008)\citenamefont {Zhang}, \citenamefont {Liu},\ and\ \citenamefont {Zhang}}]{Zhang:2007uh}%
  \BibitemOpen
  \bibfield  {author} {\bibinfo {author} {\bibfnamefont {J.}~\bibnamefont {Zhang}}, \bibinfo {author} {\bibfnamefont {H.}~\bibnamefont {Liu}}, \ and\ \bibinfo {author} {\bibfnamefont {X.}~\bibnamefont {Zhang}},\ }\href {\doibase 10.1016/j.physletb.2007.10.086} {\bibfield  {journal} {\bibinfo  {journal} {Phys. Lett. B}\ }\textbf {\bibinfo {volume} {659}},\ \bibinfo {pages} {26} (\bibinfo {year} {2008})},\ \Eprint {http://arxiv.org/abs/0705.4145} {arXiv:0705.4145 [astro-ph]} \BibitemShut {NoStop}%
\bibitem [{\citenamefont {Salvatelli}\ \emph {et~al.}(2013)\citenamefont {Salvatelli}, \citenamefont {Marchini}, \citenamefont {Lopez-Honorez},\ and\ \citenamefont {Mena}}]{Salvatelli:2013wra}%
  \BibitemOpen
  \bibfield  {author} {\bibinfo {author} {\bibfnamefont {V.}~\bibnamefont {Salvatelli}}, \bibinfo {author} {\bibfnamefont {A.}~\bibnamefont {Marchini}}, \bibinfo {author} {\bibfnamefont {L.}~\bibnamefont {Lopez-Honorez}}, \ and\ \bibinfo {author} {\bibfnamefont {O.}~\bibnamefont {Mena}},\ }\href {\doibase 10.1103/PhysRevD.88.023531} {\bibfield  {journal} {\bibinfo  {journal} {Phys. Rev. D}\ }\textbf {\bibinfo {volume} {88}},\ \bibinfo {pages} {023531} (\bibinfo {year} {2013})},\ \Eprint {http://arxiv.org/abs/1304.7119} {arXiv:1304.7119 [astro-ph.CO]} \BibitemShut {NoStop}%
\bibitem [{\citenamefont {Li}\ \emph {et~al.}(2016)\citenamefont {Li}, \citenamefont {Zhang},\ and\ \citenamefont {Zhang}}]{Li:2015vla}%
  \BibitemOpen
  \bibfield  {author} {\bibinfo {author} {\bibfnamefont {Y.-H.}\ \bibnamefont {Li}}, \bibinfo {author} {\bibfnamefont {J.-F.}\ \bibnamefont {Zhang}}, \ and\ \bibinfo {author} {\bibfnamefont {X.}~\bibnamefont {Zhang}},\ }\href {\doibase 10.1103/PhysRevD.93.023002} {\bibfield  {journal} {\bibinfo  {journal} {Phys. Rev. D}\ }\textbf {\bibinfo {volume} {93}},\ \bibinfo {pages} {023002} (\bibinfo {year} {2016})},\ \Eprint {http://arxiv.org/abs/1506.06349} {arXiv:1506.06349 [astro-ph.CO]} \BibitemShut {NoStop}%
\bibitem [{\citenamefont {Murgia}\ \emph {et~al.}(2016)\citenamefont {Murgia}, \citenamefont {Gariazzo},\ and\ \citenamefont {Fornengo}}]{Murgia:2016ccp}%
  \BibitemOpen
  \bibfield  {author} {\bibinfo {author} {\bibfnamefont {R.}~\bibnamefont {Murgia}}, \bibinfo {author} {\bibfnamefont {S.}~\bibnamefont {Gariazzo}}, \ and\ \bibinfo {author} {\bibfnamefont {N.}~\bibnamefont {Fornengo}},\ }\href {\doibase 10.1088/1475-7516/2016/04/014} {\bibfield  {journal} {\bibinfo  {journal} {JCAP}\ }\textbf {\bibinfo {volume} {04}},\ \bibinfo {pages} {014} (\bibinfo {year} {2016})},\ \Eprint {http://arxiv.org/abs/1602.01765} {arXiv:1602.01765 [astro-ph.CO]} \BibitemShut {NoStop}%
\bibitem [{\citenamefont {Kumar}\ and\ \citenamefont {Nunes}(2017)}]{Kumar:2017dnp}%
  \BibitemOpen
  \bibfield  {author} {\bibinfo {author} {\bibfnamefont {S.}~\bibnamefont {Kumar}}\ and\ \bibinfo {author} {\bibfnamefont {R.~C.}\ \bibnamefont {Nunes}},\ }\href {\doibase 10.1103/PhysRevD.96.103511} {\bibfield  {journal} {\bibinfo  {journal} {Phys. Rev. D}\ }\textbf {\bibinfo {volume} {96}},\ \bibinfo {pages} {103511} (\bibinfo {year} {2017})},\ \Eprint {http://arxiv.org/abs/1702.02143} {arXiv:1702.02143 [astro-ph.CO]} \BibitemShut {NoStop}%
\bibitem [{\citenamefont {Di~Valentino}\ \emph {et~al.}(2017)\citenamefont {Di~Valentino}, \citenamefont {Melchiorri},\ and\ \citenamefont {Mena}}]{DiValentino:2017iww}%
  \BibitemOpen
  \bibfield  {author} {\bibinfo {author} {\bibfnamefont {E.}~\bibnamefont {Di~Valentino}}, \bibinfo {author} {\bibfnamefont {A.}~\bibnamefont {Melchiorri}}, \ and\ \bibinfo {author} {\bibfnamefont {O.}~\bibnamefont {Mena}},\ }\href {\doibase 10.1103/PhysRevD.96.043503} {\bibfield  {journal} {\bibinfo  {journal} {Phys. Rev. D}\ }\textbf {\bibinfo {volume} {96}},\ \bibinfo {pages} {043503} (\bibinfo {year} {2017})},\ \Eprint {http://arxiv.org/abs/1704.08342} {arXiv:1704.08342 [astro-ph.CO]} \BibitemShut {NoStop}%
\bibitem [{\citenamefont {Di~Valentino}(2021)}]{DiValentino:2020vnx}%
  \BibitemOpen
  \bibfield  {author} {\bibinfo {author} {\bibfnamefont {E.}~\bibnamefont {Di~Valentino}},\ }\href {\doibase 10.1093/mnras/stab187} {\bibfield  {journal} {\bibinfo  {journal} {Mon. Not. Roy. Astron. Soc.}\ }\textbf {\bibinfo {volume} {502}},\ \bibinfo {pages} {2065} (\bibinfo {year} {2021})},\ \Eprint {http://arxiv.org/abs/2011.00246} {arXiv:2011.00246 [astro-ph.CO]} \BibitemShut {NoStop}%
\bibitem [{\citenamefont {Gao}\ \emph {et~al.}(2021)\citenamefont {Gao}, \citenamefont {Zhao}, \citenamefont {Xue},\ and\ \citenamefont {Zhang}}]{Gao:2021xnk}%
  \BibitemOpen
  \bibfield  {author} {\bibinfo {author} {\bibfnamefont {L.-Y.}\ \bibnamefont {Gao}}, \bibinfo {author} {\bibfnamefont {Z.-W.}\ \bibnamefont {Zhao}}, \bibinfo {author} {\bibfnamefont {S.-S.}\ \bibnamefont {Xue}}, \ and\ \bibinfo {author} {\bibfnamefont {X.}~\bibnamefont {Zhang}},\ }\href {\doibase 10.1088/1475-7516/2021/07/005} {\bibfield  {journal} {\bibinfo  {journal} {JCAP}\ }\textbf {\bibinfo {volume} {07}},\ \bibinfo {pages} {005} (\bibinfo {year} {2021})},\ \Eprint {http://arxiv.org/abs/2101.10714} {arXiv:2101.10714 [astro-ph.CO]} \BibitemShut {NoStop}%
\bibitem [{\citenamefont {Pan}\ and\ \citenamefont {Yang}(2023)}]{Pan:2023mie}%
  \BibitemOpen
  \bibfield  {author} {\bibinfo {author} {\bibfnamefont {S.}~\bibnamefont {Pan}}\ and\ \bibinfo {author} {\bibfnamefont {W.}~\bibnamefont {Yang}},\ }\href {\doibase 10.1007/978-981-99-0177-7\_29} {\  (\bibinfo {year} {2023}),\ 10.1007/978-981-99-0177-7\_29},\ \Eprint {http://arxiv.org/abs/2310.07260} {arXiv:2310.07260 [astro-ph.CO]} \BibitemShut {NoStop}%
\bibitem [{\citenamefont {Forconi}\ \emph {et~al.}(2024)\citenamefont {Forconi}, \citenamefont {Giar{\`e}}, \citenamefont {Mena}, \citenamefont {Ruchika}, \citenamefont {Di~Valentino}, \citenamefont {Melchiorri},\ and\ \citenamefont {Nunes}}]{Forconi:2023hsj}%
  \BibitemOpen
  \bibfield  {author} {\bibinfo {author} {\bibfnamefont {M.}~\bibnamefont {Forconi}}, \bibinfo {author} {\bibfnamefont {W.}~\bibnamefont {Giar{\`e}}}, \bibinfo {author} {\bibfnamefont {O.}~\bibnamefont {Mena}}, \bibinfo {author} {\bibnamefont {Ruchika}}, \bibinfo {author} {\bibfnamefont {E.}~\bibnamefont {Di~Valentino}}, \bibinfo {author} {\bibfnamefont {A.}~\bibnamefont {Melchiorri}}, \ and\ \bibinfo {author} {\bibfnamefont {R.~C.}\ \bibnamefont {Nunes}},\ }\href {\doibase 10.1088/1475-7516/2024/05/097} {\bibfield  {journal} {\bibinfo  {journal} {JCAP}\ }\textbf {\bibinfo {volume} {05}},\ \bibinfo {pages} {097} (\bibinfo {year} {2024})},\ \Eprint {http://arxiv.org/abs/2312.11074} {arXiv:2312.11074 [astro-ph.CO]} \BibitemShut {NoStop}%
\bibitem [{\citenamefont {Pourtsidou}\ and\ \citenamefont {Tram}(2016)}]{Pourtsidou:2016ico}%
  \BibitemOpen
  \bibfield  {author} {\bibinfo {author} {\bibfnamefont {A.}~\bibnamefont {Pourtsidou}}\ and\ \bibinfo {author} {\bibfnamefont {T.}~\bibnamefont {Tram}},\ }\href {\doibase 10.1103/PhysRevD.94.043518} {\bibfield  {journal} {\bibinfo  {journal} {Phys. Rev. D}\ }\textbf {\bibinfo {volume} {94}},\ \bibinfo {pages} {043518} (\bibinfo {year} {2016})},\ \Eprint {http://arxiv.org/abs/1604.04222} {arXiv:1604.04222 [astro-ph.CO]} \BibitemShut {NoStop}%
\bibitem [{\citenamefont {Nunes}\ and\ \citenamefont {Di~Valentino}(2021)}]{Nunes:2021zzi}%
  \BibitemOpen
  \bibfield  {author} {\bibinfo {author} {\bibfnamefont {R.~C.}\ \bibnamefont {Nunes}}\ and\ \bibinfo {author} {\bibfnamefont {E.}~\bibnamefont {Di~Valentino}},\ }\href {\doibase 10.1103/PhysRevD.104.063529} {\bibfield  {journal} {\bibinfo  {journal} {Phys. Rev. D}\ }\textbf {\bibinfo {volume} {104}},\ \bibinfo {pages} {063529} (\bibinfo {year} {2021})},\ \Eprint {http://arxiv.org/abs/2107.09151} {arXiv:2107.09151 [astro-ph.CO]} \BibitemShut {NoStop}%
\bibitem [{\citenamefont {Wang}\ \emph {et~al.}(2022)\citenamefont {Wang}, \citenamefont {Zhang}, \citenamefont {He}, \citenamefont {Zhang},\ and\ \citenamefont {Zhang}}]{Wang:2021kxc}%
  \BibitemOpen
  \bibfield  {author} {\bibinfo {author} {\bibfnamefont {L.-F.}\ \bibnamefont {Wang}}, \bibinfo {author} {\bibfnamefont {J.-H.}\ \bibnamefont {Zhang}}, \bibinfo {author} {\bibfnamefont {D.-Z.}\ \bibnamefont {He}}, \bibinfo {author} {\bibfnamefont {J.-F.}\ \bibnamefont {Zhang}}, \ and\ \bibinfo {author} {\bibfnamefont {X.}~\bibnamefont {Zhang}},\ }\href {\doibase 10.1093/mnras/stac1468} {\bibfield  {journal} {\bibinfo  {journal} {Mon. Not. Roy. Astron. Soc.}\ }\textbf {\bibinfo {volume} {514}},\ \bibinfo {pages} {1433} (\bibinfo {year} {2022})},\ \Eprint {http://arxiv.org/abs/2102.09331} {arXiv:2102.09331 [astro-ph.CO]} \BibitemShut {NoStop}%
\bibitem [{\citenamefont {Lucca}\ and\ \citenamefont {Hooper}(2020)}]{Lucca:2020zjb}%
  \BibitemOpen
  \bibfield  {author} {\bibinfo {author} {\bibfnamefont {M.}~\bibnamefont {Lucca}}\ and\ \bibinfo {author} {\bibfnamefont {D.~C.}\ \bibnamefont {Hooper}},\ }\href {\doibase 10.1103/PhysRevD.102.123502} {\bibfield  {journal} {\bibinfo  {journal} {Phys. Rev. D}\ }\textbf {\bibinfo {volume} {102}},\ \bibinfo {pages} {123502} (\bibinfo {year} {2020})},\ \Eprint {http://arxiv.org/abs/2002.06127} {arXiv:2002.06127 [astro-ph.CO]} \BibitemShut {NoStop}%
\bibitem [{\citenamefont {Zhai}\ \emph {et~al.}(2023)\citenamefont {Zhai}, \citenamefont {Giar{\`e}}, \citenamefont {van~de Bruck}, \citenamefont {Di~Valentino}, \citenamefont {Mena},\ and\ \citenamefont {Nunes}}]{Zhai:2023yny}%
  \BibitemOpen
  \bibfield  {author} {\bibinfo {author} {\bibfnamefont {Y.}~\bibnamefont {Zhai}}, \bibinfo {author} {\bibfnamefont {W.}~\bibnamefont {Giar{\`e}}}, \bibinfo {author} {\bibfnamefont {C.}~\bibnamefont {van~de Bruck}}, \bibinfo {author} {\bibfnamefont {E.}~\bibnamefont {Di~Valentino}}, \bibinfo {author} {\bibfnamefont {O.}~\bibnamefont {Mena}}, \ and\ \bibinfo {author} {\bibfnamefont {R.~C.}\ \bibnamefont {Nunes}},\ }\href {\doibase 10.1088/1475-7516/2023/07/032} {\bibfield  {journal} {\bibinfo  {journal} {JCAP}\ }\textbf {\bibinfo {volume} {07}},\ \bibinfo {pages} {032} (\bibinfo {year} {2023})},\ \Eprint {http://arxiv.org/abs/2303.08201} {arXiv:2303.08201 [astro-ph.CO]} \BibitemShut {NoStop}%
\bibitem [{\citenamefont {Becker}\ \emph {et~al.}(2021)\citenamefont {Becker}, \citenamefont {Hooper}, \citenamefont {Kahlhoefer}, \citenamefont {Lesgourgues},\ and\ \citenamefont {Sch{\"o}neberg}}]{Becker:2020hzj}%
  \BibitemOpen
  \bibfield  {author} {\bibinfo {author} {\bibfnamefont {N.}~\bibnamefont {Becker}}, \bibinfo {author} {\bibfnamefont {D.~C.}\ \bibnamefont {Hooper}}, \bibinfo {author} {\bibfnamefont {F.}~\bibnamefont {Kahlhoefer}}, \bibinfo {author} {\bibfnamefont {J.}~\bibnamefont {Lesgourgues}}, \ and\ \bibinfo {author} {\bibfnamefont {N.}~\bibnamefont {Sch{\"o}neberg}},\ }\href {\doibase 10.1088/1475-7516/2021/02/019} {\bibfield  {journal} {\bibinfo  {journal} {JCAP}\ }\textbf {\bibinfo {volume} {02}},\ \bibinfo {pages} {019} (\bibinfo {year} {2021})},\ \Eprint {http://arxiv.org/abs/2010.04074} {arXiv:2010.04074 [astro-ph.CO]} \BibitemShut {NoStop}%
\bibitem [{\citenamefont {Hoerning}\ \emph {et~al.}(2025)\citenamefont {Hoerning}, \citenamefont {Landim}, \citenamefont {Ponte}, \citenamefont {Rolim}, \citenamefont {Abdalla},\ and\ \citenamefont {Abdalla}}]{Hoerning:2023hks}%
  \BibitemOpen
  \bibfield  {author} {\bibinfo {author} {\bibfnamefont {G.~A.}\ \bibnamefont {Hoerning}}, \bibinfo {author} {\bibfnamefont {R.~G.}\ \bibnamefont {Landim}}, \bibinfo {author} {\bibfnamefont {L.~O.}\ \bibnamefont {Ponte}}, \bibinfo {author} {\bibfnamefont {R.~P.}\ \bibnamefont {Rolim}}, \bibinfo {author} {\bibfnamefont {F.~B.}\ \bibnamefont {Abdalla}}, \ and\ \bibinfo {author} {\bibfnamefont {E.}~\bibnamefont {Abdalla}},\ }\href {\doibase 10.1103/6zrh-8fmv} {\bibfield  {journal} {\bibinfo  {journal} {Phys. Rev. D}\ }\textbf {\bibinfo {volume} {112}},\ \bibinfo {pages} {023523} (\bibinfo {year} {2025})},\ \Eprint {http://arxiv.org/abs/2308.05807} {arXiv:2308.05807 [astro-ph.CO]} \BibitemShut {NoStop}%
\bibitem [{\citenamefont {Giar{\`e}}\ \emph {et~al.}(2024{\natexlab{b}})\citenamefont {Giar{\`e}}, \citenamefont {Zhai}, \citenamefont {Pan}, \citenamefont {Di~Valentino}, \citenamefont {Nunes},\ and\ \citenamefont {van~de Bruck}}]{Giare:2024ytc}%
  \BibitemOpen
  \bibfield  {author} {\bibinfo {author} {\bibfnamefont {W.}~\bibnamefont {Giar{\`e}}}, \bibinfo {author} {\bibfnamefont {Y.}~\bibnamefont {Zhai}}, \bibinfo {author} {\bibfnamefont {S.}~\bibnamefont {Pan}}, \bibinfo {author} {\bibfnamefont {E.}~\bibnamefont {Di~Valentino}}, \bibinfo {author} {\bibfnamefont {R.~C.}\ \bibnamefont {Nunes}}, \ and\ \bibinfo {author} {\bibfnamefont {C.}~\bibnamefont {van~de Bruck}},\ }\href {\doibase 10.1103/PhysRevD.110.063527} {\bibfield  {journal} {\bibinfo  {journal} {Phys. Rev. D}\ }\textbf {\bibinfo {volume} {110}},\ \bibinfo {pages} {063527} (\bibinfo {year} {2024}{\natexlab{b}})},\ \Eprint {http://arxiv.org/abs/2404.02110} {arXiv:2404.02110 [astro-ph.CO]} \BibitemShut {NoStop}%
\bibitem [{\citenamefont {Escamilla}\ \emph {et~al.}(2023)\citenamefont {Escamilla}, \citenamefont {Akarsu}, \citenamefont {Di~Valentino},\ and\ \citenamefont {Vazquez}}]{Escamilla:2023shf}%
  \BibitemOpen
  \bibfield  {author} {\bibinfo {author} {\bibfnamefont {L.~A.}\ \bibnamefont {Escamilla}}, \bibinfo {author} {\bibfnamefont {O.}~\bibnamefont {Akarsu}}, \bibinfo {author} {\bibfnamefont {E.}~\bibnamefont {Di~Valentino}}, \ and\ \bibinfo {author} {\bibfnamefont {J.~A.}\ \bibnamefont {Vazquez}},\ }\href {\doibase 10.1088/1475-7516/2023/11/051} {\bibfield  {journal} {\bibinfo  {journal} {JCAP}\ }\textbf {\bibinfo {volume} {11}},\ \bibinfo {pages} {051} (\bibinfo {year} {2023})},\ \Eprint {http://arxiv.org/abs/2305.16290} {arXiv:2305.16290 [astro-ph.CO]} \BibitemShut {NoStop}%
\bibitem [{\citenamefont {Di~Valentino}\ \emph {et~al.}(2020)\citenamefont {Di~Valentino}, \citenamefont {Melchiorri}, \citenamefont {Mena},\ and\ \citenamefont {Vagnozzi}}]{DiValentino:2019ffd}%
  \BibitemOpen
  \bibfield  {author} {\bibinfo {author} {\bibfnamefont {E.}~\bibnamefont {Di~Valentino}}, \bibinfo {author} {\bibfnamefont {A.}~\bibnamefont {Melchiorri}}, \bibinfo {author} {\bibfnamefont {O.}~\bibnamefont {Mena}}, \ and\ \bibinfo {author} {\bibfnamefont {S.}~\bibnamefont {Vagnozzi}},\ }\href {\doibase 10.1016/j.dark.2020.100666} {\bibfield  {journal} {\bibinfo  {journal} {Phys. Dark Univ.}\ }\textbf {\bibinfo {volume} {30}},\ \bibinfo {pages} {100666} (\bibinfo {year} {2020})},\ \Eprint {http://arxiv.org/abs/1908.04281} {arXiv:1908.04281 [astro-ph.CO]} \BibitemShut {NoStop}%
\bibitem [{\citenamefont {Li}\ \emph {et~al.}(2024{\natexlab{a}})\citenamefont {Li}, \citenamefont {Wu}, \citenamefont {Du}, \citenamefont {Jin}, \citenamefont {Li}, \citenamefont {Zhang},\ and\ \citenamefont {Zhang}}]{Li:2024qso}%
  \BibitemOpen
  \bibfield  {author} {\bibinfo {author} {\bibfnamefont {T.-N.}\ \bibnamefont {Li}}, \bibinfo {author} {\bibfnamefont {P.-J.}\ \bibnamefont {Wu}}, \bibinfo {author} {\bibfnamefont {G.-H.}\ \bibnamefont {Du}}, \bibinfo {author} {\bibfnamefont {S.-J.}\ \bibnamefont {Jin}}, \bibinfo {author} {\bibfnamefont {H.-L.}\ \bibnamefont {Li}}, \bibinfo {author} {\bibfnamefont {J.-F.}\ \bibnamefont {Zhang}}, \ and\ \bibinfo {author} {\bibfnamefont {X.}~\bibnamefont {Zhang}},\ }\href {\doibase 10.3847/1538-4357/ad87f0} {\bibfield  {journal} {\bibinfo  {journal} {Astrophys. J.}\ }\textbf {\bibinfo {volume} {976}},\ \bibinfo {pages} {1} (\bibinfo {year} {2024}{\natexlab{a}})},\ \Eprint {http://arxiv.org/abs/2407.14934} {arXiv:2407.14934 [astro-ph.CO]} \BibitemShut {NoStop}%
\bibitem [{\citenamefont {Halder}\ \emph {et~al.}(2024)\citenamefont {Halder}, \citenamefont {de~Haro}, \citenamefont {Saha},\ and\ \citenamefont {Pan}}]{Halder:2024uao}%
  \BibitemOpen
  \bibfield  {author} {\bibinfo {author} {\bibfnamefont {S.}~\bibnamefont {Halder}}, \bibinfo {author} {\bibfnamefont {J.}~\bibnamefont {de~Haro}}, \bibinfo {author} {\bibfnamefont {T.}~\bibnamefont {Saha}}, \ and\ \bibinfo {author} {\bibfnamefont {S.}~\bibnamefont {Pan}},\ }\href {\doibase 10.1103/PhysRevD.109.083522} {\bibfield  {journal} {\bibinfo  {journal} {Phys. Rev. D}\ }\textbf {\bibinfo {volume} {109}},\ \bibinfo {pages} {083522} (\bibinfo {year} {2024})},\ \Eprint {http://arxiv.org/abs/2403.01397} {arXiv:2403.01397 [gr-qc]} \BibitemShut {NoStop}%
\bibitem [{\citenamefont {Castello}\ \emph {et~al.}(2024)\citenamefont {Castello}, \citenamefont {Mancarella}, \citenamefont {Grimm}, \citenamefont {Sobral-Blanco}, \citenamefont {Tutusaus},\ and\ \citenamefont {Bonvin}}]{Castello:2023zjr}%
  \BibitemOpen
  \bibfield  {author} {\bibinfo {author} {\bibfnamefont {S.}~\bibnamefont {Castello}}, \bibinfo {author} {\bibfnamefont {M.}~\bibnamefont {Mancarella}}, \bibinfo {author} {\bibfnamefont {N.}~\bibnamefont {Grimm}}, \bibinfo {author} {\bibfnamefont {D.}~\bibnamefont {Sobral-Blanco}}, \bibinfo {author} {\bibfnamefont {I.}~\bibnamefont {Tutusaus}}, \ and\ \bibinfo {author} {\bibfnamefont {C.}~\bibnamefont {Bonvin}},\ }\href {\doibase 10.1088/1475-7516/2024/05/003} {\bibfield  {journal} {\bibinfo  {journal} {JCAP}\ }\textbf {\bibinfo {volume} {05}},\ \bibinfo {pages} {003} (\bibinfo {year} {2024})},\ \Eprint {http://arxiv.org/abs/2311.14425} {arXiv:2311.14425 [astro-ph.CO]} \BibitemShut {NoStop}%
\bibitem [{\citenamefont {Yao}\ and\ \citenamefont {Meng}(2023)}]{Yao:2023jau}%
  \BibitemOpen
  \bibfield  {author} {\bibinfo {author} {\bibfnamefont {Y.-H.}\ \bibnamefont {Yao}}\ and\ \bibinfo {author} {\bibfnamefont {X.-H.}\ \bibnamefont {Meng}},\ }\href {\doibase 10.1016/j.dark.2022.101165} {\bibfield  {journal} {\bibinfo  {journal} {Phys. Dark Univ.}\ }\textbf {\bibinfo {volume} {39}},\ \bibinfo {pages} {101165} (\bibinfo {year} {2023})},\ \Eprint {http://arxiv.org/abs/2207.05955} {arXiv:2207.05955 [astro-ph.CO]} \BibitemShut {NoStop}%
\bibitem [{\citenamefont {Li}\ \emph {et~al.}(2024{\natexlab{b}})\citenamefont {Li}, \citenamefont {Jin}, \citenamefont {Li}, \citenamefont {Zhang},\ and\ \citenamefont {Zhang}}]{Li:2023gtu}%
  \BibitemOpen
  \bibfield  {author} {\bibinfo {author} {\bibfnamefont {T.-N.}\ \bibnamefont {Li}}, \bibinfo {author} {\bibfnamefont {S.-J.}\ \bibnamefont {Jin}}, \bibinfo {author} {\bibfnamefont {H.-L.}\ \bibnamefont {Li}}, \bibinfo {author} {\bibfnamefont {J.-F.}\ \bibnamefont {Zhang}}, \ and\ \bibinfo {author} {\bibfnamefont {X.}~\bibnamefont {Zhang}},\ }\href {\doibase 10.3847/1538-4357/ad1bc9} {\bibfield  {journal} {\bibinfo  {journal} {Astrophys. J.}\ }\textbf {\bibinfo {volume} {963}},\ \bibinfo {pages} {52} (\bibinfo {year} {2024}{\natexlab{b}})},\ \Eprint {http://arxiv.org/abs/2310.15879} {arXiv:2310.15879 [astro-ph.CO]} \BibitemShut {NoStop}%
\bibitem [{\citenamefont {Mishra}\ \emph {et~al.}(2023)\citenamefont {Mishra}, \citenamefont {Pacif}, \citenamefont {Kumar},\ and\ \citenamefont {Bamba}}]{Mishra:2023ueo}%
  \BibitemOpen
  \bibfield  {author} {\bibinfo {author} {\bibfnamefont {K.~R.}\ \bibnamefont {Mishra}}, \bibinfo {author} {\bibfnamefont {S.~K.~J.}\ \bibnamefont {Pacif}}, \bibinfo {author} {\bibfnamefont {R.}~\bibnamefont {Kumar}}, \ and\ \bibinfo {author} {\bibfnamefont {K.}~\bibnamefont {Bamba}},\ }\href {\doibase 10.1016/j.dark.2023.101211} {\bibfield  {journal} {\bibinfo  {journal} {Phys. Dark Univ.}\ }\textbf {\bibinfo {volume} {40}},\ \bibinfo {pages} {101211} (\bibinfo {year} {2023})},\ \Eprint {http://arxiv.org/abs/2301.08743} {arXiv:2301.08743 [gr-qc]} \BibitemShut {NoStop}%
\bibitem [{\citenamefont {Nunes}\ \emph {et~al.}(2016)\citenamefont {Nunes}, \citenamefont {Pan},\ and\ \citenamefont {Saridakis}}]{Nunes:2016dlj}%
  \BibitemOpen
  \bibfield  {author} {\bibinfo {author} {\bibfnamefont {R.~C.}\ \bibnamefont {Nunes}}, \bibinfo {author} {\bibfnamefont {S.}~\bibnamefont {Pan}}, \ and\ \bibinfo {author} {\bibfnamefont {E.~N.}\ \bibnamefont {Saridakis}},\ }\href {\doibase 10.1103/PhysRevD.94.023508} {\bibfield  {journal} {\bibinfo  {journal} {Phys. Rev. D}\ }\textbf {\bibinfo {volume} {94}},\ \bibinfo {pages} {023508} (\bibinfo {year} {2016})},\ \Eprint {http://arxiv.org/abs/1605.01712} {arXiv:1605.01712 [astro-ph.CO]} \BibitemShut {NoStop}%
\bibitem [{\citenamefont {Silva}\ \emph {et~al.}(2025)\citenamefont {Silva}, \citenamefont {Sabogal}, \citenamefont {Scherer}, \citenamefont {Nunes}, \citenamefont {Di~Valentino},\ and\ \citenamefont {Kumar}}]{Silva:2025hxw}%
  \BibitemOpen
  \bibfield  {author} {\bibinfo {author} {\bibfnamefont {E.}~\bibnamefont {Silva}}, \bibinfo {author} {\bibfnamefont {M.~A.}\ \bibnamefont {Sabogal}}, \bibinfo {author} {\bibfnamefont {M.}~\bibnamefont {Scherer}}, \bibinfo {author} {\bibfnamefont {R.~C.}\ \bibnamefont {Nunes}}, \bibinfo {author} {\bibfnamefont {E.}~\bibnamefont {Di~Valentino}}, \ and\ \bibinfo {author} {\bibfnamefont {S.}~\bibnamefont {Kumar}},\ }\href {\doibase 10.1103/qqc6-76z4} {\bibfield  {journal} {\bibinfo  {journal} {Phys. Rev. D}\ }\textbf {\bibinfo {volume} {111}},\ \bibinfo {pages} {123511} (\bibinfo {year} {2025})},\ \Eprint {http://arxiv.org/abs/2503.23225} {arXiv:2503.23225 [astro-ph.CO]} \BibitemShut {NoStop}%
\bibitem [{\citenamefont {van~der Westhuizen}\ \emph {et~al.}(2025)\citenamefont {van~der Westhuizen}, \citenamefont {Abebe},\ and\ \citenamefont {Di~Valentino}}]{vanderWesthuizen:2025rip}%
  \BibitemOpen
  \bibfield  {author} {\bibinfo {author} {\bibfnamefont {M.}~\bibnamefont {van~der Westhuizen}}, \bibinfo {author} {\bibfnamefont {A.}~\bibnamefont {Abebe}}, \ and\ \bibinfo {author} {\bibfnamefont {E.}~\bibnamefont {Di~Valentino}},\ }\href {\doibase 10.1016/j.dark.2025.102121} {\bibfield  {journal} {\bibinfo  {journal} {Phys. Dark Univ.}\ }\textbf {\bibinfo {volume} {50}},\ \bibinfo {pages} {102121} (\bibinfo {year} {2025})},\ \Eprint {http://arxiv.org/abs/2509.04496} {arXiv:2509.04496 [gr-qc]} \BibitemShut {NoStop}%
\bibitem [{\citenamefont {Zhang}\ \emph {et~al.}(2025)\citenamefont {Zhang}, \citenamefont {Li}, \citenamefont {Du}, \citenamefont {Zhou}, \citenamefont {Gao}, \citenamefont {Zhang},\ and\ \citenamefont {Zhang}}]{Zhang:2025dwu}%
  \BibitemOpen
  \bibfield  {author} {\bibinfo {author} {\bibfnamefont {Y.-M.}\ \bibnamefont {Zhang}}, \bibinfo {author} {\bibfnamefont {T.-N.}\ \bibnamefont {Li}}, \bibinfo {author} {\bibfnamefont {G.-H.}\ \bibnamefont {Du}}, \bibinfo {author} {\bibfnamefont {S.-H.}\ \bibnamefont {Zhou}}, \bibinfo {author} {\bibfnamefont {L.-Y.}\ \bibnamefont {Gao}}, \bibinfo {author} {\bibfnamefont {J.-F.}\ \bibnamefont {Zhang}}, \ and\ \bibinfo {author} {\bibfnamefont {X.}~\bibnamefont {Zhang}},\ }\href@noop {} {\  (\bibinfo {year} {2025})},\ \Eprint {http://arxiv.org/abs/2510.12627} {arXiv:2510.12627 [astro-ph.CO]} \BibitemShut {NoStop}%
\bibitem [{\citenamefont {Wang}\ \emph {et~al.}(2026)\citenamefont {Wang}, \citenamefont {Cai}, \citenamefont {Guo},\ and\ \citenamefont {Wang}}]{Wang:2025znm}%
  \BibitemOpen
  \bibfield  {author} {\bibinfo {author} {\bibfnamefont {J.-Q.}\ \bibnamefont {Wang}}, \bibinfo {author} {\bibfnamefont {R.-G.}\ \bibnamefont {Cai}}, \bibinfo {author} {\bibfnamefont {Z.-K.}\ \bibnamefont {Guo}}, \ and\ \bibinfo {author} {\bibfnamefont {S.-J.}\ \bibnamefont {Wang}},\ }\href {\doibase 10.1103/r6cx-8ghz} {\bibfield  {journal} {\bibinfo  {journal} {Phys. Rev. D}\ }\textbf {\bibinfo {volume} {113}},\ \bibinfo {pages} {083534} (\bibinfo {year} {2026})},\ \Eprint {http://arxiv.org/abs/2508.01759} {arXiv:2508.01759 [astro-ph.CO]} \BibitemShut {NoStop}%
\bibitem [{\citenamefont {Li}\ \emph {et~al.}(2025{\natexlab{b}})\citenamefont {Li}, \citenamefont {Du}, \citenamefont {Li}, \citenamefont {Li}, \citenamefont {Ling}, \citenamefont {Zhang},\ and\ \citenamefont {Zhang}}]{Li:2025muv}%
  \BibitemOpen
  \bibfield  {author} {\bibinfo {author} {\bibfnamefont {T.-N.}\ \bibnamefont {Li}}, \bibinfo {author} {\bibfnamefont {G.-H.}\ \bibnamefont {Du}}, \bibinfo {author} {\bibfnamefont {Y.-H.}\ \bibnamefont {Li}}, \bibinfo {author} {\bibfnamefont {Y.}~\bibnamefont {Li}}, \bibinfo {author} {\bibfnamefont {J.-L.}\ \bibnamefont {Ling}}, \bibinfo {author} {\bibfnamefont {J.-F.}\ \bibnamefont {Zhang}}, \ and\ \bibinfo {author} {\bibfnamefont {X.}~\bibnamefont {Zhang}},\ }\href@noop {} {\  (\bibinfo {year} {2025}{\natexlab{b}})},\ \Eprint {http://arxiv.org/abs/2510.11363} {arXiv:2510.11363 [astro-ph.CO]} \BibitemShut {NoStop}%
\bibitem [{\citenamefont {Li}\ \emph {et~al.}(2026{\natexlab{b}})\citenamefont {Li}, \citenamefont {Du}, \citenamefont {Li}, \citenamefont {Wu}, \citenamefont {Jin}, \citenamefont {Zhang},\ and\ \citenamefont {Zhang}}]{Li:2025owk}%
  \BibitemOpen
  \bibfield  {author} {\bibinfo {author} {\bibfnamefont {T.-N.}\ \bibnamefont {Li}}, \bibinfo {author} {\bibfnamefont {G.-H.}\ \bibnamefont {Du}}, \bibinfo {author} {\bibfnamefont {Y.-H.}\ \bibnamefont {Li}}, \bibinfo {author} {\bibfnamefont {P.-J.}\ \bibnamefont {Wu}}, \bibinfo {author} {\bibfnamefont {S.-J.}\ \bibnamefont {Jin}}, \bibinfo {author} {\bibfnamefont {J.-F.}\ \bibnamefont {Zhang}}, \ and\ \bibinfo {author} {\bibfnamefont {X.}~\bibnamefont {Zhang}},\ }\href {\doibase 10.1007/s11433-025-2771-5} {\bibfield  {journal} {\bibinfo  {journal} {Sci. China Phys. Mech. Astron.}\ }\textbf {\bibinfo {volume} {69}},\ \bibinfo {pages} {210413} (\bibinfo {year} {2026}{\natexlab{b}})},\ \Eprint {http://arxiv.org/abs/2501.07361} {arXiv:2501.07361 [astro-ph.CO]} \BibitemShut {NoStop}%
\bibitem [{\citenamefont {Lyu}\ \emph {et~al.}(2026)\citenamefont {Lyu}, \citenamefont {Cai}, \citenamefont {Wang},\ and\ \citenamefont {Zeng}}]{Lyu:2025nsd}%
  \BibitemOpen
  \bibfield  {author} {\bibinfo {author} {\bibfnamefont {Z.-H.}\ \bibnamefont {Lyu}}, \bibinfo {author} {\bibfnamefont {R.-G.}\ \bibnamefont {Cai}}, \bibinfo {author} {\bibfnamefont {S.-J.}\ \bibnamefont {Wang}}, \ and\ \bibinfo {author} {\bibfnamefont {X.-X.}\ \bibnamefont {Zeng}},\ }\href {\doibase 10.1103/rmgx-rp87} {\bibfield  {journal} {\bibinfo  {journal} {Phys. Rev. D}\ }\textbf {\bibinfo {volume} {113}},\ \bibinfo {pages} {083041} (\bibinfo {year} {2026})},\ \Eprint {http://arxiv.org/abs/2511.16244} {arXiv:2511.16244 [astro-ph.CO]} \BibitemShut {NoStop}%
\bibitem [{\citenamefont {Yang}\ \emph {et~al.}(2025)\citenamefont {Yang}, \citenamefont {Zhang}, \citenamefont {Mena}, \citenamefont {Pan},\ and\ \citenamefont {Di~Valentino}}]{Yang:2025uyv}%
  \BibitemOpen
  \bibfield  {author} {\bibinfo {author} {\bibfnamefont {W.}~\bibnamefont {Yang}}, \bibinfo {author} {\bibfnamefont {S.}~\bibnamefont {Zhang}}, \bibinfo {author} {\bibfnamefont {O.}~\bibnamefont {Mena}}, \bibinfo {author} {\bibfnamefont {S.}~\bibnamefont {Pan}}, \ and\ \bibinfo {author} {\bibfnamefont {E.}~\bibnamefont {Di~Valentino}},\ }\href@noop {} {\  (\bibinfo {year} {2025})},\ \Eprint {http://arxiv.org/abs/2508.19109} {arXiv:2508.19109 [astro-ph.CO]} \BibitemShut {NoStop}%
\bibitem [{\citenamefont {Pan}\ \emph {et~al.}(2026)\citenamefont {Pan}, \citenamefont {Paul}, \citenamefont {Saridakis},\ and\ \citenamefont {Yang}}]{Pan:2025qwy}%
  \BibitemOpen
  \bibfield  {author} {\bibinfo {author} {\bibfnamefont {S.}~\bibnamefont {Pan}}, \bibinfo {author} {\bibfnamefont {S.}~\bibnamefont {Paul}}, \bibinfo {author} {\bibfnamefont {E.~N.}\ \bibnamefont {Saridakis}}, \ and\ \bibinfo {author} {\bibfnamefont {W.}~\bibnamefont {Yang}},\ }\href {\doibase 10.1103/5y21-k39n} {\bibfield  {journal} {\bibinfo  {journal} {Phys. Rev. D}\ }\textbf {\bibinfo {volume} {113}},\ \bibinfo {pages} {023515} (\bibinfo {year} {2026})},\ \Eprint {http://arxiv.org/abs/2504.00994} {arXiv:2504.00994 [astro-ph.CO]} \BibitemShut {NoStop}%
\bibitem [{\citenamefont {Chimento}\ \emph {et~al.}(2003)\citenamefont {Chimento}, \citenamefont {Jakubi}, \citenamefont {Pavon},\ and\ \citenamefont {Zimdahl}}]{Chimento:2003iea}%
  \BibitemOpen
  \bibfield  {author} {\bibinfo {author} {\bibfnamefont {L.~P.}\ \bibnamefont {Chimento}}, \bibinfo {author} {\bibfnamefont {A.~S.}\ \bibnamefont {Jakubi}}, \bibinfo {author} {\bibfnamefont {D.}~\bibnamefont {Pavon}}, \ and\ \bibinfo {author} {\bibfnamefont {W.}~\bibnamefont {Zimdahl}},\ }\href {\doibase 10.1103/PhysRevD.67.083513} {\bibfield  {journal} {\bibinfo  {journal} {Phys. Rev. D}\ }\textbf {\bibinfo {volume} {67}},\ \bibinfo {pages} {083513} (\bibinfo {year} {2003})},\ \Eprint {http://arxiv.org/abs/astro-ph/0303145} {arXiv:astro-ph/0303145} \BibitemShut {NoStop}%
\bibitem [{\citenamefont {Zhang}(2005{\natexlab{a}})}]{Zhang:2005rg}%
  \BibitemOpen
  \bibfield  {author} {\bibinfo {author} {\bibfnamefont {X.}~\bibnamefont {Zhang}},\ }\href {\doibase 10.1142/S0217732305017597} {\bibfield  {journal} {\bibinfo  {journal} {Mod. Phys. Lett. A}\ }\textbf {\bibinfo {volume} {20}},\ \bibinfo {pages} {2575} (\bibinfo {year} {2005}{\natexlab{a}})},\ \Eprint {http://arxiv.org/abs/astro-ph/0503072} {arXiv:astro-ph/0503072} \BibitemShut {NoStop}%
\bibitem [{\citenamefont {Zhang}(2005{\natexlab{b}})}]{Zhang:2005rj}%
  \BibitemOpen
  \bibfield  {author} {\bibinfo {author} {\bibfnamefont {X.}~\bibnamefont {Zhang}},\ }\href {\doibase 10.1016/j.physletb.2005.02.022} {\bibfield  {journal} {\bibinfo  {journal} {Phys. Lett. B}\ }\textbf {\bibinfo {volume} {611}},\ \bibinfo {pages} {1} (\bibinfo {year} {2005}{\natexlab{b}})},\ \Eprint {http://arxiv.org/abs/astro-ph/0503075} {arXiv:astro-ph/0503075} \BibitemShut {NoStop}%
\bibitem [{\citenamefont {Dutta}\ \emph {et~al.}(2017)\citenamefont {Dutta}, \citenamefont {Khyllep},\ and\ \citenamefont {Tamanini}}]{Dutta:2017kch}%
  \BibitemOpen
  \bibfield  {author} {\bibinfo {author} {\bibfnamefont {J.}~\bibnamefont {Dutta}}, \bibinfo {author} {\bibfnamefont {W.}~\bibnamefont {Khyllep}}, \ and\ \bibinfo {author} {\bibfnamefont {N.}~\bibnamefont {Tamanini}},\ }\href {\doibase 10.1103/PhysRevD.95.023515} {\bibfield  {journal} {\bibinfo  {journal} {Phys. Rev. D}\ }\textbf {\bibinfo {volume} {95}},\ \bibinfo {pages} {023515} (\bibinfo {year} {2017})},\ \Eprint {http://arxiv.org/abs/1701.00744} {arXiv:1701.00744 [gr-qc]} \BibitemShut {NoStop}%
\bibitem [{\citenamefont {Yang}\ \emph {et~al.}(2018)\citenamefont {Yang}, \citenamefont {Pan}, \citenamefont {Di~Valentino}, \citenamefont {Nunes}, \citenamefont {Vagnozzi},\ and\ \citenamefont {Mota}}]{Yang:2018euj}%
  \BibitemOpen
  \bibfield  {author} {\bibinfo {author} {\bibfnamefont {W.}~\bibnamefont {Yang}}, \bibinfo {author} {\bibfnamefont {S.}~\bibnamefont {Pan}}, \bibinfo {author} {\bibfnamefont {E.}~\bibnamefont {Di~Valentino}}, \bibinfo {author} {\bibfnamefont {R.~C.}\ \bibnamefont {Nunes}}, \bibinfo {author} {\bibfnamefont {S.}~\bibnamefont {Vagnozzi}}, \ and\ \bibinfo {author} {\bibfnamefont {D.~F.}\ \bibnamefont {Mota}},\ }\href {\doibase 10.1088/1475-7516/2018/09/019} {\bibfield  {journal} {\bibinfo  {journal} {JCAP}\ }\textbf {\bibinfo {volume} {09}},\ \bibinfo {pages} {019} (\bibinfo {year} {2018})},\ \Eprint {http://arxiv.org/abs/1805.08252} {arXiv:1805.08252 [astro-ph.CO]} \BibitemShut {NoStop}%
\bibitem [{\citenamefont {Guo}\ \emph {et~al.}(2019)\citenamefont {Guo}, \citenamefont {Zhang},\ and\ \citenamefont {Zhang}}]{Guo:2018ans}%
  \BibitemOpen
  \bibfield  {author} {\bibinfo {author} {\bibfnamefont {R.-Y.}\ \bibnamefont {Guo}}, \bibinfo {author} {\bibfnamefont {J.-F.}\ \bibnamefont {Zhang}}, \ and\ \bibinfo {author} {\bibfnamefont {X.}~\bibnamefont {Zhang}},\ }\href {\doibase 10.1088/1475-7516/2019/02/054} {\bibfield  {journal} {\bibinfo  {journal} {JCAP}\ }\textbf {\bibinfo {volume} {02}},\ \bibinfo {pages} {054} (\bibinfo {year} {2019})},\ \Eprint {http://arxiv.org/abs/1809.02340} {arXiv:1809.02340 [astro-ph.CO]} \BibitemShut {NoStop}%
\bibitem [{\citenamefont {Feng}\ \emph {et~al.}(2020)\citenamefont {Feng}, \citenamefont {He}, \citenamefont {Li}, \citenamefont {Zhang},\ and\ \citenamefont {Zhang}}]{Feng:2019jqa}%
  \BibitemOpen
  \bibfield  {author} {\bibinfo {author} {\bibfnamefont {L.}~\bibnamefont {Feng}}, \bibinfo {author} {\bibfnamefont {D.-Z.}\ \bibnamefont {He}}, \bibinfo {author} {\bibfnamefont {H.-L.}\ \bibnamefont {Li}}, \bibinfo {author} {\bibfnamefont {J.-F.}\ \bibnamefont {Zhang}}, \ and\ \bibinfo {author} {\bibfnamefont {X.}~\bibnamefont {Zhang}},\ }\href {\doibase 10.1007/s11433-019-1511-8} {\bibfield  {journal} {\bibinfo  {journal} {Sci. China Phys. Mech. Astron.}\ }\textbf {\bibinfo {volume} {63}},\ \bibinfo {pages} {290404} (\bibinfo {year} {2020})},\ \Eprint {http://arxiv.org/abs/1910.03872} {arXiv:1910.03872 [astro-ph.CO]} \BibitemShut {NoStop}%
\bibitem [{\citenamefont {Li}\ \emph {et~al.}(2020)\citenamefont {Li}, \citenamefont {Zhang},\ and\ \citenamefont {Zhang}}]{Li:2020gtk}%
  \BibitemOpen
  \bibfield  {author} {\bibinfo {author} {\bibfnamefont {H.-L.}\ \bibnamefont {Li}}, \bibinfo {author} {\bibfnamefont {J.-F.}\ \bibnamefont {Zhang}}, \ and\ \bibinfo {author} {\bibfnamefont {X.}~\bibnamefont {Zhang}},\ }\href {\doibase 10.1088/1572-9494/abb7c9} {\bibfield  {journal} {\bibinfo  {journal} {Commun. Theor. Phys.}\ }\textbf {\bibinfo {volume} {72}},\ \bibinfo {pages} {125401} (\bibinfo {year} {2020})},\ \Eprint {http://arxiv.org/abs/2005.12041} {arXiv:2005.12041 [astro-ph.CO]} \BibitemShut {NoStop}%
\bibitem [{\citenamefont {Gao}\ \emph {et~al.}(2024)\citenamefont {Gao}, \citenamefont {Xue},\ and\ \citenamefont {Zhang}}]{Gao:2022ahg}%
  \BibitemOpen
  \bibfield  {author} {\bibinfo {author} {\bibfnamefont {L.-Y.}\ \bibnamefont {Gao}}, \bibinfo {author} {\bibfnamefont {S.-S.}\ \bibnamefont {Xue}}, \ and\ \bibinfo {author} {\bibfnamefont {X.}~\bibnamefont {Zhang}},\ }\href {\doibase 10.1088/1674-1137/ad2b52} {\bibfield  {journal} {\bibinfo  {journal} {Chin. Phys. C}\ }\textbf {\bibinfo {volume} {48}},\ \bibinfo {pages} {051001} (\bibinfo {year} {2024})},\ \Eprint {http://arxiv.org/abs/2212.13146} {arXiv:2212.13146 [astro-ph.CO]} \BibitemShut {NoStop}%
\bibitem [{\citenamefont {Di~Valentino}\ \emph {et~al.}(2021{\natexlab{a}})\citenamefont {Di~Valentino}, \citenamefont {Mena}, \citenamefont {Pan}, \citenamefont {Visinelli}, \citenamefont {Yang}, \citenamefont {Melchiorri}, \citenamefont {Mota}, \citenamefont {Riess},\ and\ \citenamefont {Silk}}]{DiValentino:2021izs}%
  \BibitemOpen
  \bibfield  {author} {\bibinfo {author} {\bibfnamefont {E.}~\bibnamefont {Di~Valentino}}, \bibinfo {author} {\bibfnamefont {O.}~\bibnamefont {Mena}}, \bibinfo {author} {\bibfnamefont {S.}~\bibnamefont {Pan}}, \bibinfo {author} {\bibfnamefont {L.}~\bibnamefont {Visinelli}}, \bibinfo {author} {\bibfnamefont {W.}~\bibnamefont {Yang}}, \bibinfo {author} {\bibfnamefont {A.}~\bibnamefont {Melchiorri}}, \bibinfo {author} {\bibfnamefont {D.~F.}\ \bibnamefont {Mota}}, \bibinfo {author} {\bibfnamefont {A.~G.}\ \bibnamefont {Riess}}, \ and\ \bibinfo {author} {\bibfnamefont {J.}~\bibnamefont {Silk}},\ }\href {\doibase 10.1088/1361-6382/ac086d} {\bibfield  {journal} {\bibinfo  {journal} {Class. Quant. Grav.}\ }\textbf {\bibinfo {volume} {38}},\ \bibinfo {pages} {153001} (\bibinfo {year} {2021}{\natexlab{a}})},\ \Eprint {http://arxiv.org/abs/2103.01183} {arXiv:2103.01183 [astro-ph.CO]} \BibitemShut {NoStop}%
\bibitem [{\citenamefont {Lucca}(2021)}]{Lucca:2021dxo}%
  \BibitemOpen
  \bibfield  {author} {\bibinfo {author} {\bibfnamefont {M.}~\bibnamefont {Lucca}},\ }\href {\doibase 10.1016/j.dark.2021.100899} {\bibfield  {journal} {\bibinfo  {journal} {Phys. Dark Univ.}\ }\textbf {\bibinfo {volume} {34}},\ \bibinfo {pages} {100899} (\bibinfo {year} {2021})},\ \Eprint {http://arxiv.org/abs/2105.09249} {arXiv:2105.09249 [astro-ph.CO]} \BibitemShut {NoStop}%
\bibitem [{\citenamefont {Giar{\`e}}\ \emph {et~al.}(2024{\natexlab{c}})\citenamefont {Giar{\`e}}, \citenamefont {Sabogal}, \citenamefont {Nunes},\ and\ \citenamefont {Di~Valentino}}]{Giare:2024smz}%
  \BibitemOpen
  \bibfield  {author} {\bibinfo {author} {\bibfnamefont {W.}~\bibnamefont {Giar{\`e}}}, \bibinfo {author} {\bibfnamefont {M.~A.}\ \bibnamefont {Sabogal}}, \bibinfo {author} {\bibfnamefont {R.~C.}\ \bibnamefont {Nunes}}, \ and\ \bibinfo {author} {\bibfnamefont {E.}~\bibnamefont {Di~Valentino}},\ }\href {\doibase 10.1103/PhysRevLett.133.251003} {\bibfield  {journal} {\bibinfo  {journal} {Phys. Rev. Lett.}\ }\textbf {\bibinfo {volume} {133}},\ \bibinfo {pages} {251003} (\bibinfo {year} {2024}{\natexlab{c}})},\ \Eprint {http://arxiv.org/abs/2404.15232} {arXiv:2404.15232 [astro-ph.CO]} \BibitemShut {NoStop}%
\bibitem [{\citenamefont {Wetterich}(1995)}]{Wetterich:1994bg}%
  \BibitemOpen
  \bibfield  {author} {\bibinfo {author} {\bibfnamefont {C.}~\bibnamefont {Wetterich}},\ }\href@noop {} {\bibfield  {journal} {\bibinfo  {journal} {Astron. Astrophys.}\ }\textbf {\bibinfo {volume} {301}},\ \bibinfo {pages} {321} (\bibinfo {year} {1995})},\ \Eprint {http://arxiv.org/abs/hep-th/9408025} {arXiv:hep-th/9408025} \BibitemShut {NoStop}%
\bibitem [{\citenamefont {Pettorino}\ and\ \citenamefont {Baccigalupi}(2008)}]{Pettorino:2008ez}%
  \BibitemOpen
  \bibfield  {author} {\bibinfo {author} {\bibfnamefont {V.}~\bibnamefont {Pettorino}}\ and\ \bibinfo {author} {\bibfnamefont {C.}~\bibnamefont {Baccigalupi}},\ }\href {\doibase 10.1103/PhysRevD.77.103003} {\bibfield  {journal} {\bibinfo  {journal} {Phys. Rev. D}\ }\textbf {\bibinfo {volume} {77}},\ \bibinfo {pages} {103003} (\bibinfo {year} {2008})},\ \Eprint {http://arxiv.org/abs/0802.1086} {arXiv:0802.1086 [astro-ph]} \BibitemShut {NoStop}%
\bibitem [{\citenamefont {Wetterich}(2015)}]{Wetterich:2014bma}%
  \BibitemOpen
  \bibfield  {author} {\bibinfo {author} {\bibfnamefont {C.}~\bibnamefont {Wetterich}},\ }\href {\doibase 10.1007/978-3-319-10070-8_3} {\bibfield  {journal} {\bibinfo  {journal} {Lect. Notes Phys.}\ }\textbf {\bibinfo {volume} {892}},\ \bibinfo {pages} {57} (\bibinfo {year} {2015})},\ \Eprint {http://arxiv.org/abs/1402.5031} {arXiv:1402.5031 [astro-ph.CO]} \BibitemShut {NoStop}%
\bibitem [{\citenamefont {Li}\ and\ \citenamefont {Zhang}(2023)}]{Li:2023fdk}%
  \BibitemOpen
  \bibfield  {author} {\bibinfo {author} {\bibfnamefont {Y.-H.}\ \bibnamefont {Li}}\ and\ \bibinfo {author} {\bibfnamefont {X.}~\bibnamefont {Zhang}},\ }\href {\doibase 10.1088/1475-7516/2023/09/046} {\bibfield  {journal} {\bibinfo  {journal} {JCAP}\ }\textbf {\bibinfo {volume} {09}},\ \bibinfo {pages} {046} (\bibinfo {year} {2023})},\ \Eprint {http://arxiv.org/abs/2306.01593} {arXiv:2306.01593 [astro-ph.CO]} \BibitemShut {NoStop}%
\bibitem [{\citenamefont {Amendola}(1999)}]{Amendola:1999qq}%
  \BibitemOpen
  \bibfield  {author} {\bibinfo {author} {\bibfnamefont {L.}~\bibnamefont {Amendola}},\ }\href {\doibase 10.1103/PhysRevD.60.043501} {\bibfield  {journal} {\bibinfo  {journal} {Phys. Rev. D}\ }\textbf {\bibinfo {volume} {60}},\ \bibinfo {pages} {043501} (\bibinfo {year} {1999})},\ \Eprint {http://arxiv.org/abs/astro-ph/9904120} {arXiv:astro-ph/9904120} \BibitemShut {NoStop}%
\bibitem [{\citenamefont {Billyard}\ and\ \citenamefont {Coley}(2000)}]{Billyard:2000bh}%
  \BibitemOpen
  \bibfield  {author} {\bibinfo {author} {\bibfnamefont {A.~P.}\ \bibnamefont {Billyard}}\ and\ \bibinfo {author} {\bibfnamefont {A.~A.}\ \bibnamefont {Coley}},\ }\href {\doibase 10.1103/PhysRevD.61.083503} {\bibfield  {journal} {\bibinfo  {journal} {Phys. Rev. D}\ }\textbf {\bibinfo {volume} {61}},\ \bibinfo {pages} {083503} (\bibinfo {year} {2000})},\ \Eprint {http://arxiv.org/abs/astro-ph/9908224} {arXiv:astro-ph/9908224} \BibitemShut {NoStop}%
\bibitem [{\citenamefont {Li}\ \emph {et~al.}(2014)\citenamefont {Li}, \citenamefont {Zhang},\ and\ \citenamefont {Zhang}}]{Li:2014eha}%
  \BibitemOpen
  \bibfield  {author} {\bibinfo {author} {\bibfnamefont {Y.-H.}\ \bibnamefont {Li}}, \bibinfo {author} {\bibfnamefont {J.-F.}\ \bibnamefont {Zhang}}, \ and\ \bibinfo {author} {\bibfnamefont {X.}~\bibnamefont {Zhang}},\ }\href {\doibase 10.1103/PhysRevD.90.063005} {\bibfield  {journal} {\bibinfo  {journal} {Phys. Rev. D}\ }\textbf {\bibinfo {volume} {90}},\ \bibinfo {pages} {063005} (\bibinfo {year} {2014})},\ \Eprint {http://arxiv.org/abs/1404.5220} {arXiv:1404.5220 [astro-ph.CO]} \BibitemShut {NoStop}%
\bibitem [{\citenamefont {Lewis}\ \emph {et~al.}(2000)\citenamefont {Lewis}, \citenamefont {Challinor},\ and\ \citenamefont {Lasenby}}]{Lewis:1999bs}%
  \BibitemOpen
  \bibfield  {author} {\bibinfo {author} {\bibfnamefont {A.}~\bibnamefont {Lewis}}, \bibinfo {author} {\bibfnamefont {A.}~\bibnamefont {Challinor}}, \ and\ \bibinfo {author} {\bibfnamefont {A.}~\bibnamefont {Lasenby}},\ }\href {\doibase 10.1086/309179} {\bibfield  {journal} {\bibinfo  {journal} {Astrophys. J.}\ }\textbf {\bibinfo {volume} {538}},\ \bibinfo {pages} {473} (\bibinfo {year} {2000})},\ \Eprint {http://arxiv.org/abs/astro-ph/9911177} {arXiv:astro-ph/9911177} \BibitemShut {NoStop}%
\bibitem [{\citenamefont {Torrado}\ and\ \citenamefont {Lewis}(2021)}]{Torrado:2020dgo}%
  \BibitemOpen
  \bibfield  {author} {\bibinfo {author} {\bibfnamefont {J.}~\bibnamefont {Torrado}}\ and\ \bibinfo {author} {\bibfnamefont {A.}~\bibnamefont {Lewis}},\ }\href {\doibase 10.1088/1475-7516/2021/05/057} {\bibfield  {journal} {\bibinfo  {journal} {JCAP}\ }\textbf {\bibinfo {volume} {05}},\ \bibinfo {pages} {057} (\bibinfo {year} {2021})},\ \Eprint {http://arxiv.org/abs/2005.05290} {arXiv:2005.05290 [astro-ph.IM]} \BibitemShut {NoStop}%
\bibitem [{\citenamefont {Efstathiou}\ and\ \citenamefont {Gratton}(2019)}]{Efstathiou:2019mdh}%
  \BibitemOpen
  \bibfield  {author} {\bibinfo {author} {\bibfnamefont {G.}~\bibnamefont {Efstathiou}}\ and\ \bibinfo {author} {\bibfnamefont {S.}~\bibnamefont {Gratton}},\ }\href {\doibase 10.21105/astro.1910.00483} {\  (\bibinfo {year} {2019}),\ 10.21105/astro.1910.00483},\ \Eprint {http://arxiv.org/abs/1910.00483} {arXiv:1910.00483 [astro-ph.CO]} \BibitemShut {NoStop}%
\bibitem [{\citenamefont {Rosenberg}\ \emph {et~al.}(2022)\citenamefont {Rosenberg}, \citenamefont {Gratton},\ and\ \citenamefont {Efstathiou}}]{Rosenberg:2022sdy}%
  \BibitemOpen
  \bibfield  {author} {\bibinfo {author} {\bibfnamefont {E.}~\bibnamefont {Rosenberg}}, \bibinfo {author} {\bibfnamefont {S.}~\bibnamefont {Gratton}}, \ and\ \bibinfo {author} {\bibfnamefont {G.}~\bibnamefont {Efstathiou}},\ }\href {\doibase 10.1093/mnras/stac2744} {\bibfield  {journal} {\bibinfo  {journal} {Mon. Not. Roy. Astron. Soc.}\ }\textbf {\bibinfo {volume} {517}},\ \bibinfo {pages} {4620} (\bibinfo {year} {2022})},\ \Eprint {http://arxiv.org/abs/2205.10869} {arXiv:2205.10869 [astro-ph.CO]} \BibitemShut {NoStop}%
\bibitem [{\citenamefont {Louis}\ \emph {et~al.}(2025)\citenamefont {Louis} \emph {et~al.}}]{ACT:2025fju}%
  \BibitemOpen
  \bibfield  {author} {\bibinfo {author} {\bibfnamefont {T.}~\bibnamefont {Louis}} \emph {et~al.} (\bibinfo {collaboration} {Atacama Cosmology Telescope}),\ }\href {\doibase 10.1088/1475-7516/2025/11/062} {\bibfield  {journal} {\bibinfo  {journal} {JCAP}\ }\textbf {\bibinfo {volume} {11}},\ \bibinfo {pages} {062} (\bibinfo {year} {2025})},\ \Eprint {http://arxiv.org/abs/2503.14452} {arXiv:2503.14452 [astro-ph.CO]} \BibitemShut {NoStop}%
\bibitem [{\citenamefont {Camphuis}\ \emph {et~al.}(2026)\citenamefont {Camphuis} \emph {et~al.}}]{SPT-3G:2025bzu}%
  \BibitemOpen
  \bibfield  {author} {\bibinfo {author} {\bibfnamefont {E.}~\bibnamefont {Camphuis}} \emph {et~al.} (\bibinfo {collaboration} {SPT-3G}),\ }\href {\doibase 10.1103/7wt3-9v2y} {\bibfield  {journal} {\bibinfo  {journal} {Phys. Rev. D}\ }\textbf {\bibinfo {volume} {113}},\ \bibinfo {pages} {083504} (\bibinfo {year} {2026})},\ \Eprint {http://arxiv.org/abs/2506.20707} {arXiv:2506.20707 [astro-ph.CO]} \BibitemShut {NoStop}%
\bibitem [{\citenamefont {Qu}\ \emph {et~al.}(2026)\citenamefont {Qu} \emph {et~al.}}]{ACT:2025qjh}%
  \BibitemOpen
  \bibfield  {author} {\bibinfo {author} {\bibfnamefont {F.~J.}\ \bibnamefont {Qu}} \emph {et~al.} (\bibinfo {collaboration} {ACT, SPT-3G}),\ }\href {\doibase 10.1103/k5yr-3h6d} {\bibfield  {journal} {\bibinfo  {journal} {Phys. Rev. Lett.}\ }\textbf {\bibinfo {volume} {136}},\ \bibinfo {pages} {021001} (\bibinfo {year} {2026})},\ \Eprint {http://arxiv.org/abs/2504.20038} {arXiv:2504.20038 [astro-ph.CO]} \BibitemShut {NoStop}%
\bibitem [{\citenamefont {Carron}\ \emph {et~al.}(2022)\citenamefont {Carron}, \citenamefont {Mirmelstein},\ and\ \citenamefont {Lewis}}]{Carron:2022eyg}%
  \BibitemOpen
  \bibfield  {author} {\bibinfo {author} {\bibfnamefont {J.}~\bibnamefont {Carron}}, \bibinfo {author} {\bibfnamefont {M.}~\bibnamefont {Mirmelstein}}, \ and\ \bibinfo {author} {\bibfnamefont {A.}~\bibnamefont {Lewis}},\ }\href {\doibase 10.1088/1475-7516/2022/09/039} {\bibfield  {journal} {\bibinfo  {journal} {JCAP}\ }\textbf {\bibinfo {volume} {09}},\ \bibinfo {pages} {039} (\bibinfo {year} {2022})},\ \Eprint {http://arxiv.org/abs/2206.07773} {arXiv:2206.07773 [astro-ph.CO]} \BibitemShut {NoStop}%
\bibitem [{\citenamefont {Qu}\ \emph {et~al.}(2024)\citenamefont {Qu} \emph {et~al.}}]{ACT:2023dou}%
  \BibitemOpen
  \bibfield  {author} {\bibinfo {author} {\bibfnamefont {F.~J.}\ \bibnamefont {Qu}} \emph {et~al.} (\bibinfo {collaboration} {ACT}),\ }\href {\doibase 10.3847/1538-4357/acfe06} {\bibfield  {journal} {\bibinfo  {journal} {Astrophys. J.}\ }\textbf {\bibinfo {volume} {962}},\ \bibinfo {pages} {112} (\bibinfo {year} {2024})},\ \Eprint {http://arxiv.org/abs/2304.05202} {arXiv:2304.05202 [astro-ph.CO]} \BibitemShut {NoStop}%
\bibitem [{\citenamefont {Madhavacheril}\ \emph {et~al.}(2024)\citenamefont {Madhavacheril} \emph {et~al.}}]{ACT:2023kun}%
  \BibitemOpen
  \bibfield  {author} {\bibinfo {author} {\bibfnamefont {M.~S.}\ \bibnamefont {Madhavacheril}} \emph {et~al.} (\bibinfo {collaboration} {ACT}),\ }\href {\doibase 10.3847/1538-4357/acff5f} {\bibfield  {journal} {\bibinfo  {journal} {Astrophys. J.}\ }\textbf {\bibinfo {volume} {962}},\ \bibinfo {pages} {113} (\bibinfo {year} {2024})},\ \Eprint {http://arxiv.org/abs/2304.05203} {arXiv:2304.05203 [astro-ph.CO]} \BibitemShut {NoStop}%
\bibitem [{\citenamefont {Brout}\ \emph {et~al.}(2022)\citenamefont {Brout} \emph {et~al.}}]{Brout:2022vxf}%
  \BibitemOpen
  \bibfield  {author} {\bibinfo {author} {\bibfnamefont {D.}~\bibnamefont {Brout}} \emph {et~al.},\ }\href {\doibase 10.3847/1538-4357/ac8e04} {\bibfield  {journal} {\bibinfo  {journal} {Astrophys. J.}\ }\textbf {\bibinfo {volume} {938}},\ \bibinfo {pages} {110} (\bibinfo {year} {2022})},\ \Eprint {http://arxiv.org/abs/2202.04077} {arXiv:2202.04077 [astro-ph.CO]} \BibitemShut {NoStop}%
\bibitem [{\citenamefont {Abbott}\ \emph {et~al.}(2024)\citenamefont {Abbott} \emph {et~al.}}]{DES:2024jxu}%
  \BibitemOpen
  \bibfield  {author} {\bibinfo {author} {\bibfnamefont {T.~M.~C.}\ \bibnamefont {Abbott}} \emph {et~al.} (\bibinfo {collaboration} {DES}),\ }\href {\doibase 10.3847/2041-8213/ad6f9f} {\bibfield  {journal} {\bibinfo  {journal} {Astrophys. J. Lett.}\ }\textbf {\bibinfo {volume} {973}},\ \bibinfo {pages} {L14} (\bibinfo {year} {2024})},\ \Eprint {http://arxiv.org/abs/2401.02929} {arXiv:2401.02929 [astro-ph.CO]} \BibitemShut {NoStop}%
\bibitem [{\citenamefont {Pettorino}\ \emph {et~al.}(2012)\citenamefont {Pettorino}, \citenamefont {Amendola}, \citenamefont {Baccigalupi},\ and\ \citenamefont {Quercellini}}]{Pettorino:2012ts}%
  \BibitemOpen
  \bibfield  {author} {\bibinfo {author} {\bibfnamefont {V.}~\bibnamefont {Pettorino}}, \bibinfo {author} {\bibfnamefont {L.}~\bibnamefont {Amendola}}, \bibinfo {author} {\bibfnamefont {C.}~\bibnamefont {Baccigalupi}}, \ and\ \bibinfo {author} {\bibfnamefont {C.}~\bibnamefont {Quercellini}},\ }\href {\doibase 10.1103/PhysRevD.86.103507} {\bibfield  {journal} {\bibinfo  {journal} {Phys. Rev. D}\ }\textbf {\bibinfo {volume} {86}},\ \bibinfo {pages} {103507} (\bibinfo {year} {2012})},\ \Eprint {http://arxiv.org/abs/1207.3293} {arXiv:1207.3293 [astro-ph.CO]} \BibitemShut {NoStop}%
\bibitem [{\citenamefont {Pettorino}(2013)}]{Pettorino:2013oxa}%
  \BibitemOpen
  \bibfield  {author} {\bibinfo {author} {\bibfnamefont {V.}~\bibnamefont {Pettorino}},\ }\href {\doibase 10.1103/PhysRevD.88.063519} {\bibfield  {journal} {\bibinfo  {journal} {Phys. Rev. D}\ }\textbf {\bibinfo {volume} {88}},\ \bibinfo {pages} {063519} (\bibinfo {year} {2013})},\ \Eprint {http://arxiv.org/abs/1305.7457} {arXiv:1305.7457 [astro-ph.CO]} \BibitemShut {NoStop}%
\bibitem [{\citenamefont {Clemson}\ \emph {et~al.}(2012)\citenamefont {Clemson}, \citenamefont {Koyama}, \citenamefont {Zhao}, \citenamefont {Maartens},\ and\ \citenamefont {Valiviita}}]{Clemson:2011an}%
  \BibitemOpen
  \bibfield  {author} {\bibinfo {author} {\bibfnamefont {T.}~\bibnamefont {Clemson}}, \bibinfo {author} {\bibfnamefont {K.}~\bibnamefont {Koyama}}, \bibinfo {author} {\bibfnamefont {G.-B.}\ \bibnamefont {Zhao}}, \bibinfo {author} {\bibfnamefont {R.}~\bibnamefont {Maartens}}, \ and\ \bibinfo {author} {\bibfnamefont {J.}~\bibnamefont {Valiviita}},\ }\href {\doibase 10.1103/PhysRevD.85.043007} {\bibfield  {journal} {\bibinfo  {journal} {Phys. Rev. D}\ }\textbf {\bibinfo {volume} {85}},\ \bibinfo {pages} {043007} (\bibinfo {year} {2012})},\ \Eprint {http://arxiv.org/abs/1109.6234} {arXiv:1109.6234 [astro-ph.CO]} \BibitemShut {NoStop}%
\bibitem [{\citenamefont {Majerotto}\ \emph {et~al.}(2009)\citenamefont {Majerotto}, \citenamefont {Valiviita},\ and\ \citenamefont {Maartens}}]{Majerotto:2009zz}%
  \BibitemOpen
  \bibfield  {author} {\bibinfo {author} {\bibfnamefont {E.}~\bibnamefont {Majerotto}}, \bibinfo {author} {\bibfnamefont {J.}~\bibnamefont {Valiviita}}, \ and\ \bibinfo {author} {\bibfnamefont {R.}~\bibnamefont {Maartens}},\ }\href {\doibase 10.1016/j.nuclphysbps.2009.07.089} {\bibfield  {journal} {\bibinfo  {journal} {Nucl. Phys. B Proc. Suppl.}\ }\textbf {\bibinfo {volume} {194}},\ \bibinfo {pages} {260} (\bibinfo {year} {2009})}\BibitemShut {NoStop}%
\bibitem [{\citenamefont {Hu}\ and\ \citenamefont {Sugiyama}(1996)}]{Hu:1995en}%
  \BibitemOpen
  \bibfield  {author} {\bibinfo {author} {\bibfnamefont {W.}~\bibnamefont {Hu}}\ and\ \bibinfo {author} {\bibfnamefont {N.}~\bibnamefont {Sugiyama}},\ }\href {\doibase 10.1086/177989} {\bibfield  {journal} {\bibinfo  {journal} {Astrophys. J.}\ }\textbf {\bibinfo {volume} {471}},\ \bibinfo {pages} {542} (\bibinfo {year} {1996})},\ \Eprint {http://arxiv.org/abs/astro-ph/9510117} {arXiv:astro-ph/9510117} \BibitemShut {NoStop}%
\bibitem [{\citenamefont {Verde}\ \emph {et~al.}(2019)\citenamefont {Verde}, \citenamefont {Treu},\ and\ \citenamefont {Riess}}]{Verde:2019ivm}%
  \BibitemOpen
  \bibfield  {author} {\bibinfo {author} {\bibfnamefont {L.}~\bibnamefont {Verde}}, \bibinfo {author} {\bibfnamefont {T.}~\bibnamefont {Treu}}, \ and\ \bibinfo {author} {\bibfnamefont {A.~G.}\ \bibnamefont {Riess}},\ }\href {\doibase 10.1038/s41550-019-0902-0} {\bibfield  {journal} {\bibinfo  {journal} {Nature Astron.}\ }\textbf {\bibinfo {volume} {3}},\ \bibinfo {pages} {891} (\bibinfo {year} {2019})},\ \Eprint {http://arxiv.org/abs/1907.10625} {arXiv:1907.10625 [astro-ph.CO]} \BibitemShut {NoStop}%
\bibitem [{\citenamefont {Di~Valentino}\ \emph {et~al.}(2021{\natexlab{b}})\citenamefont {Di~Valentino} \emph {et~al.}}]{DiValentino:2020zio}%
  \BibitemOpen
  \bibfield  {author} {\bibinfo {author} {\bibfnamefont {E.}~\bibnamefont {Di~Valentino}} \emph {et~al.},\ }\href {\doibase 10.1016/j.astropartphys.2021.102605} {\bibfield  {journal} {\bibinfo  {journal} {Astropart. Phys.}\ }\textbf {\bibinfo {volume} {131}},\ \bibinfo {pages} {102605} (\bibinfo {year} {2021}{\natexlab{b}})},\ \Eprint {http://arxiv.org/abs/2008.11284} {arXiv:2008.11284 [astro-ph.CO]} \BibitemShut {NoStop}%
\bibitem [{\citenamefont {Perivolaropoulos}\ and\ \citenamefont {Skara}(2022)}]{Perivolaropoulos:2021jda}%
  \BibitemOpen
  \bibfield  {author} {\bibinfo {author} {\bibfnamefont {L.}~\bibnamefont {Perivolaropoulos}}\ and\ \bibinfo {author} {\bibfnamefont {F.}~\bibnamefont {Skara}},\ }\href {\doibase 10.1016/j.newar.2022.101659} {\bibfield  {journal} {\bibinfo  {journal} {New Astron. Rev.}\ }\textbf {\bibinfo {volume} {95}},\ \bibinfo {pages} {101659} (\bibinfo {year} {2022})},\ \Eprint {http://arxiv.org/abs/2105.05208} {arXiv:2105.05208 [astro-ph.CO]} \BibitemShut {NoStop}%
\bibitem [{\citenamefont {Sch{\"o}neberg}\ \emph {et~al.}(2022)\citenamefont {Sch{\"o}neberg}, \citenamefont {Franco~Abell{\'a}n}, \citenamefont {P{\'e}rez~S{\'a}nchez}, \citenamefont {Witte}, \citenamefont {Poulin},\ and\ \citenamefont {Lesgourgues}}]{Schoneberg:2021qvd}%
  \BibitemOpen
  \bibfield  {author} {\bibinfo {author} {\bibfnamefont {N.}~\bibnamefont {Sch{\"o}neberg}}, \bibinfo {author} {\bibfnamefont {G.}~\bibnamefont {Franco~Abell{\'a}n}}, \bibinfo {author} {\bibfnamefont {A.}~\bibnamefont {P{\'e}rez~S{\'a}nchez}}, \bibinfo {author} {\bibfnamefont {S.~J.}\ \bibnamefont {Witte}}, \bibinfo {author} {\bibfnamefont {V.}~\bibnamefont {Poulin}}, \ and\ \bibinfo {author} {\bibfnamefont {J.}~\bibnamefont {Lesgourgues}},\ }\href {\doibase 10.1016/j.physrep.2022.07.001} {\bibfield  {journal} {\bibinfo  {journal} {Phys. Rept.}\ }\textbf {\bibinfo {volume} {984}},\ \bibinfo {pages} {1} (\bibinfo {year} {2022})},\ \Eprint {http://arxiv.org/abs/2107.10291} {arXiv:2107.10291 [astro-ph.CO]} \BibitemShut {NoStop}%
\bibitem [{\citenamefont {Shah}\ \emph {et~al.}(2021)\citenamefont {Shah}, \citenamefont {Lemos},\ and\ \citenamefont {Lahav}}]{Shah:2021onj}%
  \BibitemOpen
  \bibfield  {author} {\bibinfo {author} {\bibfnamefont {P.}~\bibnamefont {Shah}}, \bibinfo {author} {\bibfnamefont {P.}~\bibnamefont {Lemos}}, \ and\ \bibinfo {author} {\bibfnamefont {O.}~\bibnamefont {Lahav}},\ }\href {\doibase 10.1007/s00159-021-00137-4} {\bibfield  {journal} {\bibinfo  {journal} {Astron. Astrophys. Rev.}\ }\textbf {\bibinfo {volume} {29}},\ \bibinfo {pages} {9} (\bibinfo {year} {2021})},\ \Eprint {http://arxiv.org/abs/2109.01161} {arXiv:2109.01161 [astro-ph.CO]} \BibitemShut {NoStop}%
\bibitem [{\citenamefont {Abdalla}\ \emph {et~al.}(2022)\citenamefont {Abdalla} \emph {et~al.}}]{Abdalla:2022yfr}%
  \BibitemOpen
  \bibfield  {author} {\bibinfo {author} {\bibfnamefont {E.}~\bibnamefont {Abdalla}} \emph {et~al.},\ }\href {\doibase 10.1016/j.jheap.2022.04.002} {\bibfield  {journal} {\bibinfo  {journal} {JHEAp}\ }\textbf {\bibinfo {volume} {34}},\ \bibinfo {pages} {49} (\bibinfo {year} {2022})},\ \Eprint {http://arxiv.org/abs/2203.06142} {arXiv:2203.06142 [astro-ph.CO]} \BibitemShut {NoStop}%
\bibitem [{\citenamefont {Di~Valentino}(2022)}]{DiValentino:2022fjm}%
  \BibitemOpen
  \bibfield  {author} {\bibinfo {author} {\bibfnamefont {E.}~\bibnamefont {Di~Valentino}},\ }\href {\doibase 10.3390/universe8080399} {\bibfield  {journal} {\bibinfo  {journal} {Universe}\ }\textbf {\bibinfo {volume} {8}},\ \bibinfo {pages} {399} (\bibinfo {year} {2022})}\BibitemShut {NoStop}%
\bibitem [{\citenamefont {Kamionkowski}\ and\ \citenamefont {Riess}(2023)}]{Kamionkowski:2022pkx}%
  \BibitemOpen
  \bibfield  {author} {\bibinfo {author} {\bibfnamefont {M.}~\bibnamefont {Kamionkowski}}\ and\ \bibinfo {author} {\bibfnamefont {A.~G.}\ \bibnamefont {Riess}},\ }\href {\doibase 10.1146/annurev-nucl-111422-024107} {\bibfield  {journal} {\bibinfo  {journal} {Ann. Rev. Nucl. Part. Sci.}\ }\textbf {\bibinfo {volume} {73}},\ \bibinfo {pages} {153} (\bibinfo {year} {2023})},\ \Eprint {http://arxiv.org/abs/2211.04492} {arXiv:2211.04492 [astro-ph.CO]} \BibitemShut {NoStop}%
\bibitem [{\citenamefont {Giar{\`e}}(2023)}]{Giare:2023xoc}%
  \BibitemOpen
  \bibfield  {author} {\bibinfo {author} {\bibfnamefont {W.}~\bibnamefont {Giar{\`e}}},\ }\href {\doibase 10.1007/978-981-99-0177-7\_36} {\  (\bibinfo {year} {2023}),\ 10.1007/978-981-99-0177-7\_36},\ \Eprint {http://arxiv.org/abs/2305.16919} {arXiv:2305.16919 [astro-ph.CO]} \BibitemShut {NoStop}%
\bibitem [{\citenamefont {Hu}\ and\ \citenamefont {Wang}(2023)}]{Hu:2023jqc}%
  \BibitemOpen
  \bibfield  {author} {\bibinfo {author} {\bibfnamefont {J.-P.}\ \bibnamefont {Hu}}\ and\ \bibinfo {author} {\bibfnamefont {F.-Y.}\ \bibnamefont {Wang}},\ }\href {\doibase 10.3390/universe9020094} {\bibfield  {journal} {\bibinfo  {journal} {Universe}\ }\textbf {\bibinfo {volume} {9}},\ \bibinfo {pages} {94} (\bibinfo {year} {2023})},\ \Eprint {http://arxiv.org/abs/2302.05709} {arXiv:2302.05709 [astro-ph.CO]} \BibitemShut {NoStop}%
\bibitem [{\citenamefont {Di~Valentino}\ \emph {et~al.}(2025)\citenamefont {Di~Valentino} \emph {et~al.}}]{CosmoVerseNetwork:2025alb}%
  \BibitemOpen
  \bibfield  {author} {\bibinfo {author} {\bibfnamefont {E.}~\bibnamefont {Di~Valentino}} \emph {et~al.} (\bibinfo {collaboration} {CosmoVerse Network}),\ }\href {\doibase 10.1016/j.dark.2025.101965} {\bibfield  {journal} {\bibinfo  {journal} {Phys. Dark Univ.}\ }\textbf {\bibinfo {volume} {49}},\ \bibinfo {pages} {101965} (\bibinfo {year} {2025})},\ \Eprint {http://arxiv.org/abs/2504.01669} {arXiv:2504.01669 [astro-ph.CO]} \BibitemShut {NoStop}%
\bibitem [{\citenamefont {Vagnozzi}(2020)}]{Vagnozzi:2019ezj}%
  \BibitemOpen
  \bibfield  {author} {\bibinfo {author} {\bibfnamefont {S.}~\bibnamefont {Vagnozzi}},\ }\href {\doibase 10.1103/PhysRevD.102.023518} {\bibfield  {journal} {\bibinfo  {journal} {Phys. Rev. D}\ }\textbf {\bibinfo {volume} {102}},\ \bibinfo {pages} {023518} (\bibinfo {year} {2020})},\ \Eprint {http://arxiv.org/abs/1907.07569} {arXiv:1907.07569 [astro-ph.CO]} \BibitemShut {NoStop}%
\bibitem [{\citenamefont {Vagnozzi}(2023)}]{Vagnozzi:2023nrq}%
  \BibitemOpen
  \bibfield  {author} {\bibinfo {author} {\bibfnamefont {S.}~\bibnamefont {Vagnozzi}},\ }\href {\doibase 10.3390/universe9090393} {\bibfield  {journal} {\bibinfo  {journal} {Universe}\ }\textbf {\bibinfo {volume} {9}},\ \bibinfo {pages} {393} (\bibinfo {year} {2023})},\ \Eprint {http://arxiv.org/abs/2308.16628} {arXiv:2308.16628 [astro-ph.CO]} \BibitemShut {NoStop}%
\bibitem [{\citenamefont {Loverde}\ and\ \citenamefont {Weiner}(2024)}]{Loverde:2024nfi}%
  \BibitemOpen
  \bibfield  {author} {\bibinfo {author} {\bibfnamefont {M.}~\bibnamefont {Loverde}}\ and\ \bibinfo {author} {\bibfnamefont {Z.~J.}\ \bibnamefont {Weiner}},\ }\href {\doibase 10.1088/1475-7516/2024/12/048} {\bibfield  {journal} {\bibinfo  {journal} {JCAP}\ }\textbf {\bibinfo {volume} {12}},\ \bibinfo {pages} {048} (\bibinfo {year} {2024})},\ \Eprint {http://arxiv.org/abs/2410.00090} {arXiv:2410.00090 [astro-ph.CO]} \BibitemShut {NoStop}%
\bibitem [{\citenamefont {Colg{\'a}in}\ and\ \citenamefont {Sheikh-Jabbari}(2025)}]{Colgain:2024mtg}%
  \BibitemOpen
  \bibfield  {author} {\bibinfo {author} {\bibfnamefont {E.~{\'O}.}\ \bibnamefont {Colg{\'a}in}}\ and\ \bibinfo {author} {\bibfnamefont {M.~M.}\ \bibnamefont {Sheikh-Jabbari}},\ }\href {\doibase 10.1093/mnrasl/slaf042} {\bibfield  {journal} {\bibinfo  {journal} {Mon. Not. Roy. Astron. Soc.}\ }\textbf {\bibinfo {volume} {542}},\ \bibinfo {pages} {L24} (\bibinfo {year} {2025})},\ \Eprint {http://arxiv.org/abs/2412.12905} {arXiv:2412.12905 [astro-ph.CO]} \BibitemShut {NoStop}%
\bibitem [{\citenamefont {Elbers}\ \emph {et~al.}(2025{\natexlab{a}})\citenamefont {Elbers} \emph {et~al.}}]{Elbers:2025vlz}%
  \BibitemOpen
  \bibfield  {author} {\bibinfo {author} {\bibfnamefont {W.}~\bibnamefont {Elbers}} \emph {et~al.},\ }\href {\doibase 10.1103/w9pk-xsk7} {\bibfield  {journal} {\bibinfo  {journal} {Phys. Rev. D}\ }\textbf {\bibinfo {volume} {112}},\ \bibinfo {pages} {083513} (\bibinfo {year} {2025}{\natexlab{a}})},\ \Eprint {http://arxiv.org/abs/2503.14744} {arXiv:2503.14744 [astro-ph.CO]} \BibitemShut {NoStop}%
\bibitem [{\citenamefont {Lynch}\ and\ \citenamefont {Knox}(2025)}]{Lynch:2025ine}%
  \BibitemOpen
  \bibfield  {author} {\bibinfo {author} {\bibfnamefont {G.~P.}\ \bibnamefont {Lynch}}\ and\ \bibinfo {author} {\bibfnamefont {L.}~\bibnamefont {Knox}},\ }\href {\doibase 10.1103/613p-pph2} {\bibfield  {journal} {\bibinfo  {journal} {Phys. Rev. D}\ }\textbf {\bibinfo {volume} {112}},\ \bibinfo {pages} {083543} (\bibinfo {year} {2025})},\ \Eprint {http://arxiv.org/abs/2503.14470} {arXiv:2503.14470 [astro-ph.CO]} \BibitemShut {NoStop}%
\bibitem [{\citenamefont {Sailer}\ \emph {et~al.}(2026)\citenamefont {Sailer}, \citenamefont {Farren}, \citenamefont {Ferraro},\ and\ \citenamefont {White}}]{Sailer:2025lxj}%
  \BibitemOpen
  \bibfield  {author} {\bibinfo {author} {\bibfnamefont {N.}~\bibnamefont {Sailer}}, \bibinfo {author} {\bibfnamefont {G.~S.}\ \bibnamefont {Farren}}, \bibinfo {author} {\bibfnamefont {S.}~\bibnamefont {Ferraro}}, \ and\ \bibinfo {author} {\bibfnamefont {M.}~\bibnamefont {White}},\ }\href {\doibase 10.1103/6r54-8lv4} {\bibfield  {journal} {\bibinfo  {journal} {Phys. Rev. Lett.}\ }\textbf {\bibinfo {volume} {136}},\ \bibinfo {pages} {081002} (\bibinfo {year} {2026})},\ \Eprint {http://arxiv.org/abs/2504.16932} {arXiv:2504.16932 [astro-ph.CO]} \BibitemShut {NoStop}%
\bibitem [{\citenamefont {Jhaveri}\ \emph {et~al.}(2025)\citenamefont {Jhaveri}, \citenamefont {Karwal},\ and\ \citenamefont {Hu}}]{Jhaveri:2025neg}%
  \BibitemOpen
  \bibfield  {author} {\bibinfo {author} {\bibfnamefont {T.}~\bibnamefont {Jhaveri}}, \bibinfo {author} {\bibfnamefont {T.}~\bibnamefont {Karwal}}, \ and\ \bibinfo {author} {\bibfnamefont {W.}~\bibnamefont {Hu}},\ }\href {\doibase 10.1103/6vd2-rbfn} {\bibfield  {journal} {\bibinfo  {journal} {Phys. Rev. D}\ }\textbf {\bibinfo {volume} {112}},\ \bibinfo {pages} {043541} (\bibinfo {year} {2025})},\ \Eprint {http://arxiv.org/abs/2504.21813} {arXiv:2504.21813 [astro-ph.CO]} \BibitemShut {NoStop}%
\bibitem [{\citenamefont {Du}\ \emph {et~al.}(2025{\natexlab{a}})\citenamefont {Du}, \citenamefont {Wu}, \citenamefont {Li},\ and\ \citenamefont {Zhang}}]{Du:2024pai}%
  \BibitemOpen
  \bibfield  {author} {\bibinfo {author} {\bibfnamefont {G.-H.}\ \bibnamefont {Du}}, \bibinfo {author} {\bibfnamefont {P.-J.}\ \bibnamefont {Wu}}, \bibinfo {author} {\bibfnamefont {T.-N.}\ \bibnamefont {Li}}, \ and\ \bibinfo {author} {\bibfnamefont {X.}~\bibnamefont {Zhang}},\ }\href {\doibase 10.1140/epjc/s10052-025-14094-0} {\bibfield  {journal} {\bibinfo  {journal} {Eur. Phys. J. C}\ }\textbf {\bibinfo {volume} {85}},\ \bibinfo {pages} {392} (\bibinfo {year} {2025}{\natexlab{a}})},\ \Eprint {http://arxiv.org/abs/2407.15640} {arXiv:2407.15640 [astro-ph.CO]} \BibitemShut {NoStop}%
\bibitem [{\citenamefont {Craig}\ \emph {et~al.}(2024)\citenamefont {Craig}, \citenamefont {Green}, \citenamefont {Meyers},\ and\ \citenamefont {Rajendran}}]{Craig:2024tky}%
  \BibitemOpen
  \bibfield  {author} {\bibinfo {author} {\bibfnamefont {N.}~\bibnamefont {Craig}}, \bibinfo {author} {\bibfnamefont {D.}~\bibnamefont {Green}}, \bibinfo {author} {\bibfnamefont {J.}~\bibnamefont {Meyers}}, \ and\ \bibinfo {author} {\bibfnamefont {S.}~\bibnamefont {Rajendran}},\ }\href {\doibase 10.1007/JHEP09(2024)097} {\bibfield  {journal} {\bibinfo  {journal} {JHEP}\ }\textbf {\bibinfo {volume} {09}},\ \bibinfo {pages} {097} (\bibinfo {year} {2024})},\ \Eprint {http://arxiv.org/abs/2405.00836} {arXiv:2405.00836 [astro-ph.CO]} \BibitemShut {NoStop}%
\bibitem [{\citenamefont {Naredo-Tuero}\ \emph {et~al.}(2024)\citenamefont {Naredo-Tuero}, \citenamefont {Escudero}, \citenamefont {Fern{\'a}ndez-Mart{\'\i}nez}, \citenamefont {Marcano},\ and\ \citenamefont {Poulin}}]{Naredo-Tuero:2024sgf}%
  \BibitemOpen
  \bibfield  {author} {\bibinfo {author} {\bibfnamefont {D.}~\bibnamefont {Naredo-Tuero}}, \bibinfo {author} {\bibfnamefont {M.}~\bibnamefont {Escudero}}, \bibinfo {author} {\bibfnamefont {E.}~\bibnamefont {Fern{\'a}ndez-Mart{\'\i}nez}}, \bibinfo {author} {\bibfnamefont {X.}~\bibnamefont {Marcano}}, \ and\ \bibinfo {author} {\bibfnamefont {V.}~\bibnamefont {Poulin}},\ }\href {\doibase 10.1103/PhysRevD.110.123537} {\bibfield  {journal} {\bibinfo  {journal} {Phys. Rev. D}\ }\textbf {\bibinfo {volume} {110}},\ \bibinfo {pages} {123537} (\bibinfo {year} {2024})},\ \Eprint {http://arxiv.org/abs/2407.13831} {arXiv:2407.13831 [astro-ph.CO]} \BibitemShut {NoStop}%
\bibitem [{\citenamefont {Green}\ and\ \citenamefont {Meyers}(2025)}]{Green:2024xbb}%
  \BibitemOpen
  \bibfield  {author} {\bibinfo {author} {\bibfnamefont {D.}~\bibnamefont {Green}}\ and\ \bibinfo {author} {\bibfnamefont {J.}~\bibnamefont {Meyers}},\ }\href {\doibase 10.1103/PhysRevD.111.083507} {\bibfield  {journal} {\bibinfo  {journal} {Phys. Rev. D}\ }\textbf {\bibinfo {volume} {111}},\ \bibinfo {pages} {083507} (\bibinfo {year} {2025})},\ \Eprint {http://arxiv.org/abs/2407.07878} {arXiv:2407.07878 [astro-ph.CO]} \BibitemShut {NoStop}%
\bibitem [{\citenamefont {Elbers}\ \emph {et~al.}(2025{\natexlab{b}})\citenamefont {Elbers}, \citenamefont {Frenk}, \citenamefont {Jenkins}, \citenamefont {Li},\ and\ \citenamefont {Pascoli}}]{Elbers:2024sha}%
  \BibitemOpen
  \bibfield  {author} {\bibinfo {author} {\bibfnamefont {W.}~\bibnamefont {Elbers}}, \bibinfo {author} {\bibfnamefont {C.~S.}\ \bibnamefont {Frenk}}, \bibinfo {author} {\bibfnamefont {A.}~\bibnamefont {Jenkins}}, \bibinfo {author} {\bibfnamefont {B.}~\bibnamefont {Li}}, \ and\ \bibinfo {author} {\bibfnamefont {S.}~\bibnamefont {Pascoli}},\ }\href {\doibase 10.1103/PhysRevD.111.063534} {\bibfield  {journal} {\bibinfo  {journal} {Phys. Rev. D}\ }\textbf {\bibinfo {volume} {111}},\ \bibinfo {pages} {063534} (\bibinfo {year} {2025}{\natexlab{b}})},\ \Eprint {http://arxiv.org/abs/2407.10965} {arXiv:2407.10965 [astro-ph.CO]} \BibitemShut {NoStop}%
\bibitem [{\citenamefont {Du}\ \emph {et~al.}(2026)\citenamefont {Du}, \citenamefont {Li}, \citenamefont {Wu}, \citenamefont {Feng}, \citenamefont {Zhou}, \citenamefont {Zhang},\ and\ \citenamefont {Zhang}}]{Du:2025iow}%
  \BibitemOpen
  \bibfield  {author} {\bibinfo {author} {\bibfnamefont {G.-H.}\ \bibnamefont {Du}}, \bibinfo {author} {\bibfnamefont {T.-N.}\ \bibnamefont {Li}}, \bibinfo {author} {\bibfnamefont {P.-J.}\ \bibnamefont {Wu}}, \bibinfo {author} {\bibfnamefont {L.}~\bibnamefont {Feng}}, \bibinfo {author} {\bibfnamefont {S.-H.}\ \bibnamefont {Zhou}}, \bibinfo {author} {\bibfnamefont {J.-F.}\ \bibnamefont {Zhang}}, \ and\ \bibinfo {author} {\bibfnamefont {X.}~\bibnamefont {Zhang}},\ }\href {\doibase 10.1140/epjc/s10052-025-15237-z} {\bibfield  {journal} {\bibinfo  {journal} {Eur. Phys. J. C}\ }\textbf {\bibinfo {volume} {86}},\ \bibinfo {pages} {110} (\bibinfo {year} {2026})},\ \Eprint {http://arxiv.org/abs/2501.10785} {arXiv:2501.10785 [astro-ph.CO]} \BibitemShut {NoStop}%
\bibitem [{\citenamefont {Giar{\`e}}\ \emph {et~al.}(2025{\natexlab{b}})\citenamefont {Giar{\`e}}, \citenamefont {Mena}, \citenamefont {Specogna},\ and\ \citenamefont {Di~Valentino}}]{Giare:2025ath}%
  \BibitemOpen
  \bibfield  {author} {\bibinfo {author} {\bibfnamefont {W.}~\bibnamefont {Giar{\`e}}}, \bibinfo {author} {\bibfnamefont {O.}~\bibnamefont {Mena}}, \bibinfo {author} {\bibfnamefont {E.}~\bibnamefont {Specogna}}, \ and\ \bibinfo {author} {\bibfnamefont {E.}~\bibnamefont {Di~Valentino}},\ }\href {\doibase 10.1103/njfc-pd1w} {\bibfield  {journal} {\bibinfo  {journal} {Phys. Rev. D}\ }\textbf {\bibinfo {volume} {112}},\ \bibinfo {pages} {103520} (\bibinfo {year} {2025}{\natexlab{b}})},\ \Eprint {http://arxiv.org/abs/2507.01848} {arXiv:2507.01848 [astro-ph.CO]} \BibitemShut {NoStop}%
\bibitem [{\citenamefont {Du}\ \emph {et~al.}(2025{\natexlab{b}})\citenamefont {Du}, \citenamefont {Li}, \citenamefont {Wu}, \citenamefont {Zhang},\ and\ \citenamefont {Zhang}}]{Du:2025xes}%
  \BibitemOpen
  \bibfield  {author} {\bibinfo {author} {\bibfnamefont {G.-H.}\ \bibnamefont {Du}}, \bibinfo {author} {\bibfnamefont {T.-N.}\ \bibnamefont {Li}}, \bibinfo {author} {\bibfnamefont {P.-J.}\ \bibnamefont {Wu}}, \bibinfo {author} {\bibfnamefont {J.-F.}\ \bibnamefont {Zhang}}, \ and\ \bibinfo {author} {\bibfnamefont {X.}~\bibnamefont {Zhang}},\ }\href@noop {} {\  (\bibinfo {year} {2025}{\natexlab{b}})},\ \Eprint {http://arxiv.org/abs/2507.16589} {arXiv:2507.16589 [astro-ph.CO]} \BibitemShut {NoStop}%
\bibitem [{\citenamefont {Zhou}\ \emph {et~al.}(2025)\citenamefont {Zhou}, \citenamefont {Li}, \citenamefont {Du}, \citenamefont {Jiang}, \citenamefont {Zhang},\ and\ \citenamefont {Zhang}}]{Zhou:2025nkb}%
  \BibitemOpen
  \bibfield  {author} {\bibinfo {author} {\bibfnamefont {S.-H.}\ \bibnamefont {Zhou}}, \bibinfo {author} {\bibfnamefont {T.-N.}\ \bibnamefont {Li}}, \bibinfo {author} {\bibfnamefont {G.-H.}\ \bibnamefont {Du}}, \bibinfo {author} {\bibfnamefont {J.-Q.}\ \bibnamefont {Jiang}}, \bibinfo {author} {\bibfnamefont {J.-F.}\ \bibnamefont {Zhang}}, \ and\ \bibinfo {author} {\bibfnamefont {X.}~\bibnamefont {Zhang}},\ }\href {\doibase 10.1103/mtdg-hbqt} {\bibfield  {journal} {\bibinfo  {journal} {Phys. Rev. D}\ }\textbf {\bibinfo {volume} {112}},\ \bibinfo {pages} {123532} (\bibinfo {year} {2025})},\ \Eprint {http://arxiv.org/abs/2509.10836} {arXiv:2509.10836 [astro-ph.CO]} \BibitemShut {NoStop}%
\bibitem [{\citenamefont {Lewis}\ and\ \citenamefont {Challinor}(2006)}]{Lewis:2006fu}%
  \BibitemOpen
  \bibfield  {author} {\bibinfo {author} {\bibfnamefont {A.}~\bibnamefont {Lewis}}\ and\ \bibinfo {author} {\bibfnamefont {A.}~\bibnamefont {Challinor}},\ }\href {\doibase 10.1016/j.physrep.2006.03.002} {\bibfield  {journal} {\bibinfo  {journal} {Phys. Rept.}\ }\textbf {\bibinfo {volume} {429}},\ \bibinfo {pages} {1} (\bibinfo {year} {2006})},\ \Eprint {http://arxiv.org/abs/astro-ph/0601594} {arXiv:astro-ph/0601594} \BibitemShut {NoStop}%
\bibitem [{\citenamefont {G{\'o}mez-Valent}\ \emph {et~al.}(2026)\citenamefont {G{\'o}mez-Valent}, \citenamefont {Zheng},\ and\ \citenamefont {Amendola}}]{Gomez-Valent:2026ept}%
  \BibitemOpen
  \bibfield  {author} {\bibinfo {author} {\bibfnamefont {A.}~\bibnamefont {G{\'o}mez-Valent}}, \bibinfo {author} {\bibfnamefont {Z.}~\bibnamefont {Zheng}}, \ and\ \bibinfo {author} {\bibfnamefont {L.}~\bibnamefont {Amendola}},\ }\href@noop {} {\  (\bibinfo {year} {2026})},\ \Eprint {http://arxiv.org/abs/2604.12032} {arXiv:2604.12032 [astro-ph.CO]} \BibitemShut {NoStop}%
\bibitem [{\citenamefont {Avila}\ \emph {et~al.}(2022)\citenamefont {Avila}, \citenamefont {Bernui}, \citenamefont {Bonilla},\ and\ \citenamefont {Nunes}}]{Avila:2022xad}%
  \BibitemOpen
  \bibfield  {author} {\bibinfo {author} {\bibfnamefont {F.}~\bibnamefont {Avila}}, \bibinfo {author} {\bibfnamefont {A.}~\bibnamefont {Bernui}}, \bibinfo {author} {\bibfnamefont {A.}~\bibnamefont {Bonilla}}, \ and\ \bibinfo {author} {\bibfnamefont {R.~C.}\ \bibnamefont {Nunes}},\ }\href {\doibase 10.1140/epjc/s10052-022-10561-0} {\bibfield  {journal} {\bibinfo  {journal} {Eur. Phys. J. C}\ }\textbf {\bibinfo {volume} {82}},\ \bibinfo {pages} {594} (\bibinfo {year} {2022})},\ \Eprint {http://arxiv.org/abs/2201.07829} {arXiv:2201.07829 [astro-ph.CO]} \BibitemShut {NoStop}%
\bibitem [{\citenamefont {Sabogal}\ \emph {et~al.}(2024)\citenamefont {Sabogal}, \citenamefont {Silva}, \citenamefont {Nunes}, \citenamefont {Kumar}, \citenamefont {Di~Valentino},\ and\ \citenamefont {Giar{\`e}}}]{Sabogal:2024yha}%
  \BibitemOpen
  \bibfield  {author} {\bibinfo {author} {\bibfnamefont {M.~A.}\ \bibnamefont {Sabogal}}, \bibinfo {author} {\bibfnamefont {E.}~\bibnamefont {Silva}}, \bibinfo {author} {\bibfnamefont {R.~C.}\ \bibnamefont {Nunes}}, \bibinfo {author} {\bibfnamefont {S.}~\bibnamefont {Kumar}}, \bibinfo {author} {\bibfnamefont {E.}~\bibnamefont {Di~Valentino}}, \ and\ \bibinfo {author} {\bibfnamefont {W.}~\bibnamefont {Giar{\`e}}},\ }\href {\doibase 10.1103/PhysRevD.110.123508} {\bibfield  {journal} {\bibinfo  {journal} {Phys. Rev. D}\ }\textbf {\bibinfo {volume} {110}},\ \bibinfo {pages} {123508} (\bibinfo {year} {2024})},\ \Eprint {http://arxiv.org/abs/2408.12403} {arXiv:2408.12403 [astro-ph.CO]} \BibitemShut {NoStop}%
\end{thebibliography}%

%%%%%%%%%%%%%%%%% APPENDICES %%%%%%%%%%%%%%%%%%%%%

% \widetext
\clearpage
\onecolumngrid

\begin{center}
\huge{\textsc{Supplementary Material}}  
\end{center}

\vspace{0.5 cm}

\section{Theoretical Method}\label{appendixA}

Here, we describe in detail the calculation methods for the CQ and CF scenarios. At the background level, the energy exchange between the dark sectors relates the time evolution of their energy densities to the energy transfer rate $Q$. The individual continuity equations governing the background evolution are
\begin{align}
\rho'_{\mathrm{de}} + 3\mathcal{H}(1+w)\rho_{\mathrm{de}} &= aQ, \label{eq:cont_de} \\
\rho'_{\mathrm{c}} + 3\mathcal{H}\rho_{\mathrm{c}} &= -aQ, \label{eq:cont_c}
\end{align}
where the prime denotes the derivative with respect to conformal time, $\mathcal{H}$ is the conformal Hubble parameter, and $w = p_{\mathrm{de}}/\rho_{\mathrm{de}}$ is the EoS parameter of DE. A positive $Q$ indicates energy transfer from CDM to DE.

At the linear perturbation level, the conservation equations for the density perturbation $\delta\rho_I$ and the velocity $v_I$ (where $I = \mathrm{de}, \mathrm{c}$) are given by~\cite{Li:2023fdk}
\begin{align}
&{\delta\rho_I'} + 3\mathcal{H}({\delta \rho_I}+ {\delta p_I}) + (\rho_I+p_I)(k{v}_I + 3 H_L') = a(\delta Q_I+AQ_I), \\
&[(\rho_I + p_I)\theta_I]' + 4\mathcal{H}(\rho_I + p_I)\theta_I - { \delta p_I } + \frac{2}{3}c_K p_I {\Pi_I} - (\rho_I + p_I) {A} = a(Q_I\theta+f_{k,I}). \label{eqn:conservation2}
\end{align}
For notational convenience, we define $\theta_I \equiv (v_I-B)/k$ and $f_{k,I} \equiv f_I/k$. Here, $A$, $B$, and $H_L$ characterize the scalar metric perturbations, $\Pi_I$ denotes the anisotropic stress, and $c_K \equiv 1-3K/k^2$ accounts for the spatial curvature $K$. Note that the variable $\theta$ in Eq.~\eqref{eqn:conservation2} represents the velocity perturbation of the total cosmic fluid and is identical to the one used in Eqs.~\eqref{eq:unified_dQ} and \eqref{eq:unified_fk}. Analogous to the background case, the conservation of total energy-momentum dictates that $Q = Q_{\mathrm{de}} = -Q_{\mathrm{c}}$, $\delta Q = \delta Q_{\mathrm{de}} = -\delta Q_{\mathrm{c}}$, and $f = f_{\mathrm{de}} = -f_{\mathrm{c}}$, assuming positive values correspond to transfer from CDM to DE. 

In the CQ scenario, the total Lagrangian including the coupling term can be written as
\begin{equation}
\mathcal{L} = -\frac{1}{2}\partial^{\mu}\phi\partial_{\mu}\phi - U(\phi) - m(\phi)\bar{\psi}\psi + \mathcal{L}_{\mathrm{kin},\psi},
\end{equation}
where $U(\phi)$ is the potential of the quintessence field, and the mass of the DM field $\psi$ is modulated by $\phi$ through the function $m(\phi)$. The energy-momentum transfer vector $Q_{\mu}$ is derived from the variation of the action as
\begin{equation}
Q_{\mu} = \frac{\partial \ln m(\phi)}{\partial \phi} \rho_{\mathrm{c}} \partial_{\mu} \phi.
\end{equation}
We consider a widely studied exponential coupling form $m(\phi) \propto e^{-\beta\sqrt{\kappa}\phi}$, where $\kappa = 8\pi G$ and $\beta$ is the coupling constant~\cite{Pettorino:2008ez,Pettorino:2012ts,Pettorino:2013oxa}. This leads to a covariant interaction form $Q_{\mu} = -\beta \rho_{\mathrm{c}} \sqrt{\kappa} \partial_{\mu}\phi$. 

In the CF scenario, note that treating DE as a fluid with interaction often triggers non-physical instabilities in the curvature perturbation on large scales due to the modification of non-adiabatic modes~\cite{Clemson:2011an,Majerotto:2009zz}. To ensure stable perturbation evolution across the entire parameter space, the parameterized post-Friedmann (PPF) framework has been extended to IDE models, known as the ePPF approach~\cite{Li:2014eha,Li:2023fdk}. By replacing the ill-defined large-scale pressure condition with a parametrized relationship between dark sector momentum densities, the ePPF approach effectively eliminates the instability while preserving accurate small-scale dynamics.

In our analysis, CDM is consistently treated as a pressureless fluid with a sound speed of $c_{s,c}^2 = 0$. In the CQ model, the DE component is described by a canonical scalar field, which naturally yields a rest-frame sound speed of $c_{s,de}^2 = 1$. In the CF model, the DE pressure perturbations are handled by the ePPF framework; thus, there is no need to define a rest-frame sound speed.

\section{Cosmological datasets and Methodologies}\label{appendixB}

As discussed in the main text, we perform a Bayesian inference to constrain the cosmological parameters using the latest high-precision observational datasets. The datasets employed in this work are summarized below:
\begin{itemize}
\item \textbf{\texttt{CMB}:} The CMB measurements are based on a combination of data from satellite and ground-based experiments. For Planck 2018 CMB data, we utilize the large-scale temperature (TT) and E-mode polarization (EE) spectra ($2 \leq \ell \leq 30$) via the \texttt{Commander} and \texttt{SimAll} likelihoods, respectively. On smaller scales, we adopt the \texttt{CamSpec} likelihood for TT ($30 \leq \ell \leq 1000$) and cross (TE)/EE ($30 \leq \ell \leq 600$) spectra~\cite{Efstathiou:2019mdh,Rosenberg:2022sdy}. We augment the Planck 2018 data with high-multipole measurements from ACT DR6 (\texttt{ACT-lite})~\cite{ACT:2025fju} and SPT-3G (\texttt{SPT-lite})~\cite{SPT-3G:2025bzu}. To strictly avoid mode double-counting when combining these datasets, we restrict the Planck TT and polarization spectra (utilizing the \texttt{Commander}, \texttt{SimAll}, and \texttt{CamSpec} likelihoods) to $\ell < 1000$ and $\ell < 600$, respectively, following the ACT methodology~\cite{ACT:2025fju}. Additionally, we include the joint CMB lensing power spectrum reconstructed from Planck, ACT, and SPT-3G maps~\cite{ACT:2025qjh,Carron:2022eyg,ACT:2023dou,ACT:2023kun}.

\item \textbf{\texttt{DESI}:} The BAO measurements from the DESI DR2 of galaxies, quasars, and the Lyman-$\alpha$ forest are presented in Table IV of Ref.~\cite{DESI:2025zgx}. These measurements provide constraints on the transverse comoving distance $D_{\mathrm{M}}/r_{\mathrm{d}}$, the angle-averaged distance $D_{\mathrm{V}}/r_{\mathrm{d}}$, and the Hubble horizon $D_{\mathrm{H}}/r_{\mathrm{d}}$, where the distances are normalized by the comoving sound horizon $r_{\mathrm{d}}$ at the drag epoch.

\item \textbf{\texttt{PantheonPlus}:} The PantheonPlus compilation aggregates 1550 spectroscopically confirmed type Ia supernovae (SNe) from 18 distinct surveys, covering the redshift interval $0.01 < z < 2.26$~\cite{Brout:2022vxf}.

\item \textbf{\texttt{DESY5}:} The DESY5 compilation consists of 1829 SNe, including 1635 high-redshift ($0.1 < z < 1.3$) photometrically classified events from the full 5-year Dark Energy Survey sample, along with a low-redshift anchor sample of 194 SNe ($0.025 < z < 0.1$)~\cite{DES:2024jxu}.

\item \textbf{\texttt{DES-Dovekie}:} The DES-Dovekie dataset is a newly reconstructed SN dataset derived from a re-calibration and thorough re-analysis of the DESY5 SN sample, which includes 1623 likely SNe (with $P_{\mathrm{SNe}} > 0.5$) and 197 low-redshift SNe, totaling 1820 SNe~\cite{DES:2025sig}.

\end{itemize}

The theoretical predictions for the background and perturbation evolution of the different models are computed using the \texttt{IDECAMB} package~\cite{Li:2023fdk}, a modified version of \texttt{CAMB} specifically tailored for IDE scenarios. Parameter inference is performed via MCMC sampling using the \texttt{Cobaya} framework~\cite{Torrado:2020dgo}. The baseline parameter set for the $\Lambda$CDM model is given by $\boldsymbol{\theta}_{\Lambda\mathrm{CDM}} = \{\Omega_{\mathrm{b}} h^2, \Omega_{\mathrm{c}} h^2, H_0, \tau, \log(10^{10}A_{\mathrm{s}}), n_{\mathrm{s}}\}$. For the IDE models, the parameter sets are extended to $\boldsymbol{\theta}_{\mathrm{CQ}} = \{\boldsymbol{\theta}_{\Lambda\mathrm{CDM}}, \alpha, \beta\}$ for the CQ scenario and $\boldsymbol{\theta}_{\mathrm{CF}} = \{\boldsymbol{\theta}_{\Lambda\mathrm{CDM}}, w, \beta\}$ for the CF scenario. Table~\ref{tab:parameter priors} summarizes the full set of free parameters and the uniform priors adopted in the analysis.

\begin{table}[t]
%\small
\caption{Flat priors on the main cosmological parameters constrained in this work.}
\begin{center}
\renewcommand{\arraystretch}{1.4}
\begin{tabular}{@{\hspace{0.8cm}}c@{\hspace{0.9cm}} c@{\hspace{0.9cm}} c @{\hspace{0.8cm}} }
\hline\hline
\textbf{Model}       & \textbf{Parameter}       & \textbf{Prior}\\
%Model       & Parameter       & Prior\\
\hline
$\Lambda$CDM        & $\Omega_{\rm b} h^2$                     & $\mathcal{U}$[0.005\,,\,0.1] \\
                    & $\Omega_{\rm c} h^2$                     & $\mathcal{U}$[0.01\,,\,0.99] \\
                    & $H_0$                                    & $\mathcal{U}$[20\,,\,100] \\
                    & $\tau$                                   & $\mathcal{U}$[0.01\,,\,0.8] \\
                    & $\log(10^{10}A_{\rm s})$                 & $\mathcal{U}$[1.61\,,\,3.91] \\
                    & $n_{\rm s}$                              & $\mathcal{U}$[0.8\,,\,1.2] \\
\hline  
CPL                 & $w_0$                                    & $\mathcal{U}$[-3\,,\,1] \\
                    & $w_a$                                    & $\mathcal{U}$[-3\,,\,2] \\
\hline                    
CQ
                   & $\alpha$                                  & $\mathcal{U}$[0\,,\,1.4] \\
                   & $\beta$                                   & $\mathcal{U}$[0\,,\,0.15] \\
\hline  
CF                & $w$                                        & $\mathcal{U}$[-3\,,\,1] \\
                  & $\beta$                                    & $\mathcal{U}$[-5\,,\,1] \\                   
\hline\hline
\end{tabular}
\label{tab:parameter priors}
\end{center}	
\end{table}

Note that we treat $H_0$ as a free sampling parameter rather than the approximate angular acoustic scale $\theta_*$. Typically, in standard analyses for the $\Lambda$CDM model, $\theta_*$ is often used as a sampling parameter, which relies on the fitting formula by Hu and Sugiyama to estimate the redshift of the last scattering surface ($z_*$)~\cite{Hu:1995en}. This approximation takes the form
\begin{equation}
    z_*^{\mathrm{fit}} \approx 1048 \left[ 1 + 0.00124 (\Omega_{\mathrm{b}} h^2)^{-0.738} \right] \left[ 1 + g_1 (\Omega_{\mathrm{m}} h^2)^{g_2} \right],
    \label{eq:zstar_fit}
\end{equation}
where the coefficients $g_1$ and $g_2$ are given by
\begin{equation}
    g_1 = \frac{0.0783 (\Omega_{\mathrm{b}} h^2)^{-0.238}}{1 + 39.5 (\Omega_{\mathrm{b}} h^2)^{0.763}}, \quad
    g_2 = \frac{0.560}{1 + 21.1 (\Omega_{\mathrm{b}} h^2)^{1.81}}.
\end{equation}
While this formula is highly accurate for the standard non-interacting background evolution, it is potentially invalid in IDE scenarios where the energy density scaling laws and the recombination history can be modified by the dark sector interaction.

Therefore, to avoid model-dependent biases, we sample $H_0$ directly. In our analysis, the acoustic angular scale $\theta_*$ is treated as a derived parameter, rigorously defined as
\begin{equation}
    \theta_* = \frac{r_{\mathrm{d}}(z_*)}{D_{\mathrm{A}}(z_*)},
    \label{eq:thetastar_def}
\end{equation}
where the sound horizon $r_{\mathrm{d}}(z_*)$ and the comoving angular diameter distance $D_{\mathrm{A}}(z_*)$ are computed by numerically integrating the modified background equations. The decoupling redshift $z_*$ is determined precisely by finding the maximum of the visibility function (i.e., optical depth $\tau(z_*) \approx 1$) rather than using the fitting formula in Eq.~\eqref{eq:zstar_fit}. This approach ensures that the geometric constraints from CMB are applied consistently without enforcing an incorrect $\theta_*$ definition on the IDE models.

\begin{figure}[htbp]
\includegraphics[scale=0.45]{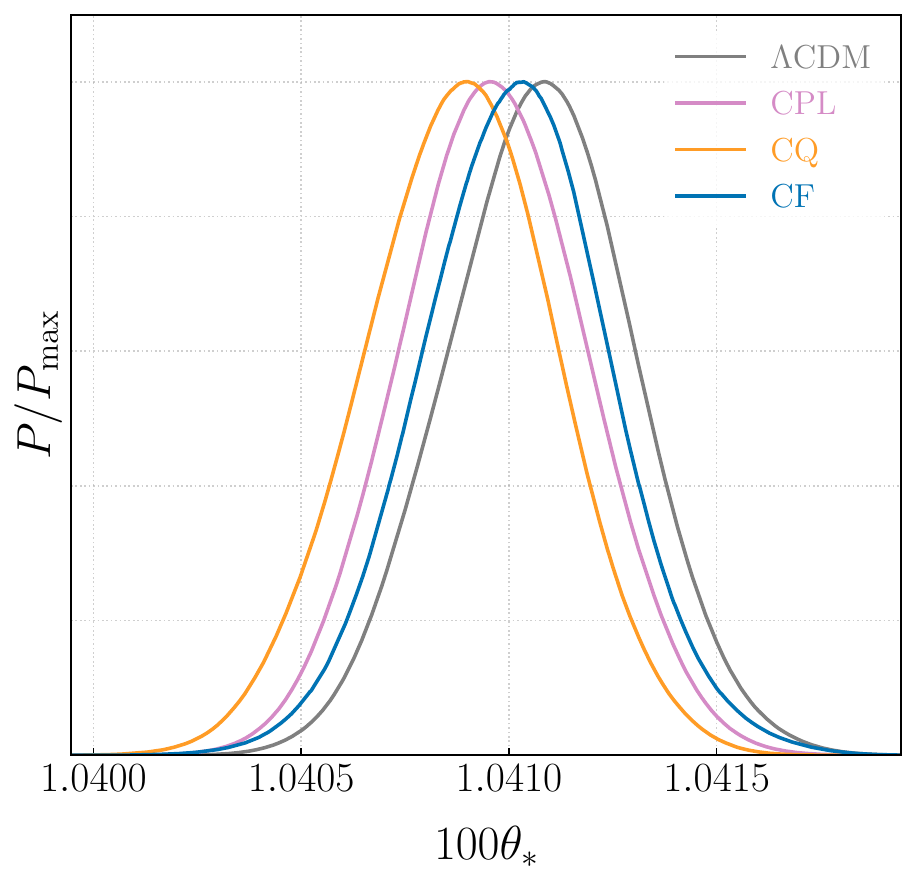}
\centering
\caption{The marginalized one-dimensional posterior distributions for $\theta_*$ in the $\Lambda$CDM, CPL, CQ, and CF models, constrained by the CMB+DESI+DES-Dovekie data. \label{fig:thetastar}}
\end{figure}

In Fig.~\ref{fig:thetastar}, we present the one-dimensional posterior distributions of $\theta_*$ for the $\Lambda$CDM, CPL, CQ, and CF models using the CMB+DESI+DES-Dovekie data. The values of $\theta_*$ are generally consistent across these models, with the extended models showing a slight decrease compared to $\Lambda$CDM, due to differences in the background evolution and expansion history.

\section{Supplementary Discussion}
\label{appendix:A}

\begin{table*}[htbp]
\centering
\caption{Fitting results (at the $1\sigma$ confidence level) in the $\Lambda$CDM, CPL, CQ, and CF models from the CMB+DESI, CMB+DESI+PantheonPlus, CMB+DESI+DESY5, and CMB+DESI+DES-Dovekie data. $H_{0}$ is in units of ${\rm km}~{\rm s}^{-1}~{\rm Mpc}^{-1}$.}
\label{tab:many_params}
\setlength{\tabcolsep}{2mm}
\renewcommand{\arraystretch}{1.4}
\scriptsize
\begin{tabular}{l c c c c}
\hline 
\hline
Model/Parameters & \texttt{CMB+DESI} & \texttt{CMB+DESI+PantheonPlus} & \texttt{CMB+DESI+DESY5} & \texttt{CMB+DESI+DES-Dovekie} \\
\hline
$\bm{\Lambda}$\textbf{CDM} &  &  &  & \\
$\Omega_\mathrm{b} h^2$ & $0.022482\pm 0.000091$ & $0.022466\pm 0.000091$ & $0.022476\pm 0.000091$ & $0.022472\pm 0.000092$ \\
$\Omega_\mathrm{c} h^2$ & $0.11805\pm 0.00060$ & $0.11837\pm 0.00057$ & $0.11821\pm 0.00059$ & $0.11825\pm 0.00059$ \\
$100\theta_\mathrm{*}$ & $1.04109\pm0.00022$ & $1.04108\pm0.00023$ & $1.04106\pm0.00022$ & $1.04108\pm0.00023$ \\
$\tau_\mathrm{reio}$ & $0.0606\pm 0.0044$ & $0.0601\pm 0.0043$ & $0.0602\pm 0.0043$ & $0.0603\pm 0.0043$ \\
$n_\mathrm{s}$ & $0.9734\pm 0.0029$ & $0.9727\pm 0.0029$ & $0.9730\pm 0.0029$ & $0.9729\pm 0.0029$ \\
$\log(10^{10} A_\mathrm{s})$ & $3.0565\pm 0.0084$ & $3.0556\pm 0.0083$ & $3.0560\pm 0.0083$ & $3.0562\pm 0.0083$ \\
$H_0$ & $68.12\pm 0.25$ & $68.05\pm 0.24$ & $67.98\pm 0.24$ & $68.03\pm 0.25$  \\
$\Omega_{\mathrm{m}}$ & $0.3043\pm 0.0033$ & $0.3052\pm 0.0033$ & $0.3062\pm 0.0032$ & $0.3055\pm 0.0033$ \\
$\sigma_8$ & $0.8117\pm 0.0036$ & $0.8123\pm 0.0035$ & $0.8119\pm 0.0035$ & $0.8121\pm 0.0035$ \\
$S_8$ & $0.8174\pm 0.0064$ & $0.8205\pm 0.0062$ & $0.8190\pm 0.0064$ & $0.8196\pm 0.0064$ \\
$r_\mathrm{d}$ [Mpc] & $147.50\pm 0.18$ & $147.43\pm 0.17$ & $147.46\pm 0.17$ & $147.45\pm 0.17$ \\

\hline
$\bm{\textbf{CPL}}$ &  &  &  & \\
$\Omega_\mathrm{b} h^2$ & $0.022429\pm 0.000092$ & $0.022434\pm 0.000095$ & $0.022440\pm 0.000092$ & $0.022436\pm 0.000092$ \\
$\Omega_\mathrm{c} h^2$ & $0.11962\pm 0.00073$ & $0.11937\pm 0.00072$ & $0.11925\pm 0.00071$ & $0.11936\pm 0.00072$ \\
$100\theta_\mathrm{*}$ & $1.04093\pm 0.00023$ & $1.04097\pm 0.00022$ & $1.04095\pm 0.00023$ & $1.04096\pm 0.00022$ \\
$\tau_\mathrm{reio}$ & $0.0557\pm 0.0044$ & $0.0566\pm 0.0044$ & $0.0570\pm 0.0044$ & $0.0567\pm 0.0044$ \\
$n_\mathrm{s}$ & $0.9700\pm 0.0029$ & $0.9705\pm 0.0030$ & $0.9708\pm 0.0030$ & $0.9705\pm 0.0030$ \\
$\log(10^{10} A_\mathrm{s})$ & $3.0459\pm 0.0086$ & $3.0479\pm 0.0085$ & $3.0488\pm 0.0084$ & $3.0481\pm 0.0085$ \\
$H_0$ & $63.80^{+1.70}_{-2.00}$ & $67.66\pm 0.59$ & $66.86\pm 0.55$ & $67.45\pm 0.53$  \\
$\Omega_{\mathrm{m}}$ & $0.3510\pm 0.0210$ & $0.3110\pm 0.0056$ & $0.3187\pm 0.0054$ & $0.3132\pm 0.0052$  \\
$w_0$ & $-0.430\pm 0.210$ & $-0.833\pm 0.055$ & $-0.749\pm 0.057$ & $-0.802\pm 0.055$  \\
$w_a$ & $-1.72\pm 0.57$ & $-0.65\pm 0.20$ & $-0.88^{+0.23}_{-0.19}$ & $-0.75\pm 0.21$  \\
$\sigma_8$ & $0.785\pm 0.017$ & $0.8101\pm 0.0070$ & $0.8160\pm 0.0073$ & $0.8149\pm 0.0068$ \\
$S_8$ & $0.8486\pm 0.0108$ & $0.8349\pm 0.0071$ & $0.8307\pm 0.0071$ & $0.8325\pm 0.0071$ \\
$r_\mathrm{d}$ [Mpc] & $147.14\pm 0.20$ & $147.20\pm 0.20$ & $147.22\pm 0.19$ & $147.20\pm 0.19$ \\
\hline
$\bm{\textbf{CQ}}$ &  &  &  & \\
$\Omega_\mathrm{b} h^2$ & $0.022398\pm 0.000097$ & $0.022395\pm 0.000098$ & $0.022402\pm 0.000095$ & $0.022395\pm 0.000095$ \\
$\Omega_\mathrm{c} h^2$ & $0.1157^{+0.0016}_{-0.00099}$ & $0.1142\pm 0.0014$ & $0.1153^{+0.0014}_{-0.0012}$ & $0.1150^{+0.0014}_{-0.0012}$ \\
$100\theta_\mathrm{*}$ & $1.04089\pm 0.00023$ & $1.04089\pm 0.00023$ & $1.04088\pm 0.00024$ & $1.04088\pm 0.00023$ \\
$\tau_\mathrm{reio}$ & $0.0573\pm 0.0044$ & $0.0578\pm 0.0045$ & $0.0575\pm 0.0045$ & $0.0576^{+0.0041}_{-0.0046}$ \\
$n_\mathrm{s}$ & $0.9702\pm 0.0031$ & $0.9702\pm 0.0031$ & $0.9703\pm 0.0031$ & $0.9702\pm 0.0030$ \\
$\log(10^{10} A_\mathrm{s})$ & $3.0526\pm 0.0085$ & $3.0546\pm 0.0086$ & $3.0534\pm 0.0085$ & $3.0539\pm 0.0084$ \\
$H_0$ & $68.53^{+0.85}_{-0.54}$ & $68.17\pm 0.60$ & $67.53\pm 0.56$ & $68.04\pm 0.55$  \\
$\Omega_{\mathrm{m}}$ & $0.2956^{+0.0050}_{-0.0063}$ & $0.2978\pm 0.0053$ & $0.3011\pm 0.0053$ & $0.2984^{+0.0047}_{-0.0053}$  \\
$\alpha$ & $< 0.39$ & $0.41\pm 0.19$ & $0.62\pm 0.17$ & $0.46\pm 0.18$  \\
$\beta$ & $0.0507^{+0.0091}_{-0.0075}$ & $0.0517^{+0.0092}_{-0.0078}$ & $0.0561^{+0.0094}_{-0.0068}$ & $0.0527\pm 0.0087$ \\
$\sigma_8$ & $0.8320\pm 0.0097$ & $0.8251\pm 0.0090$ & $0.8290\pm 0.0090$ & $0.8287\pm 0.0087$ \\
$S_8$ & $0.8256\pm 0.0069$ & $0.8265\pm 0.0068$ & $0.8260\pm 0.0068$ & $0.8263\pm 0.0068$ \\
$r_\mathrm{d}$ [Mpc] & $147.02\pm 0.24$ & $146.99\pm 0.24$ & $147.02\pm 0.23$ & $147.00\pm 0.23$ \\
\hline
$\bm{\textbf{CF}}$ &  &  &  & \\
$\Omega_\mathrm{b} h^2$ & $0.022478\pm 0.000093$ & $0.022466\pm 0.000097$ & $0.022457^{+0.000089}_{-0.000099}$ & $0.022446\pm 0.000090$ \\
$\Omega_\mathrm{c} h^2$ & $0.246^{+0.034}_{-0.018}$ & $0.271^{+0.023}_{-0.018}$ & $0.245^{+0.029}_{-0.023}$ & $0.257^{+0.025}_{-0.022}$ \\
$100\theta_\mathrm{*}$ & $1.04099\pm 0.00022$ & $1.04098\pm 0.00022$ & $1.04097\pm 0.00023$ & $1.04102\pm 0.00023$ \\
$\tau_\mathrm{reio}$ & $0.0575^{+0.0053}_{-0.0041}$ & $0.0558^{+0.0040}_{-0.0047}$ & $0.0564\pm 0.0044$ & $0.0558\pm 0.0045$ \\
$n_\mathrm{s}$ & $0.9715\pm 0.0032$ & $0.9708\pm 0.0030$ & $0.9711\pm 0.0030$ & $0.9709\pm 0.0030$ \\
$\log(10^{10} A_\mathrm{s})$ & $3.0480^{+0.0096}_{-0.0081}$ & $3.0445^{+0.0080}_{-0.0089}$ & $3.0461\pm 0.0086$ & $3.0447\pm 0.0087$ \\
$H_0$ & $67.20^{+1.20}_{-1.40}$ & $67.60\pm 0.57$ & $66.99^{+0.51}_{-0.58}$ & $67.46\pm 0.53$ \\
$\Omega_{\mathrm{m}}$ & $0.598^{+0.091}_{-0.057}$ & $0.587^{+0.064}_{-0.052}$ & $0.655^{+0.053}_{-0.043}$ & $0.615^{+0.057}_{-0.049}$ \\
$w$ & $-1.45^{+0.13}_{-0.15}$ & $-1.46^{+0.16}_{-0.13}$ & $-1.60\pm 0.15$ & $-1.52^{+0.16}_{-0.13}$ \\
$\beta$ & $-2.37^{+0.82}_{-0.93}$ & $-2.25^{+0.73}_{-0.63}$ & $-3.08\pm 0.70$ & $-2.59^{+0.76}_{-0.62}$ \\
$\sigma_8$ & $0.493^{+0.018}_{-0.051}$ & $0.461^{+0.017}_{-0.026}$ & $0.494^{+0.022}_{-0.041}$ & $0.478^{+0.021}_{-0.033}$ \\
$S_8$ & $0.6890^{+0.0077}_{-0.0179}$ & $0.6879^{+0.0097}_{-0.0164}$ & $0.6786^{+0.0082}_{-0.0110}$ & $0.6826^{+0.0100}_{-0.0139}$ \\
$r_\mathrm{d}$ [Mpc] & $147.29^{+0.21}_{-0.24}$ & $147.21\pm 0.21$ & $147.25\pm 0.19$ & $147.24\pm 0.19$ \\
\hline
\hline
\end{tabular}
\end{table*}

\begin{figure*}[htbp]
\includegraphics[scale=0.42]{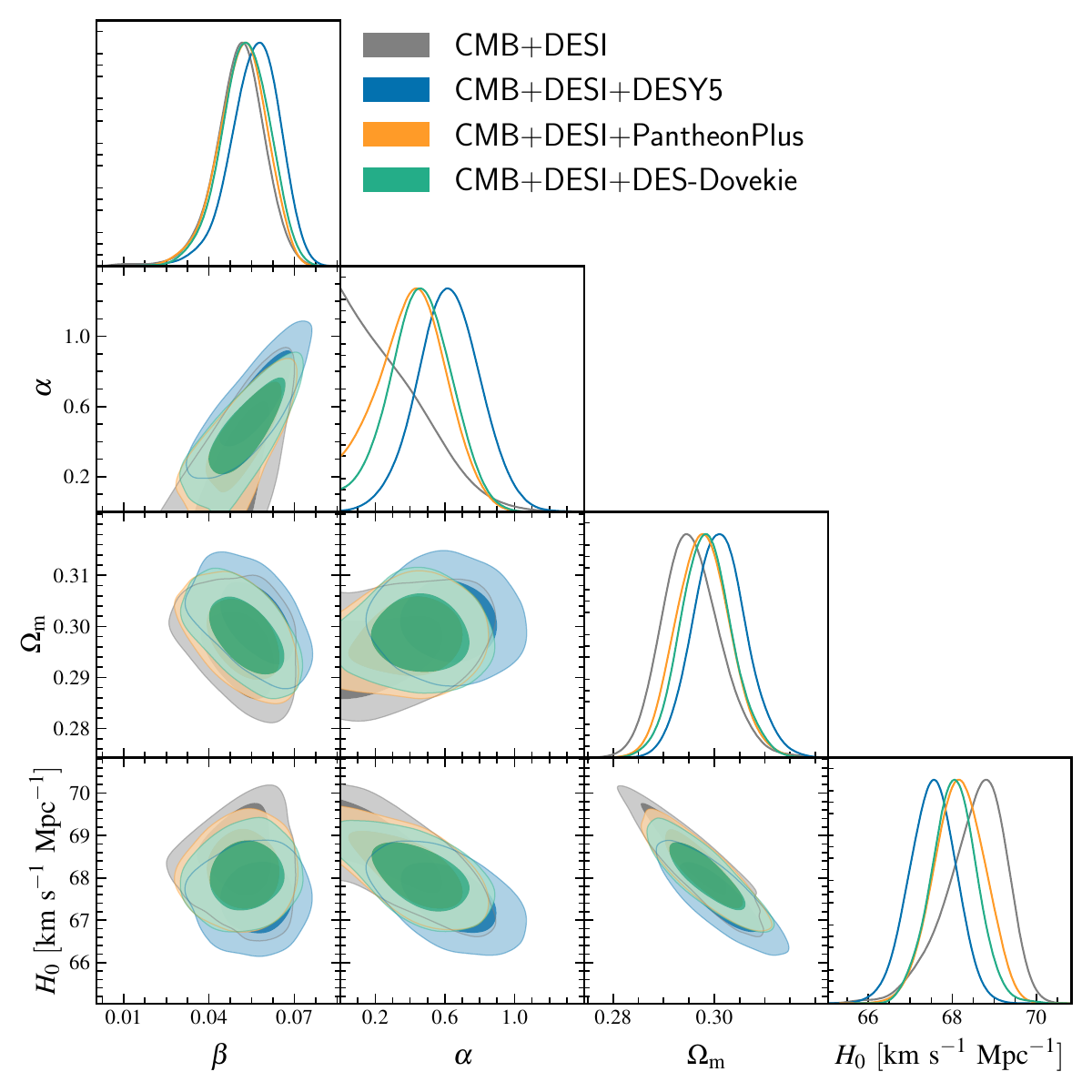}
\includegraphics[scale=0.42]{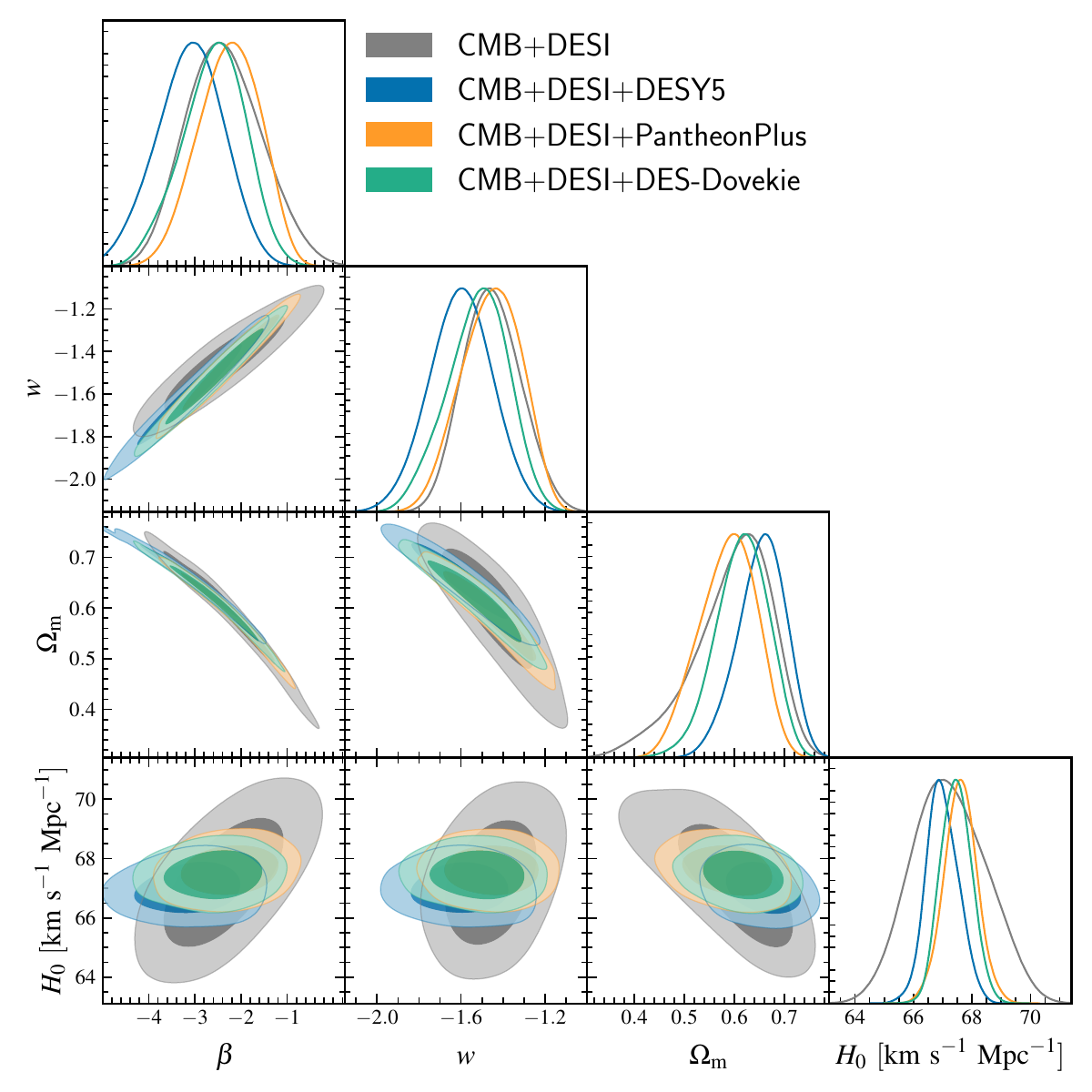}
\centering
\caption{Constraints on cosmological parameters from the CMB+DESI, CMB+DESI+PantheonPlus, CMB+DESI+DESY5, and CMB+DESI+DES-Dovekie data combinations for the CQ (left panel) and CF (right panel) models. \label{fig:TPlots} }
\end{figure*}

In this section, we discuss in more detail some additional results reported in Table~\ref{tab:many_params} and visualized in Fig.~\ref{fig:TPlots} and Fig.~\ref{fig:Omega_and_w}. In particular, Table~\ref{tab:many_params} reports the $1\sigma$ confidence level constraints on all sampled and derived parameters across the different models considered in this work. Additionally, we report and discuss the fitting results for BAO and SN data, as presented in Table~\ref{tab.BAOs}, Fig.~\ref{fig:BAO_fit}, and Fig.~\ref{fig:SN_fit}. We also assess the impact of CMB lensing information and of the inclusion of $f\sigma_8(z)$ growth-rate measurements on our results. The corresponding parameter constraints are summarized in Table~\ref{tab:newmany_params} and shown in Fig.~\ref{fig:lensing}, Fig.~\ref{fig:fsigma8}, and Fig.~\ref{fig:fsigma8_fit}.

\subsection{Constraining the additional cosmological parameters}\label{additional:par}

We begin by focusing on the inferred values of the Hubble parameter $H_0$ and the present-day matter density $\Omega_{\mathrm{m}}$. Together with the additional parameters describing the DE dynamics or the interaction in the dark sector, these quantities play a central role in determining the background evolution of the Universe within the different scenarios. Moreover, both $H_0$ and $\Omega_{\mathrm{m}}$ are of particular interest in light of current cosmological tensions. On the one hand, $H_0$ lies at the core of the well-known Hubble tension~\cite{Verde:2019ivm,DiValentino:2020zio,DiValentino:2021izs,Perivolaropoulos:2021jda,Schoneberg:2021qvd,Shah:2021onj,Abdalla:2022yfr,DiValentino:2022fjm,Kamionkowski:2022pkx,Giare:2023xoc,Hu:2023jqc,Guo:2018ans,CosmoVerseNetwork:2025alb,Vagnozzi:2019ezj,Vagnozzi:2023nrq} between early- and late-Universe measurements of the present-day expansion rate. On the other hand, recent analyses by the DESI collaboration have highlighted emerging discrepancies in the determination of $\Omega_{\mathrm{m}}$ within $\Lambda$CDM when comparing different cosmological probes~\cite{DESI:2025zgx,Loverde:2024nfi,Colgain:2024mtg,Elbers:2025vlz,Lynch:2025ine,Sailer:2025lxj,Jhaveri:2025neg}, with implications ranging from dynamical DE models to neutrino cosmology~\cite{Du:2024pai,Craig:2024tky,Naredo-Tuero:2024sgf,Green:2024xbb,Elbers:2024sha,Du:2025iow,Giare:2025ath,Du:2025xes,Zhou:2025nkb}.

Within $\Lambda$CDM, the value of $H_0$ inferred from the different data combinations is remarkably stable, clustering around $H_0 \simeq 68~{\rm km}~{\rm s}^{-1}~{\rm Mpc}^{-1}$ with small uncertainties. Similarly, the inferred values of $\Omega_{\mathrm{m}}$ remain tightly constrained and consistently centered around $\Omega_{\mathrm{m}} \simeq 0.305$.

In the CPL parametrization, the situation is more nuanced. For the CMB+DESI combination, the inferred value of $H_0$ is significantly lower, with a central value of $H_0 \simeq 63.80^{+1.70}_{-2.00}~{\rm km}~{\rm s}^{-1}~{\rm Mpc}^{-1}$, accompanied by substantially larger uncertainties. This shift is associated with a significantly higher matter density, $\Omega_{\mathrm{m}} = 0.3510 \pm 0.0210$, reflecting the well-known degeneracies between the DE EoS parameters $(w_0,w_a)$, $H_0$, and $\Omega_{\mathrm{m}}$. The inclusion of SN data substantially tightens the constraints and shifts $H_0$ back toward values around $67~{\rm km}~{\rm s}^{-1}~{\rm Mpc}^{-1}$, which nevertheless remain systematically lower than those inferred within $\Lambda$CDM. This behavior reinforces the trend already discussed in the literature, namely that CPL models, while providing an improved fit to cosmological data, do not alleviate the Hubble tension and instead tend to shift the constraints in the opposite direction to that required for its resolution. Once SN data are included, the inferred values of $\Omega_{\mathrm{m}}$ cluster around $\Omega_{\mathrm{m}} \simeq 0.31$, remaining slightly higher than in $\Lambda$CDM.

In light of these differences between the $\Lambda$CDM and CPL scenarios, it is instructive to compare this behavior with that of IDE models. The distinct phenomenology exhibited by the CQ and CF realizations manifests itself clearly in the inferred values of $H_0$ and $\Omega_{\mathrm{m}}$, as well as in their redshift evolution.

\begin{figure*}[!htp]
\includegraphics[width=0.42\textwidth]{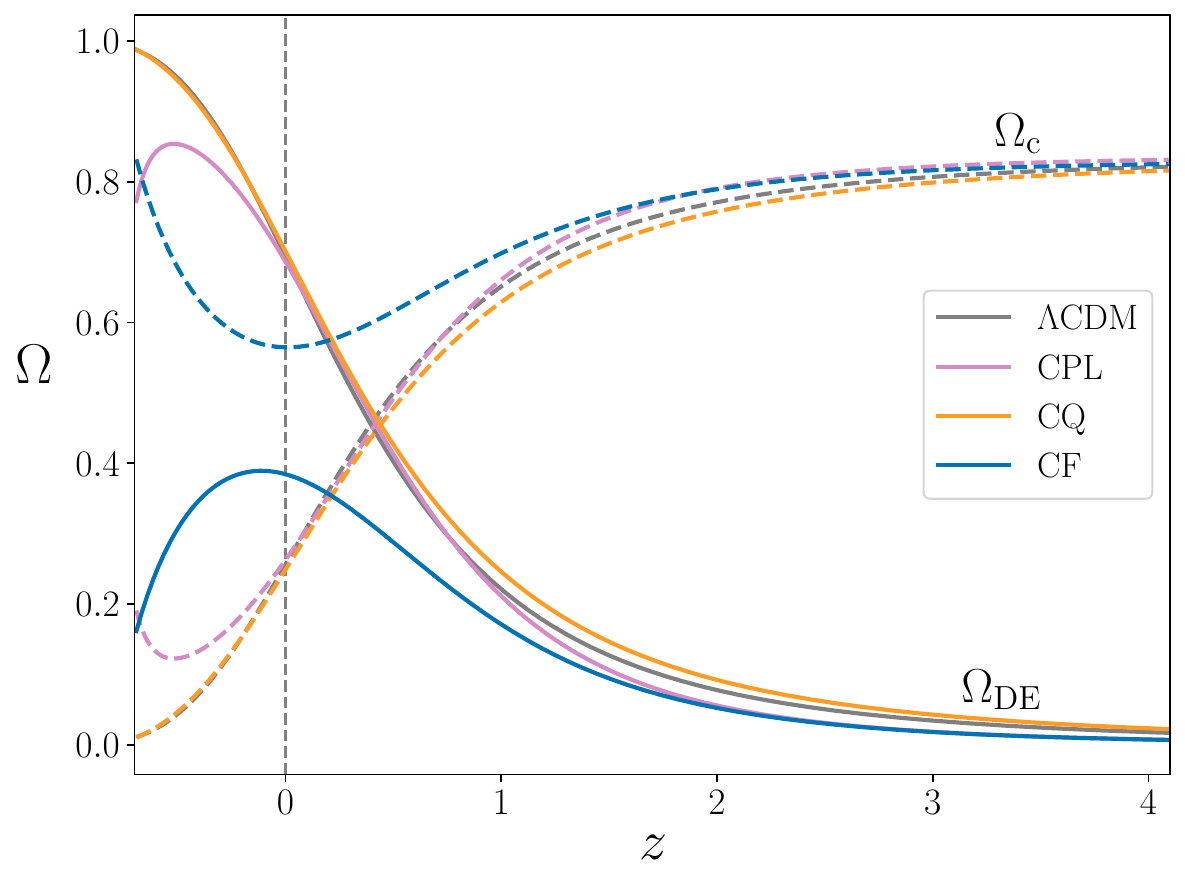} \ \hspace{1cm}
\includegraphics[width=0.43\textwidth]{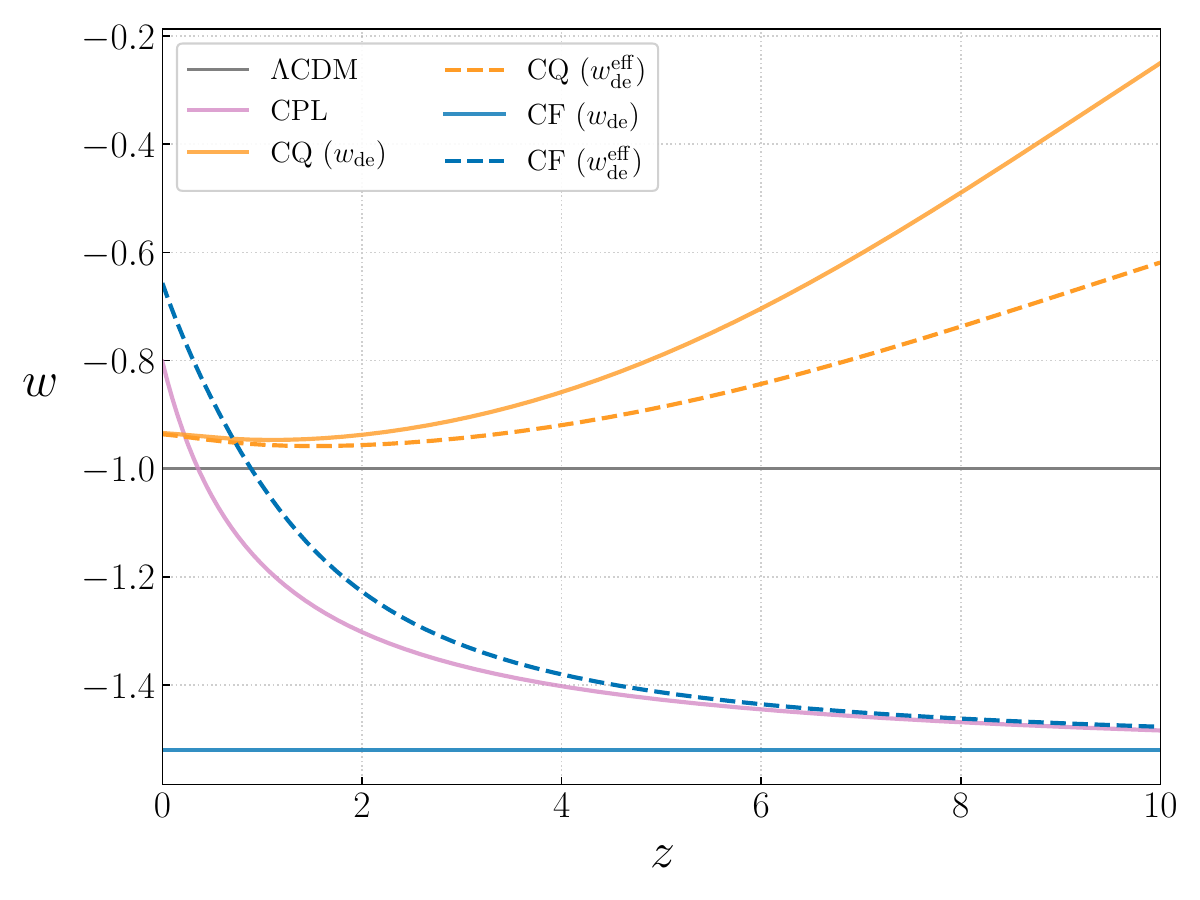}\ \hspace{1cm}
\centering \caption{\label{fig:Omega_and_w} \textbf{\textit{Left panel:}} Evolution of the DM (dashed lines) and the DE (solid lines) density fractions in the $\Lambda$CDM, CPL, CQ, and CF models. \textbf{\textit{Right panel:}} Evolution of the DE EoS (dashed lines) and the effective EoS (solid lines) in the $\Lambda$CDM, CPL, CQ, and CF models. For all models, the curves are computed using the best-fit parameters for CMB+DESI+DES-Dovekie. Note that, in the CF scenario, the phantom-divide crossing occurs in the effective EoS only. This should be understood as an effective phantom-like behavior induced by the dark-sector interaction, rather than as a phantom crossing of the intrinsic DE EoS itself.}
\end{figure*}

In the CQ scenario, the inferred values of $H_0$ remain close to those obtained in $\Lambda$CDM across all dataset combinations, with central values typically in the range $H_0 \simeq 67.5-68$~${\rm km}~{\rm s}^{-1}~{\rm Mpc}^{-1}$. Compared to $\Lambda$CDM, the uncertainties are moderately larger, similarly to what is observed in the CPL parametrization, reflecting the additional degeneracies introduced by the interaction parameter $\beta$ and the potential slope $\alpha$, as also visible in the left panel of Fig.~\ref{fig:TPlots}. Importantly, the CQ model does not drive $H_0$ toward either significantly higher or lower values. Instead, the interaction modifies the redshift evolution of the DM and DE energy densities contributing to $H(z)$, while leaving its overall normalization, and hence the inferred value of $H_0$, nearly unchanged. Concerning the matter density, the inferred values of $\Omega_{\mathrm{m}}$ in the CQ model are systematically lower than those obtained in CPL and slightly smaller, yet fully compatible, with those inferred within $\Lambda$CDM, typically lying in the range $\Omega_{\mathrm{m}} \simeq 0.298-0.301$ once SN data are included. This downward shift is driven by the continuous transfer of energy from DM to DE. As illustrated in Fig.~\ref{fig:Omega_and_w}, this interaction sustains an enhanced DE density at intermediate redshifts, $0.5 \lesssim z \lesssim 2$, while causing the DM density to dilute faster than in $\Lambda$CDM. The resulting background evolution partially overlaps with that obtained in CPL models, while remaining consistent with a quintessence-like DE EoS at all redshifts.

A qualitatively different behavior is observed in the CF scenario. In this case, the inferred values of $H_0$ are lower than those obtained in $\Lambda$CDM and CQ, typically clustering around $H_0 \simeq 67~{\rm km}~{\rm s}^{-1}~{\rm Mpc}^{-1}$, closely mirroring the behavior observed in the CPL parametrization\footnote{\parbox[t]{\linewidth}{We note that, unlike the specific IDE scenario studied in Ref.~\cite{Giare:2024smz}, which alleviates the $H_0$ tension, the CF model in our analysis does not yield a higher $H_0$. This discrepancy primarily arises from model specifications. Specifically, we do not impose a prior on the sign of the coupling parameter or on the direction of energy transfer, we treat $w$ as a free parameter, and we adopt the interaction form $Q \propto H_0 \rho_{\rm de}$ rather than $Q \propto H \rho_{\rm de}$. The discrepancy also stems from data choices, as our baseline analysis incorporates the DES-Dovekie SN data that naturally favor a lower $H_0$, whereas the primary results of Ref.~\cite{Giare:2024smz} rely on CMB and DESI BAO data.}}. In contrast, the CF model favors substantially higher values of the matter density parameter, with central values in the range $\Omega_{\mathrm{m}} \simeq 0.59-0.63$ depending on the dataset combination. These large values are driven by the negative coupling, which corresponds to a net transfer of energy from DE to DM. This behavior is clearly reflected in Fig.~\ref{fig:Omega_and_w}. In the CF model, the continuous feeding of DM from DE leads to a substantial enhancement of the DM energy density already at $z \gtrsim 2$, while the DE density is correspondingly suppressed relative to $\Lambda$CDM and CPL. At the background level, the increased matter abundance is compensated by the strongly phantom DE EoS favored in this scenario, allowing the model to reproduce the observed distance measurements despite its markedly different late-time energy budget. The strong correlations among $w$, $\Omega_{\mathrm{m}}$, and $H_0$ visible in the right panel of Fig.~\ref{fig:TPlots} further illustrate this compensation mechanism. Conversely, at the level of cosmological perturbations, the CF scenario is characterized by a strongly suppressed growth of structure, resulting in very low values of $\sigma_8 \sim 0.48$ and $S_8 \sim 0.68$ compared to those typically inferred in $\Lambda$CDM, CPL, and CQ. This behavior is a direct consequence of the same interaction mechanism that enhances the late-time matter density: the continuous energy transfer from DE to DM, combined with a strongly phantom DE EoS, leads to an expansion history that significantly increases the Hubble friction term and induces an earlier and more efficient suppression of the linear growth of matter perturbations at late times. In this regime, despite the large matter abundance, the accelerated expansion inhibits the efficient clustering of matter, resulting in a reduced growth factor and, consequently, in low values of $\sigma_8$ and $S_8$. As a result, the CF realization lives in a region of parameter space with markedly different clustering properties, despite providing an excellent fit to CMB, BAO, and SN data. While such low values of $\sigma_8$ and $S_8$ would, at face value, appear in tension with standard weak-lensing determinations, it is important to note that WL constraints are themselves model dependent to some extent, and their interpretation within interacting dark-sector scenarios requires a dedicated and self-consistent treatment of both background and perturbation dynamics. In this sense, the CF model should be regarded as an extreme but instructive realization of IDE phenomenology, illustrating how physically distinct dark-sector interactions can reproduce current background and CMB observations while predicting significantly different structure-formation histories. A comprehensive assessment of the viability of this scenario in light of WL and LSS measurements is therefore beyond the scope of the present work and is left for future dedicated analyses.

Taken together, these results indicate that the interacting models considered here do not lead to a statistically significant shift in the inferred value of $H_0$, which remains broadly consistent with early-Universe determinations. At the same time, IDE scenarios introduce additional freedom in the background evolution, providing increased flexibility to accommodate differences across datasets that, within $\Lambda$CDM, primarily translate into mismatches in the inferred value of $\Omega_{\mathrm{m}}$.

As a final consistency check, we examine the two-dimensional correlations shown in Fig.~\ref{fig:TPlots} among the interaction parameters and the key background quantities in the CQ and CF scenarios. While a preference for non-zero interactions is found in both models, the structure of the degeneracies and their physical origin are qualitatively different. Importantly, the directions of the correlations are physically consistent in both cases.

In the CQ scenario, the observed correlations reflect a relatively mild reshaping of the background evolution:
\begin{itemize}
\item \boldsymbol{$\alpha - \beta$}: The coupling parameter $\beta$ is positively correlated with the potential slope $\alpha$. A stronger interaction requires a steeper inverse power-law potential in order to sustain a viable background evolution and maintain consistency with CMB and distance measurements.
\item \boldsymbol{$\beta - \Omega_{\mathrm{m}}$}: A mild anti-correlation is observed between $\beta$ and $\Omega_{\mathrm{m}}$. As $\beta$ increases, energy is transferred from DM to DE, leading to a slightly faster dilution of the DM density and hence to marginally lower values of $\Omega_{\mathrm{m}}$ today.
\item \boldsymbol{$\beta - H_0$}: Correlations involving $H_0$ are weak. This indicates that, in the CQ model, the interaction primarily affects the redshift evolution of the energy densities entering $H(z)$, while leaving the overall normalization of the expansion rate largely unchanged. As a result, the inferred values of $H_0$ remain close to those obtained in $\Lambda$CDM.
\end{itemize}

Conversely, the correlations in the CF scenario are substantially stronger and reflect a more radical reorganization of the background dynamics driven by the phenomenological interaction $Q \propto \beta H_0 \rho_{\rm de}$:
\begin{itemize}
\item \boldsymbol{$\beta - \Omega_{\mathrm{m}}$}: A strong correlation is present between $\beta$ and $\Omega_{\mathrm{m}}$, such that increasingly negative values of $\beta$ correspond to progressively larger values of $\Omega_{\mathrm{m}}$. This behavior directly reflects the fact that more negative $\beta$ implies a stronger energy transfer from DE to DM, leading to a substantial enhancement of the DM abundance at late times.
\item \boldsymbol{$\beta - w$}: The parameters $\beta$ and $w$ are also strongly correlated. As $w \to -1$, the preferred values of $\beta$ approach smaller absolute values (i.e., $\beta \to 0$): for less phantom DE, the DE energy density is more relevant at earlier times, so that a smaller coupling is sufficient to keep the interaction term $Q \propto \beta H_0 \rho_{\rm de}$ effective. Conversely, for increasingly phantom values of $w \ll -1$, the DE density becomes less relevant in the past, and the energy transfer is pushed to very late times. In this regime, a more negative $\beta$ is required to produce a significant late-time interaction, which in turn leads to a strong enhancement of $\Omega_{\mathrm{m}}$.
\item \boldsymbol{$\Omega_{\mathrm{m}} - H_0$}: A mild anti-correlation is observed between $\Omega_{\mathrm{m}}$ and $H_0$. Larger values of the present-day matter density require lower values of $H_0$ in order to preserve a background expansion history compatible with distance measurements, reflecting the standard geometrical degeneracy in $H(z)$.
\end{itemize}

\subsection{Fitting analysis of BAO and SN data}\label{fitt:BAO and SN}

\begin{figure}[htbp]
\includegraphics[scale=0.45]{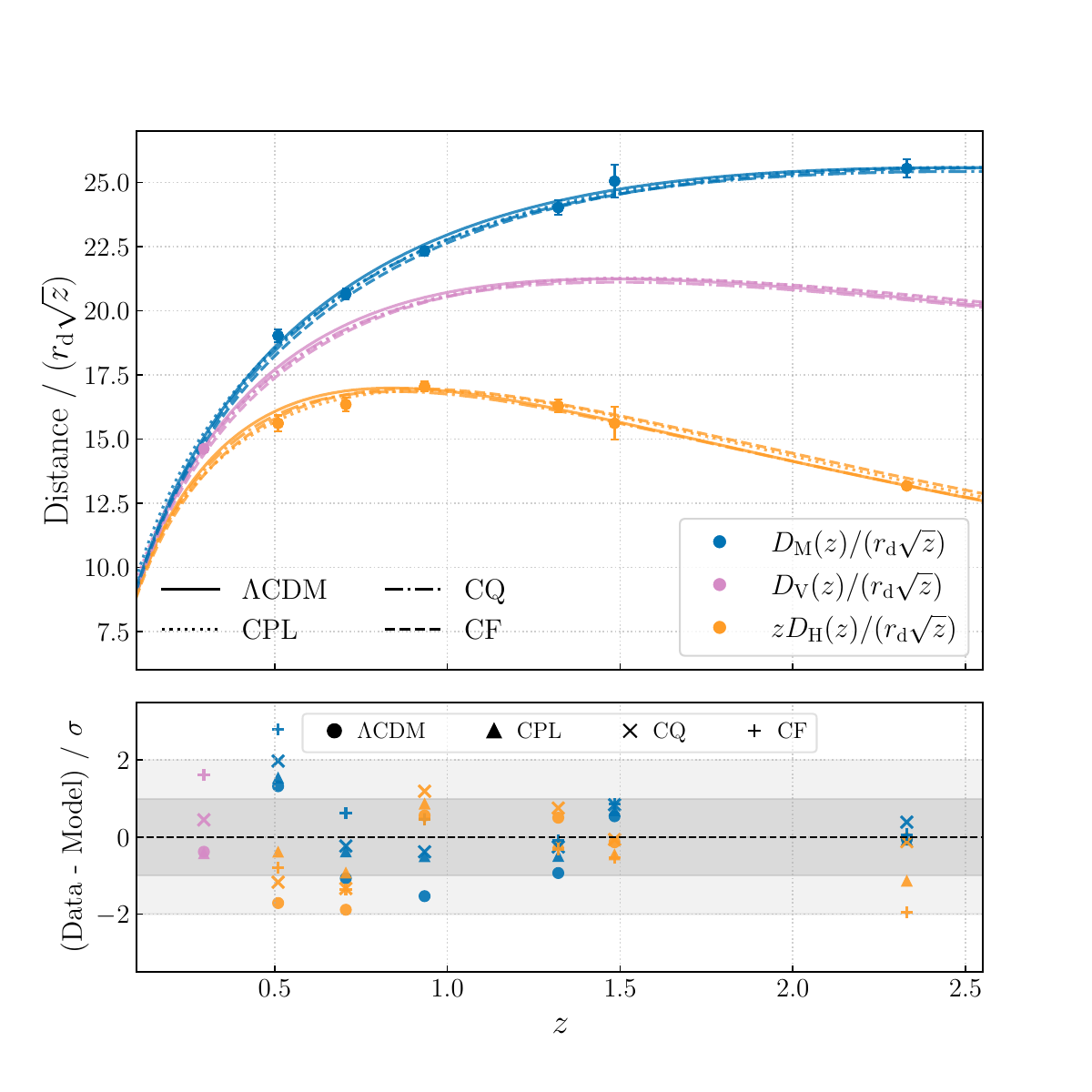}
\centering
\caption{\label{fig:BAO_fit} Best-fit predictions for (rescaled) distance-redshift relations for $\Lambda$CDM, CPL, CQ, and CF obtained from the analysis of CMB+DESI data.}
\end{figure}

To better investigate the fit of the DESI BAO data to these models, we present a comparison between the best-fit CMB+BAO predictions and the DESI BAO measurements in Fig.~\ref{fig:BAO_fit}, emphasizing a high degree of degeneracy among the CQ, CF, and CPL models. As illustrated by the residuals shown in the bottom panel of the same figure, all models provide a good description of the BAO distance measurements and improve upon the $\Lambda$CDM fit, particularly for the transverse BAO observable $D_{\mathrm{M}} / (r_{\mathrm{d}} \sqrt{z})$ and, to a lesser extent, for the radial BAO observable $z D_{\mathrm{H}} / (r_{\mathrm{d}} \sqrt{z})$ in the redshift range $0.5 \lesssim z \lesssim 1$. To further elucidate the origin of the differences discussed above, we provide a more detailed perspective on the BAO constraints. In Table~\ref{tab.BAOs} we explicitly report the residuals between the best-fit predictions of the $\Lambda$CDM, CPL, CQ, and CF models and the DESI DR2 BAO measurements on a point-by-point basis, expressed in units of observational uncertainties. This provides a quantitative counterpart to the information shown in Fig.~\ref{fig:BAO_fit}. Overall, all models remain broadly consistent with BAO observations, with most deviations lying within the $1-2\sigma$ level. Nevertheless, systematic differences can be identified. In particular, an improved fit to the transverse BAO observable $D_{\mathrm{M}}/(r_{\mathrm{d}}\sqrt{z})$ is observed in the four redshift bins between $z = 0.510$ and $z = 1.321$ for the CPL, CQ, and CF models relative to $\Lambda$CDM, with CQ typically yielding the smallest residuals. At the same redshifts, the radial BAO observable $zD_{\mathrm{H}}/(r_{\mathrm{d}}\sqrt{z})$ shows a different behavior: while CPL and CF continue to improve upon $\Lambda$CDM, CQ exhibits a reduced level of agreement, partially compensating for the gains observed in $D_{\mathrm{M}}/(r_{\mathrm{d}}\sqrt{z})$. At higher redshift, around $z \simeq 2.33$, CPL and CF slightly worsen the fit to $zD_{\mathrm{H}}/(r_{\mathrm{d}}\sqrt{z})$ compared to $\Lambda$CDM, highlighting the redshift- and observable-dependent impact of departures from the standard expansion history.

\begin{table*}[tpb!]
\caption{DESI DR2 results, together with their $1\sigma$ uncertainties, presented for three different types of BAO distance measurements. For each data point, we show the level of consistency between the best-fit predictions of the $\Lambda$CDM, CPL, CQ, and CF models obtained from the CMB+DESI dataset and the corresponding observed BAO measurement, expressed in units of observational uncertainty ($\#\sigma$).}
\label{tab.BAOs}
\centering
\renewcommand{\arraystretch}{1.5}
\resizebox{0.75\textwidth}{!}{%
\begin{tabular}{l c c c c c c}
\hline\hline
\textbf{Distance} & \textbf{Redshift} & \textbf{DESI}  & \textbf{$\Lambda$CDM ($\#\sigma$)} & \textbf{CPL ($\#\sigma$)} & \textbf{CQ ($\#\sigma$)} & \textbf{CF ($\#\sigma$)} \\ 
\hline
$D_{\mathrm{V}}(z)/(r_{\mathrm{d}} \sqrt{z})$ & 0.295 & 14.622 $\pm$ 0.140 & $-0.380$ & $+0.417$ & $+0.449$ & $+1.611$ \\
\hline
\multirow{7}{*}{$D_{\mathrm{M}}(z)/(r_{\mathrm{d}} \sqrt{z})$}
                        & 0.510 & 19.026 $\pm$ 0.236 & $+1.322$ & $+1.545$ & $+1.976$ & $+2.795$ \\
                        & 0.706 & 20.650 $\pm$ 0.214 & $-1.071$ & $-0.369$ & $-0.231$ & $+0.619$ \\
                        & 0.934 & 22.325 $\pm$ 0.290 & $-1.534$ & $-0.493$ & $-0.381$ & $+0.449$ \\
                        & 1.321 & 24.014 $\pm$ 0.480 & $-0.934$ & $-0.486$ & $-0.252$ & $-0.082$ \\
                        & 1.484 & 25.047 $\pm$ 0.632 & $+0.542$ & $+0.699$ & $+0.841$ & $+0.861$ \\
                        & 2.330 & 25.543 $\pm$ 0.250 & $-0.054$ & $-0.061$ & $+0.388$ & $+0.067$ \\
\hline
\multirow{6}{*}{$z D_{\mathrm{H}}(z)/(r_{\mathrm{d}} \sqrt{z})$}
                        & 0.510 & 15.613 $\pm$ 0.306 & $-1.712$ & $-0.376$ & $-1.171$ & $-0.787$ \\
                        & 0.706 & 16.347 $\pm$ 0.281 & $-1.886$ & $-0.920$ & $-1.334$ & $-1.352$ \\
                        & 0.934 & 17.049 $\pm$ 0.340 & $+0.560$ & $+0.873$ & $+1.191$ & $+0.469$ \\
                        & 1.321 & 16.293 $\pm$ 0.480 & $+0.504$ & $-0.118$ & $+0.755$ & $-0.310$ \\
                        & 1.484 & 15.614 $\pm$ 0.631 & $-0.133$ & $-0.446$ & $-0.062$ & $-0.535$ \\
                        & 2.330 & 13.175 $\pm$ 0.250 & $-0.037$ & $-1.129$ & $-0.119$ & $-1.950$ \\
\hline\hline
\end{tabular}%
}
\end{table*}

\begin{figure}[htbp]
\includegraphics[width=0.5\linewidth]{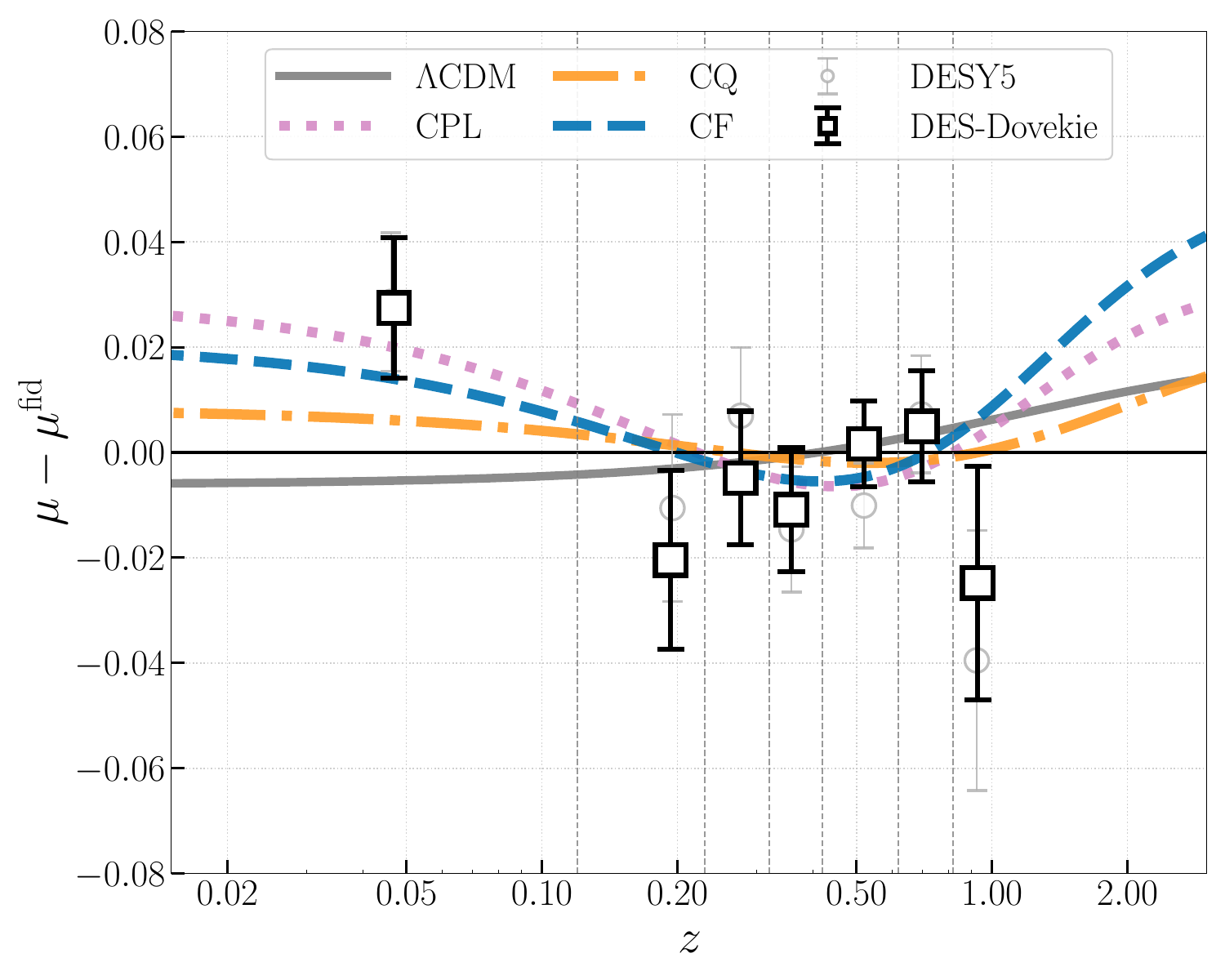}
\centering
\caption{\label{fig:SN_fit} Observables from the best-fit $\Lambda$CDM, CPL, CQ, and CF models are compared with the distance modulus $\mu$, normalized to the Planck 2018 best-fit $\Lambda$CDM cosmology. The binned distance modulus residuals for DES-Dovekie and DESY5 are shown, represented by black squares and gray circles, respectively. The bin edges for the SN bins are indicated by vertical gray dashed lines, and the SN binning method is described in Section IVC of Ref.~\cite{DESI:2025zgx}.} 
\end{figure}

More pronounced differences emerge when considering SN measurements. In Fig.~\ref{fig:SN_fit}, we show the binned distance modulus residuals, $\mu - \mu^{\rm fid}$, for the DESY5 SN sample and for the recently recalibrated DES-Dovekie sample (shown as gray circles and black squares, respectively), both normalized to the Planck 2018 best-fit $\Lambda$CDM cosmology. This recalibration has been shown to reduce the preference for dynamical DE in the context of the CPL parametrization. The figure therefore illustrates both the net effect of the recalibration on the binned SN measurements and its implications for the CPL, CQ, and CF models, whose best-fit predictions are shown relative to the normalized distance moduli. A direct comparison of the best-fit predictions shows that CPL and CF provide a better fit to the lowest-redshift bin at $z \simeq 0.05$ than the CQ model. This contributes to the more favourable values of $\Delta\chi^2_{\rm MAP}$ and $\Delta{\rm DIC}$ obtained for these two models, while CQ still performs better than $\Lambda$CDM. We also note that CQ appears to be closer to the SN measurements in the highest-redshift bin, around $z \sim 1$, compared to CPL, CF, and $\Lambda$CDM. However, this data point is affected by significantly larger uncertainties, and therefore this improvement carries less statistical weight than the low-redshift improvement observed for the other models. In the intermediate redshift range, $0.1 \lesssim z \lesssim 1$, all models remain competitive and provide comparably good fits to the SN data.

\subsection{Testing the effects of CMB lensing}\label{subsec:test_cmb_lensing}

\begin{table*}[htbp]
\centering
\caption{Fitting results (at the $1\sigma$ confidence level) in the $\Lambda$CDM, CPL, CQ, and CF models from the Planck-nl+DESI+DES-Dovekie, CMB-nl+DESI+DES-Dovekie, CMB-cut+DESI+DES-Dovekie, and CMB+DESI+DES-Dovekie+$f\sigma_8(z)$ data. $H_{0}$ is in units of ${\rm km}~{\rm s}^{-1}~{\rm Mpc}^{-1}$.}
\label{tab:newmany_params}
\setlength{\tabcolsep}{2mm}
\renewcommand{\arraystretch}{1.4}
\scriptsize
\begin{tabular}{l c c c c}
\hline \hline
Model/Parameters & \texttt{Planck-nl+DESI+DES-Dovekie} & \texttt{CMB-nl+DESI+DES-Dovekie} & \texttt{CMB-cut+DESI+DES-Dovekie} & \texttt{CMB+DESI+DES-Dovekie+$f\sigma_8(z)$} \\
\hline
$\bm{\Lambda}$\textbf{CDM} & & & & \\
$\Omega_\mathrm{b} h^2$ & $0.022310\pm 0.000120$ & $0.022484\pm 0.000092$ & $0.022486\pm 0.000094$ & $0.022479\pm 0.000091$ \\
$\Omega_\mathrm{c} h^2$ & $0.11772\pm 0.00062$ & $0.11792\pm 0.00062$ & $0.11790\pm 0.00063$ & $0.11821\pm 0.00058$ \\
$100\theta_\mathrm{*}$ & $1.04121\pm 0.00023$ & $1.04108\pm 0.00023$ & $1.04107\pm 0.00022$ & $1.04107\pm 0.00022$ \\
$\tau_\mathrm{reio}$ & $0.0549\pm 0.0077$ & $0.0585\pm 0.0044$ & $0.0584\pm 0.0045$ & $0.0598\pm 0.0043$ \\
$n_\mathrm{s}$ & $0.9684^{+0.0037}_{-0.0032}$ & $0.9734\pm 0.0030$ & $0.9737\pm 0.0030$ & $0.9729\pm 0.0029$ \\
$\log(10^{10} A_\mathrm{s})$ & $3.0380\pm 0.0160$ & $3.0515\pm 0.0098$ & $3.0510\pm 0.0100$ & $3.0553\pm 0.0082$ \\
$H_0$ & $68.13\pm 0.28$ & $68.16\pm 0.25$ & $68.17\pm 0.26$ & $68.05\pm 0.24$ \\
$\Omega_{\mathrm{m}}$ & $0.3031\pm 0.0036$ & $0.3037\pm 0.0035$ & $0.3035\pm 0.0035$ & $0.3052\pm 0.0032$ \\
$\sigma_8$ & $0.8022\pm 0.0069$ & $0.8092\pm 0.0044$ & $0.8089\pm 0.0045$ & $0.8115\pm 0.0034$ \\
$S_8$ & $0.8064\pm 0.0092$ & $0.8141\pm 0.0076$ & $0.8137\pm 0.0077$ & $0.8185\pm 0.0061$ \\
$r_\mathrm{d}$ [Mpc] & $147.77\pm 0.19$ & $147.53\pm 0.18$ & $147.53\pm 0.18$ & $147.46\pm 0.18$ \\
\hline
$\bm{\textbf{CPL}}$ & & & & \\
$\Omega_\mathrm{b} h^2$ & $0.022240\pm 0.000120$ & $0.022450\pm 0.000095$ & $0.022454\pm 0.000096$ & $0.022447\pm 0.000092$ \\
$\Omega_\mathrm{c} h^2$ & $0.11899\pm 0.00089$ & $0.11923\pm 0.00084$ & $0.11920\pm 0.00083$ & $0.11921\pm 0.00069$ \\
$100\theta_\mathrm{*}$ & $1.04107\pm 0.00024$ & $1.04096\pm 0.00023$ & $1.04096\pm 0.00023$ & $1.04098\pm 0.00023$ \\
$\tau_\mathrm{reio}$ & $0.0524\pm 0.0076$ & $0.0565\pm 0.0044$ & $0.0564\pm 0.0044$ & $0.0566\pm 0.0044$ \\
$n_\mathrm{s}$ & $0.9656\pm 0.0037$ & $0.9707\pm 0.0031$ & $0.9709\pm 0.0032$ & $0.9707\pm 0.0030$ \\
$\log(10^{10} A_\mathrm{s})$ & $3.0350\pm 0.0160$ & $3.0497\pm 0.0098$ & $3.0494\pm 0.0098$ & $3.0481\pm 0.0085$ \\
$H_0$ & $67.34\pm 0.55$ & $67.45\pm 0.54$ & $67.41\pm 0.55$ & $67.36\pm 0.53$ \\
$\Omega_{\mathrm{m}}$ & $0.3130\pm 0.0054$ & $0.3129\pm 0.0053$ & $0.3132\pm 0.0054$ & $0.3136\pm 0.0052$ \\
$w_0$ & $-0.812\pm 0.056$ & $-0.803\pm 0.056$ & $-0.802\pm 0.056$ & $-0.799\pm 0.054$ \\
$w_a$ & $-0.69^{+0.24}_{-0.21}$ & $-0.74^{+0.23}_{-0.20}$ & $-0.74^{+0.23}_{-0.20}$ & $-0.74^{+0.21}_{-0.19}$ \\
$\sigma_8$ & $0.8070\pm 0.0100$ & $0.8146\pm 0.0088$ & $0.8139\pm 0.0085$ & $0.8130\pm 0.0066$ \\
$S_8$ & $0.8240\pm 0.0110$ & $0.8319\pm 0.0095$ & $0.8315\pm 0.0094$ & $0.8313\pm 0.0069$ \\
$r_\mathrm{d}$ [Mpc] & $147.52\pm 0.22$ & $147.22\pm 0.22$ & $147.22\pm 0.22$ & $147.23\pm 0.19$ \\
\hline
$\bm{\textbf{CQ}}$ & & & & \\
$\Omega_\mathrm{b} h^2$ & $0.022170\pm 0.000140$ & $0.022400\pm 0.000100$ & $0.022410\pm 0.000100$ & $0.022408\pm 0.000096$ \\
$\Omega_\mathrm{c} h^2$ & $0.1148\pm 0.0013$ & $0.1150\pm 0.0013$ & $0.1150\pm 0.0013$ & $0.1150\pm 0.0013$ \\
$100\theta_\mathrm{*}$ & $1.04099\pm 0.00026$ & $1.04089\pm 0.00024$ & $1.04090\pm 0.00024$ & $1.04090\pm 0.00023$ \\
$\tau_\mathrm{reio}$ & $0.0528\pm 0.0078$ & $0.0572\pm 0.0046$ & $0.0571\pm 0.0046$ & $0.0575\pm 0.0045$ \\
$n_\mathrm{s}$ & $0.9657\pm 0.0039$ & $0.9706\pm 0.0033$ & $0.9709\pm 0.0033$ & $0.9706\pm 0.0030$ \\
$\log(10^{10} A_\mathrm{s})$ & $3.0380\pm 0.0160$ & $3.0540\pm 0.0100$ & $3.0530\pm 0.0100$ & $3.0533\pm 0.0085$ \\
$H_0$ & $67.80\pm 0.55$ & $68.01\pm 0.54$ & $67.99\pm 0.56$ & $67.91\pm 0.53$ \\
$\Omega_{\mathrm{m}}$ & $0.2995\pm 0.0052$ & $0.2986\pm 0.0051$ & $0.2987\pm 0.0052$ & $0.2993\pm 0.0050$ \\
$\alpha$ & $0.47\pm 0.18$ & $0.46\pm 0.18$ & $0.46\pm 0.19$ & $0.48\pm 0.17$ \\
$\beta$ & $0.0503^{+0.0130}_{-0.0082}$ & $0.0520^{+0.0110}_{-0.0078}$ & $0.0514^{+0.0110}_{-0.0080}$ & $0.0514^{+0.0099}_{-0.0073}$ \\
$\sigma_8$ & $0.817\pm 0.013$ & $0.827\pm 0.012$ & $0.826\pm 0.012$ & $0.8250\pm 0.0082$ \\
$S_8$ & $0.8160\pm 0.0110$ & $0.8253\pm 0.0095$ & $0.8243\pm 0.0095$ & $0.8241\pm 0.0065$ \\
$r_\mathrm{d}$ [Mpc] & $147.40\pm 0.28$ & $147.04^{+0.28}_{-0.26}$ & $147.06\pm 0.28$ & $147.07\pm 0.23$ \\
\hline
$\bm{\textbf{CF}}$ & & & & \\
$\Omega_\mathrm{b} h^2$ & $0.022280\pm 0.000120$ & $0.022465\pm 0.000092$ & $0.022468\pm 0.000097$ & $0.022458\pm 0.000093$ \\
$\Omega_\mathrm{c} h^2$ & $0.242^{+0.028}_{-0.035}$ & $0.261^{+0.028}_{-0.024}$ & $0.265\pm 0.027$ & $0.256\pm 0.028$ \\
$100\theta_\mathrm{*}$ & $1.04105\pm 0.00025$ & $1.04097\pm 0.00023$ & $1.04097\pm 0.00023$ & $1.04096\pm 0.00022$ \\
$\tau_\mathrm{reio}$ & $0.0522^{+0.0079}_{-0.0071}$ & $0.0562\pm 0.0045$ & $0.0553^{+0.0040}_{-0.0048}$ & $0.0563\pm 0.0046$ \\
$n_\mathrm{s}$ & $0.9662\pm 0.0037$ & $0.9708\pm 0.0031$ & $0.9710\pm 0.0031$ & $0.9712\pm 0.0031$ \\
$\log(10^{10} A_\mathrm{s})$ & $3.0360^{+0.0170}_{-0.0150}$ & $3.0480\pm 0.0100$ & $3.0462^{+0.0092}_{-0.0100}$ & $3.0453\pm 0.0089$ \\
$H_0$ & $67.37^{+0.57}_{-0.49}$ & $67.45\pm 0.53$ & $67.39\pm 0.53$ & $67.42\pm 0.53$ \\
$\Omega_{\mathrm{m}}$ & $0.584^{+0.064}_{-0.080}$ & $0.625\pm 0.059$ & $0.635\pm 0.063$ & $0.613\pm 0.062$ \\
$w$ & $-1.44^{+0.23}_{-0.12}$ & $-1.55^{+0.19}_{-0.14}$ & $-1.58^{+0.22}_{-0.17}$ & $-1.51^{+0.20}_{-0.14}$ \\
$\beta$ & $-2.24^{+1.10}_{-0.54}$ & $-2.74^{+0.91}_{-0.69}$ & $-2.90^{+1.10}_{-0.80}$ & $-2.59^{+0.95}_{-0.68}$ \\
$\sigma_8$ & $0.494^{+0.037}_{-0.044}$ & $0.474^{+0.020}_{-0.035}$ & $0.469^{+0.024}_{-0.036}$ & $0.480^{+0.026}_{-0.040}$ \\
$S_8$ & $0.685^{+0.014}_{-0.018}$ & $0.681^{+0.009}_{-0.013}$ & $0.679^{+0.009}_{-0.013}$ & $0.682^{+0.011}_{-0.016}$ \\
$r_\mathrm{d}$ [Mpc] & $147.54\pm 0.22$ & $147.25\pm 0.21$ & $147.25\pm 0.22$ & $147.28\pm 0.21$ \\
\hline \hline
\end{tabular}
\end{table*}

In our baseline analysis, we compute the lensing smoothing of the CMB temperature and polarization spectra, as well as the theoretical predictions for the CMB lensing power spectrum, with \texttt{IDECAMB} based on the background and linear perturbation evolution of a given IDE model. We emphasize that existing halo model type nonlinear prescriptions are mainly calibrated against numerical simulations of $\Lambda$CDM or smooth dynamical DE cosmologies, and have not been specifically calibrated for the interacting dark-sector scenarios investigated in this work. Therefore, for IDE models, any use of standard nonlinear prescriptions should be regarded as a potential theoretical caveat, rather than as a dedicated validation of nonlinear structure formation in IDE cosmologies. To this end, we further assess the potential impact of CMB lensing and small-scale matter power spectrum modeling on the IDE parameter constraints.

To quantitatively test the specific contribution of CMB lensing and high-$\ell$ data to the evidence supporting dark-sector interactions, we perform detailed robustness tests below by ablating and comparing different CMB datasets. First, we remove the CMB lensing power spectrum likelihood from the baseline CMB data combination, denoting this case as \texttt{CMB-nl}. Second, based on the exclusion of the CMB lensing power spectrum, we further restrict the CMB temperature and polarization spectra to $\ell < 3000$ to reduce the sensitivity to small-scale lensing smoothing, high-$\ell$ data, and related modeling uncertainties~\cite{Lewis:2006fu}; this combination is denoted as \texttt{CMB-cut}. Finally, as a more conservative test, we use solely the Planck data without the CMB lensing power spectrum, thereby simultaneously excluding the high-$\ell$ measurements from ACT and SPT; this case is denoted as \texttt{Planck-nl}.

\begin{figure*}[htbp]
\includegraphics[scale=0.45]{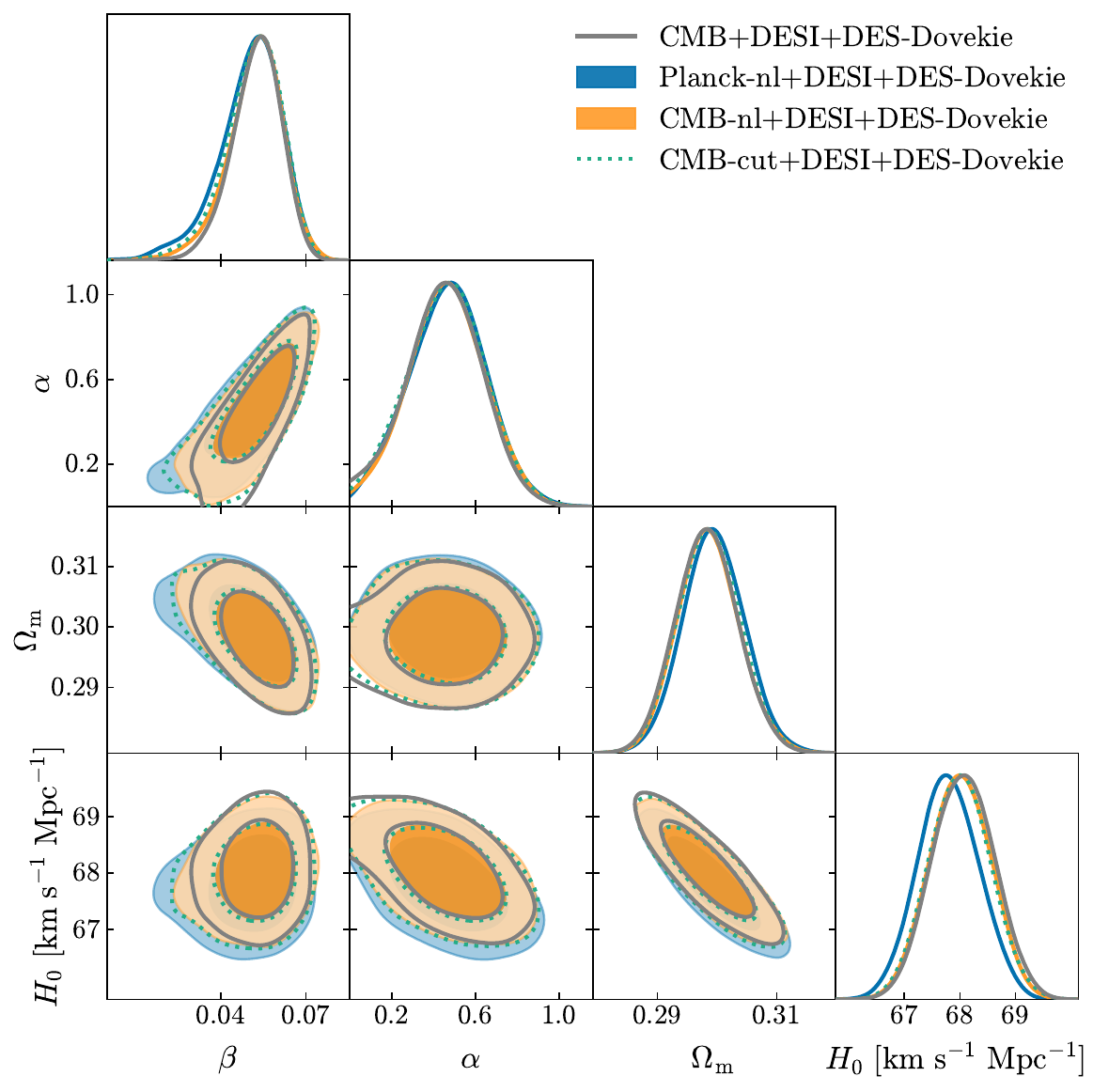}
\includegraphics[scale=0.45]{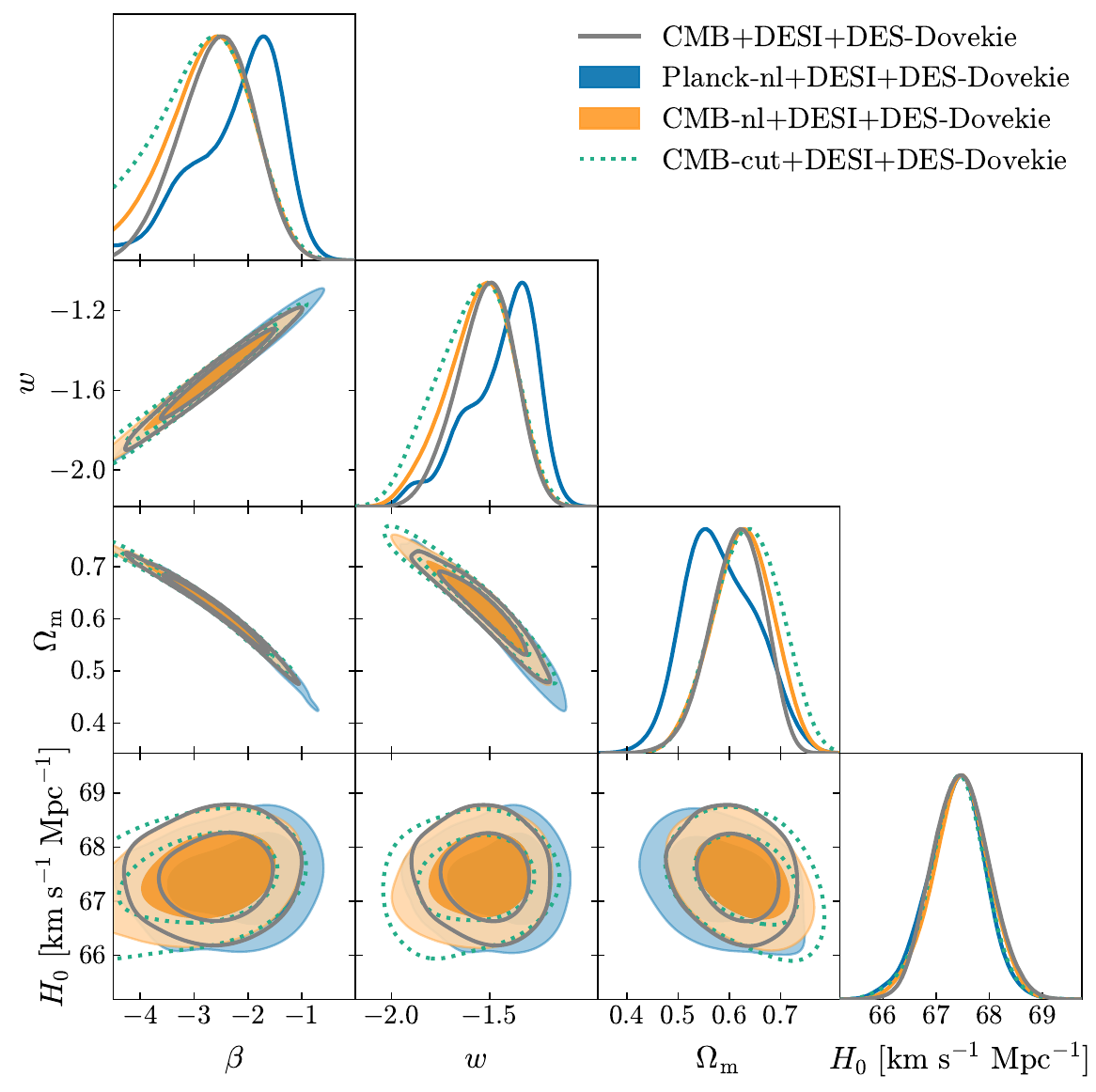}
\centering
\caption{Constraints on cosmological parameters from the CMB+DESI+DES-Dovekie, Planck-nl+DESI+DES-Dovekie, CMB-nl+DESI+DES-Dovekie, and CMB-cut+DESI+DES-Dovekie data for the CQ (left panel) and CF (right panel) models. \label{fig:lensing} }
\end{figure*}

The corresponding parameter constraints are summarized in Table~\ref{tab:newmany_params} and shown in Fig.~\ref{fig:lensing}. For the CQ model, the posterior distributions obtained using these reduced CMB data combinations remain very close to the baseline analysis results. In all three test cases, we consistently obtain a $>5\sigma$ preference for a nonzero interaction, with $\beta = 0.0520^{+0.0110}_{-0.0078}$, $\beta = 0.0514^{+0.0110}_{-0.0080}$, and $\beta = 0.0503^{+0.0130}_{-0.0082}$ for the CMB-nl+DESI+DES-Dovekie, CMB-cut+DESI+DES-Dovekie, and Planck-nl+DESI+DES-Dovekie data, respectively. This indicates that the preference for a nonzero interaction in the CQ model is not driven by the CMB lensing likelihood or high-$\ell$ data. For the CF model, the impact of these data ablations is more pronounced, which is consistent with its more extreme late-time structure growth properties. Nevertheless, even in the most conservative Planck-nl+DESI+DES-Dovekie data, the interaction parameters remain displaced from the noninteracting limit, yielding $\beta = -2.24^{+1.10}_{-0.54}$ and $w = -1.44^{+0.23}_{-0.15}$. Thus, although the statistical preference is weakened relative to the baseline analysis, the preference for a nonzero interaction does not disappear.

As an additional cross-check, we repeated the CQ analysis using the reduced Planck+DESI+PantheonPlus data combination, matching the setup considered in Ref.~\cite{Gomez-Valent:2026ept}. We obtain $\beta=0.0500^{+0.0120}_{-0.0084}$ and $\alpha=0.41\pm0.20$, showing that the preference for a nonzero coupling is already present in this conservative dataset, without relying on the ACT/SPT high-$\ell$ measurements, the CMB lensing likelihood, or the specific SN compilation adopted in our baseline analysis.\footnote{\parbox[t]{\linewidth} {In comparing with Ref.~\cite{Gomez-Valent:2026ept}, one should also account for implementation-level differences, including the scalar-field potential, the treatment of initial conditions, and the coupling-prior choice. Our analysis follows the \texttt{IDECAMB} conventions, whose implementation has been benchmarked against the IDE results of the Planck collaboration~\cite{Li:2023fdk}.}} For the CF model, the reduced-data constraints are broader, as expected given its stronger impact on late-time growth, but the interaction parameters remain displaced from the noninteracting limit.

In summary, the current preference for dark-sector interactions is not solely produced by the CMB lensing power spectrum likelihood or high-$\ell$ CMB data. At the same time, future nonlinear calibrations based on dedicated IDE numerical simulations are still required to achieve a fully self-consistent treatment of nonlinear structure formation in IDE models, especially for scenarios with extreme parameter regions such as the CF model.

\subsection{Testing the effects of $f\sigma_8(z)$ data}\label{subsec:test_fsigma8}

As discussed in Sec.~\ref{additional:par}, since the CF model prefers an unusual region of parameter space, characterized by a large present-day matter density and a strongly suppressed amplitude of structure growth, it is important to test whether this scenario is compatible with direct low-redshift growth measurements. We therefore include a compilation of $f\sigma_8(z)$ measurements from redshift-space distortion observations and assess their impact on the IDE constraints. The $f\sigma_8(z)$ data are compiled from various surveys and related literature as summarized in Ref.~\cite{Avila:2022xad}, yielding a total of 20 measurements over the redshift range $0.02<z<1.944$. The specific selection criteria for these data are detailed in Ref.~\cite{Sabogal:2024yha}.

\begin{figure*}[!htbp]
\includegraphics[scale=0.45]{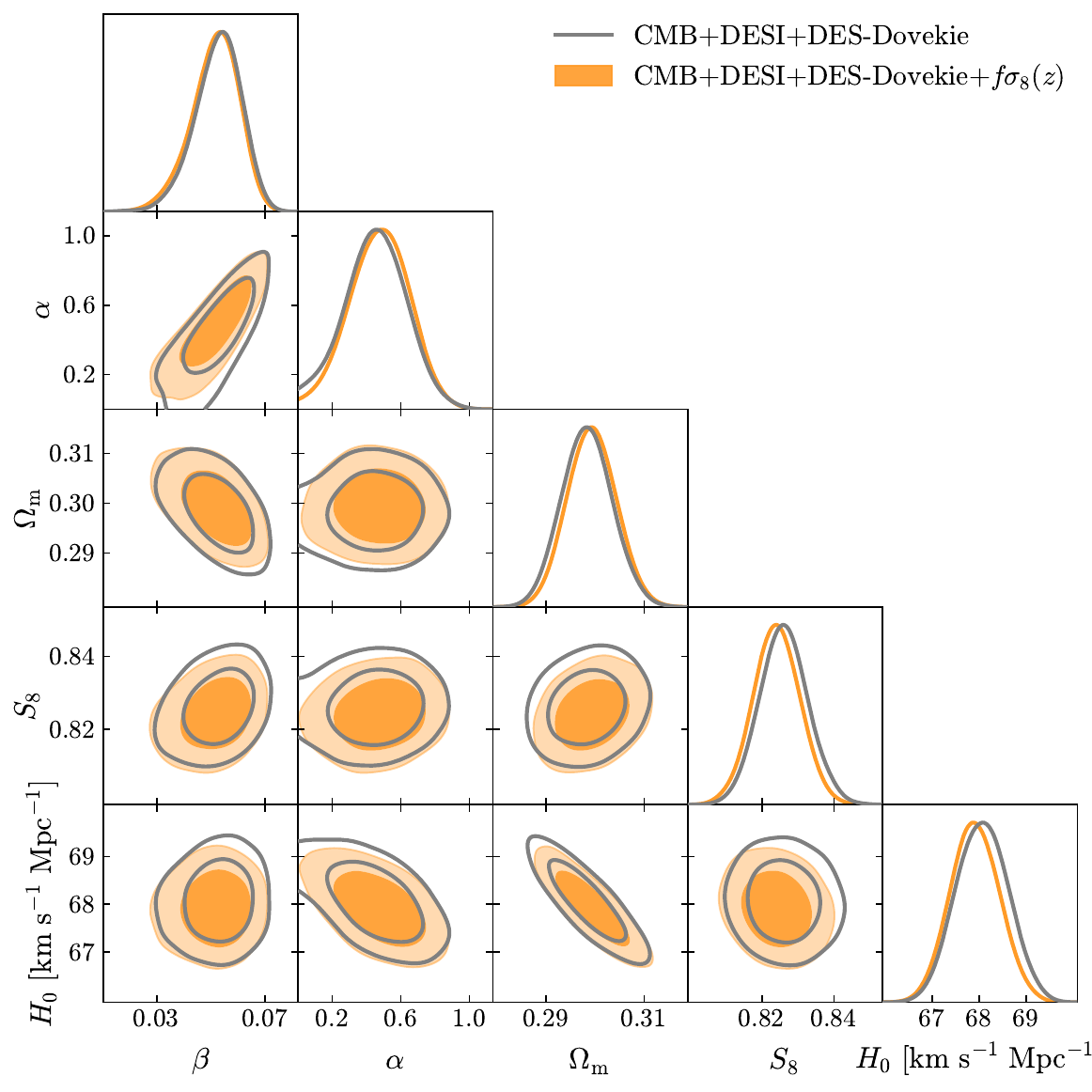}
\includegraphics[scale=0.45]{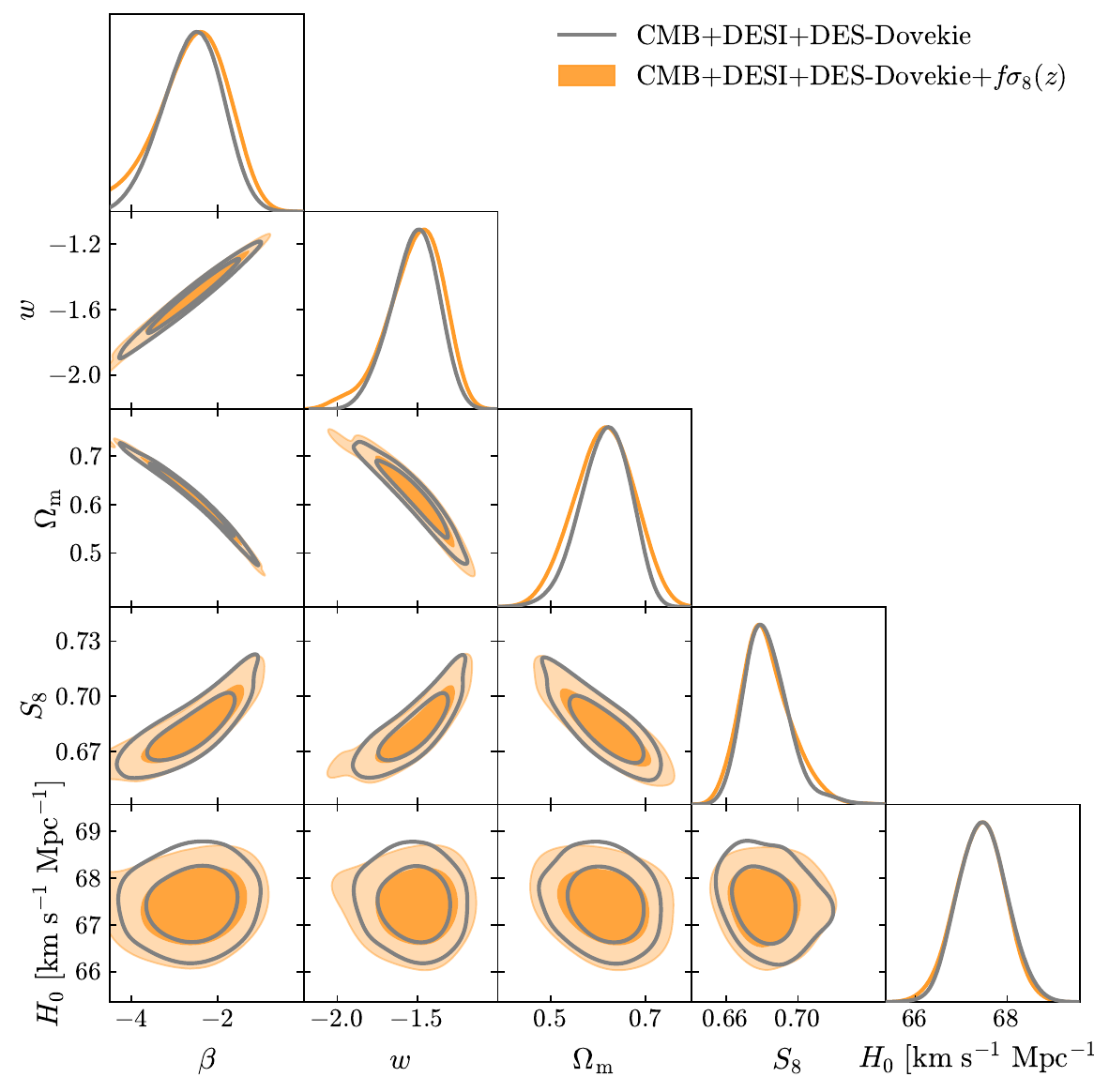}
\centering
\caption{Constraints on cosmological parameters from the CMB+DESI+DES-Dovekie and CMB+DESI+DES-Dovekie+$f\sigma_8(z)$ data for the CQ (left panel) and CF (right panel) models. \label{fig:fsigma8} }
\end{figure*}

The constraints on the CQ and CF models obtained from the CMB+DESI+DES-Dovekie+$f\sigma_8(z)$ data are shown in Fig.~\ref{fig:fsigma8} and summarized in Table~\ref{tab:newmany_params}. For the CQ model, the inclusion of $f\sigma_8(z)$ data has a negligible impact on the posterior distributions. The constraints remain very close to those obtained from the CMB+DESI+DES-Dovekie data combination alone, with $\alpha=0.48\pm0.17$ and $\beta=0.0514^{+0.0099}_{-0.0073}$. This shows that the preference for a nonzero interaction in the CQ scenario %, which exceeds the $5\sigma$ level, 
remains stable after including $f\sigma_8(z)$ measurements.

For the CF model, the inclusion of $f\sigma_8(z)$ data also does not significantly alter the parameter constraints inferred from the combined data. We obtain $w=-1.51^{+0.20}_{-0.14}$ and $\beta=-2.59^{+0.95}_{-0.68}$, in good agreement with the baseline constraints without $f\sigma_8(z)$. The inferred values of $\Omega_{\rm m}=0.613\pm0.062$, $\sigma_8=0.480^{+0.026}_{-0.040}$, and $S_8=0.682^{+0.011}_{-0.016}$ also remain in the same region as in the baseline analysis. Thus, current growth-rate data do not significantly shift the global parameter constraints of the CF model.

It is important to emphasize, however, that after including $f\sigma_8(z)$ data, the CF model provides a poorer fit to the growth-rate measurements themselves than $\Lambda$CDM, CPL, and CQ. This can be seen from the individual minimum-$\chi^2$ contributions from the $f\sigma_8(z)$ data alone, which are $\chi^2_{f\sigma_8(z)} = 9.03$, $9.02$, $9.55$, and $10.03$ for the $\Lambda$CDM, CPL, CQ, and CF models, respectively. Nevertheless, in the CMB+DESI+DES-Dovekie+$f\sigma_8(z)$ analysis, this degradation in the fit to the growth-rate data is not sufficient to offset the improvement achieved by the CF model in fitting the CMB, BAO, and SN data. Specifically, for the full data combination, the CF model still yields a significant overall improvement relative to $\Lambda$CDM, with $\Delta\chi^2 = -15.77$ and $\Delta\text{DIC} = -11.24$. Therefore, while current growth-rate measurements provide a useful test of the CF model, they are not sufficient to exclude the interaction region preferred in the combined analysis.

\begin{figure}[htbp]
\includegraphics[width=0.6\linewidth]{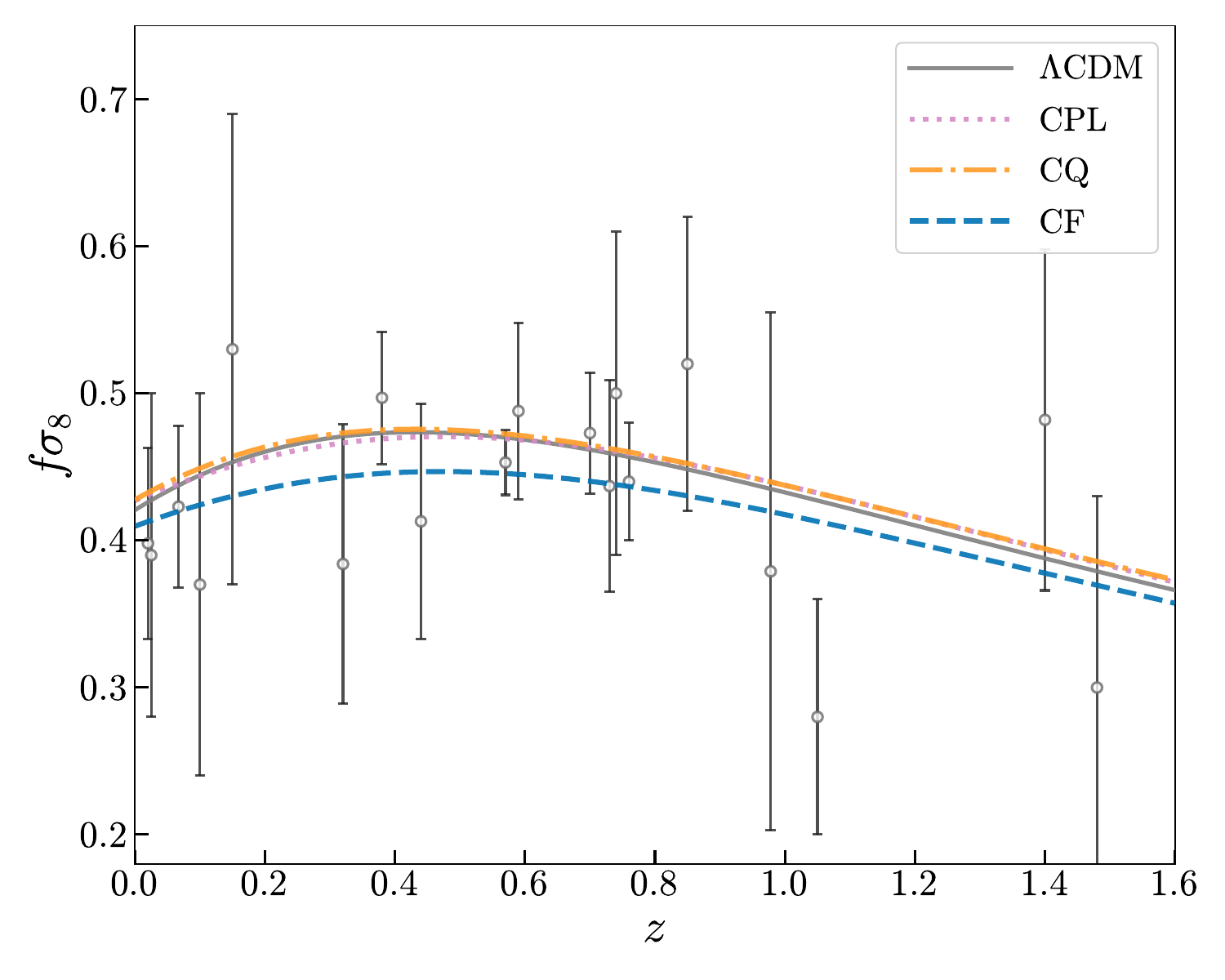}
\centering
\caption{\label{fig:fsigma8_fit} The theoretical predictions for $f\sigma_8(z)$ from the $\Lambda$CDM, CPL, CQ, and CF models, along with the compilation of $f\sigma_8(z)$ observational measurements. The theoretical curves are generated using the best-fit parameters derived from the CMB+DESI+DES-Dovekie+$f\sigma_8(z)$ data.}
\end{figure}

To further illustrate the relation between these models and the $f\sigma_8(z)$ data, we show in Fig.~\ref{fig:fsigma8_fit} the theoretical predictions for $f\sigma_8(z)$ in the $\Lambda$CDM, CPL, CQ, and CF models, together with the observational measurements. The CQ prediction remains close to those of $\Lambda$CDM and CPL, following the standard-growth branch over the redshift range probed by the data. By contrast, the CF model predicts substantially lower values of $f\sigma_8(z)$, reflecting its suppressed growth of structure. This explains why the CF model provides a poorer fit to the $f\sigma_8(z)$ data alone. However, the current precision and scatter of these measurements are not sufficient to exclude this low-growth branch in the combined analysis. This distinctive feature provides a clear target for future tests with growth-rate and large-scale-structure observations.

We conclude that current $f\sigma_8(z)$ measurements do not qualitatively change the IDE constraints. In particular, after including these growth-rate data, the combined data still prefer a nonzero dark-sector interaction in both the CQ and CF models. At the same time, $f\sigma_8(z)$ measurements provide an important low-redshift test of IDE scenarios, especially for the CF model, whose large $\Omega_{\mathrm{m}}$ and suppressed structure growth lead to distinctive predictions for the growth history. The CF model gives a slightly poorer fit to the current $f\sigma_8(z)$ data than $\Lambda$CDM, CPL, and CQ, but the present precision of these measurements is not sufficient to exclude the interaction region preferred by the combined CMB, BAO, and SN data. Future low-redshift structure probes, including redshift-space distortions, cosmic shear, lensing, and galaxy clustering, will therefore be crucial for testing this scenario. A fully self-consistent use of these probes in IDE cosmologies will require dedicated modeling of nonlinear structure formation, galaxy bias, baryonic effects, intrinsic alignments, and the modified perturbation evolution induced by the dark-sector interaction, which we leave for future work.

\end{document}